\documentclass[twocolumn,prl,superscriptaddress,floatfix,showpacs]{revtex4-1}
\usepackage[utf8]{inputenc}
\usepackage{amsmath}
\usepackage{amssymb}
\usepackage{graphicx}

\makeatletter
\usepackage{graphics}
\usepackage{epsfig}
\usepackage{color}

\def\cly{\color{red}}

\usepackage[normalem]{ulem}
\newcommand{\be}{\begin{equation}} \newcommand{\ee}{\end{equation}}
\newcommand{\bea}{\begin{eqnarray}} \newcommand{\eea}{\end{eqnarray}}

\makeatother

\begin{document}

\title{The Oslo model, hyperuniformity, and the quenched Edwards-Wilkinson model}

\author{Peter Grassberger} \affiliation{JSC, FZ J\"ulich, D-52425 J\"ulich, Germany}
\author{Deepak Dhar} \affiliation{Tata Institute for Fundamental Research, Mumbai, India} 
\author{P. K. Mohanty} \affiliation{Condensed Matter Physics  Division, Saha Institute of Nuclear Physics, Kolkata, India}


\begin{abstract}
We present simulations of the 1-dimensional Oslo rice pile model in which the critical height at each site is 
randomly reset after each toppling. We use the fact that the stationary state  of this sandpile model  is
hyperuniform 
to reach system  of sizes $ > 10^7$. Most previous simulations were seriously flawed by important finite size 
corrections. We find  that all critical exponents have values consistent with simple rationals: $\nu=4/3$ 
for the correlation length 
exponent, $D =9/4$ for the fractal dimension of avalanche clusters, and $z=10/7 $ for the dynamical exponent. In 
addition  we relate the hyperuniformity exponent to the correlation length exponent $\nu$. Finally we discuss the relationship with the quenched Edwards-Wilkinson (qEW) model, where we 
find in particular that the local roughness exponent is $\alpha_{\rm loc} = 1$.  
\end{abstract}
\maketitle

\section{Introduction}

Although self-organized critical (SOC) sandpile models \cite{Bak:1987, Ivash:1994, Ktit:2000, ddrev, redig, propp} have been studied 
intensively during the last thirty years, many of their aspects are still not well
understood. For example,  the critical exponents of avalanche distributions in the original Bak-Tang-Wiesenfeld sandpile model, on the square lattice, are still not known. The question of universality classes of different sandpile models is also not well understood \cite{Benhur,Tebaldi:1999}.  It is realized that models with stochastic toppling rules  \cite{Manna,Oslo,Zhang,Maslov-Zhang,Rossi,Mohanty:2002,Mohanty:2007} are in a different universality class than those with deterministic toppling rules. In particular, strong indications were found in \cite{BBBMH} that models with stochastic toppling rules and with continuous local 
stresses (sandpile heights) behave differently from those with discrete ones. On the other hand, as already noted by Tang and 
Bak \cite{Tang,Tang2,Grass-Manna}, 
for all these models exist also non-self organized versions, now called fixed-energy sandpiles,  that should show 
conventional (co-dimension one) critical points. The  fixed-energy sandpiles (FES)  undergo an active-absorbing state transition as a function of the mean density of particles. An important question has been if this transition is in the universality class of directed percolation.  This question is not clearly understood yet \cite{Manna,Oslo,Zhang,Maslov-Zhang,Rossi,Mohanty:2002,Mohanty:2007}.  In FES, the number of absorbing states grows exponentially  with the size of system.  This alone would 
not create a problem, as it is known that models with many absorbing states can
still be in the DP universality class -- provided they do not have  too-long-ranged 
correlations \cite{Hinrichsen}.

One problem in numerical studies is 
precisely the long-ranged correlations in the absorbing states at criticality,
called in the following ``natural critical states" (NCS). A straightforward 
strategy seems to consist in studying states remaining after large 
avalanches have died, in systems poised to the critical point. But this is not 
possible, since it is  not feasible to wait until avalanches die on very large
systems (the average CPU time per avalanche diverges with system size). Thus one has 
to do some tricks that -- unless one is sufficiently careful -- can introduce
spurious correlations in the NCS. While this problem was known quite early 
\cite{footnote3}, the first papers that tried to deal with it carefully and 
systematically \cite{Mohanty:2002,Mohanty:2007,BBBMH} were published rather recently. 
They  indicated that there exists  in fact a universality class of stochastic sandpile 
models, but it seemed to be identical 
with the DP universality class. In fact, two of us have given heuristic arguments earlier, but no proof \cite{Mohanty:2002,Mohanty:2007}, that stochastic
sandpile models in the Manna universality class will flow into the DP universality class if we add  an appropriate perturbation.

It is the purpose of the present paper to clarify the situation somewhat.  We study in detail the one-dimensional 
Oslo model \cite{oslo1}, which is one of the simplest nontrivial stochastic sandpile models. It has stochasticity in the toppling rules, 
and the critical height at each site is randomly reset after each toppling. Thus, it may be said that there is a degree of "stickiness" in the model. While the model has some interesting properties due to its unusual algebraic structure, its steady state and critical properties are not known exactly so far \cite{ddoslo}. 
We will   study the behavior of other directed Oslo-type sandpile models on the 2-dimensional square lattice in a forthcoming paper.  Here we study the 1-dimensional  Oslo model using numerical  simulations of much larger systems (and with much higher statistics) than what had 
been possible previously. As we said, simulations of FES {\it at the critical point} 
are hampered by the difficulty   of sampling from the correct 
NCS. On the other hand, precise simulations of the SOC versions are difficult, because 
the open boundary conditions lead to large finite-size corrections, unless one can 
simulate huge systems. The latter, however, is made difficult by very long transients
(during which the proper NCS has to build up). As a consequence, the largest published 
simulations of the 1-$d$ Oslo model are for systems of size $\approx 20,000$. Without 
the transients, systems larger by one or two units of magnitude would be easy to 
simulate on modern computers.

Our large-scale simulations are made possible by two technical improvements: (i) We use 
a new method of triggering avalanches in the FES that preserves all NCS correlations;
(ii) We  use initial configurations which are close to NCS 
configurations to reduce the time required to reach the NCS state.

Crucial for the latter is the observation, made first in \cite{BBBMH} and verified later 
in \cite{Hexner,Lee:2014,Dick:2015}, that NCS's of   some SOC models are `hyperuniform' \cite{Torquato,Gabrielli}.
 Consider a statistically stationary random point process on a line. Then, so long as correlations in the system die sufficiently fast with distance, using Gauss' central limit theorem,  the variance of the number $n_L$ of points in an
interval of size $L$, ${\rm Var}[n_L] \sim L$. In contrast, a periodic
distribution would have variance ${\rm Var}[n_L] \sim const$.   A point process on a line
is called hyperuniform, if the variance falls between these two limits, more precisely
\be 
    {\rm Var}[n_L] \sim L^\zeta     \label{hyperunif}
\ee
with  hyperuniformity exponent $0 < \zeta < 1$.

Notice that Eq. (\ref{hyperunif}) implies negative long range correlations, and it would 
be non-trivial to choose initial conditions which satisfy it exactly (with the correct
exponent $\zeta$), but this is not really needed. It will turn out that it is sufficient 
to use initial conditions which have (a) the right density, and (b) variances much smaller
than those for random distributions. We shall use periodic initial conditions with long 
periods (typically $\gtrsim 10^2$) which are carefully chosen so that the density is close to 
the measured one of the NCS, the period is as small as possible for the given density, 
and the distribution within one period is as uniform as possible.    We note that hyperuniformity is not a generic property of all sandpile models. While the one-dimensional undirected sandpile model does show hyperuniformity, the steady state of the prototypical BTW model on a square lattice, slowly-driven by random particle additions does not. 

Our results can be summarized very succinctly: The 1-d Oslo model is clearly not in the 
DP universality class. It is in the qEW class, and our estimates for the critical exponents 
$ \nu =4/3, D =9/4, z=10/7$
are more precise than all previous estimates for any model in the Manna and/or qEW classes.
They strongly indicate that $\it all$ critical exponents are simple rationals. Finally, we 
have clear evidence that the SOC and FES versions of the 1-d Oslo model are related to each 
other trivially, while this is still debated for the BTW model \cite{Fey1}.

In the next section we define the model and its variants -- distinguished by boundary conditions
and ways of driving. In section III, we give some  simulations details. In Sec.~IV,  we present
the main numerical data for the determination of numerical exponents of the model. In Sec.~V, we discuss the relationship to the quenched Edwards-Wilkinson model.  Sec.~VI contains  a summary of our results, and  some concluding remarks.

\section{The model and its variants}

\subsection{The original version: Open b.c. and boundary driven}

The Oslo model was invented to mimic a one-dimensional pile of non-spherical particles (rice)
\cite{Oslo}.  In the original version, particles are added one after the other at the left end 
(which is closed), so that they pile up until they fall off from the open right end. Actually, 
as we shall see, it is more convenient to formulate it entirely in terms of local slopes,
and to disregard completely the actual height of the pile. The reason is that we shall discuss
later (in Sec. \ref{surf}) a completely different interface associated with the 
local slopes, and we do not want to confuse the original height profile of the pile with it.

Because the slopes of the original pile will turn out to be {\it not} the slopes of the 
new interface, we will also change notation (even if this might look confusing at   first) 
and speak of ``stresses" instead of slopes.

Formally, the model is a one-dimensional cellular automaton where an integer $z_i \geq 0$
(the local stress) is attached to each site $i\in {1,2,\ldots L}$.    Each site has a  
threshold stress ${\color{blue} z^*},$  which   can be  either  $2$ or $3;$  
sites  $i$ with $z_i <  z^*$   are called {\it stable}, whereas those with $z_i \ge  z^*$ 
are unstable.   Initially at $t=0,$  $ z^*$  at  different  sites are chosen  (as $2$ or $3$) 
randomly and independently. Unstable sites immediately  `topple'  and reset  their  
threshold values.  For   sites  $1 < i < L$, toppling occurs as  
\be 
   z_i \to z_i-2, z_{i\pm 1} \to z_{i\pm 1}+1.  \label{toppl}
\ee
This corresponds to moving a single grain of rice from top of the pile at site $i$ to site  $i+1$.
At the boundaries, i.e. for $i=1$ and $i=L$, only the appropriate neighbor gets increased,
and the unit of stress that would go to $i= 0$ resp. $i=L+1$ gets lost \cite{footnote4}.
  It is easily verified that in this stochastic model, topplings still  have the abelian property \cite{ddrev}.

\subsection{Boundary- and bulk- driving}

 \begin{figure}
 \begin{centering}
 \includegraphics[scale=0.3]{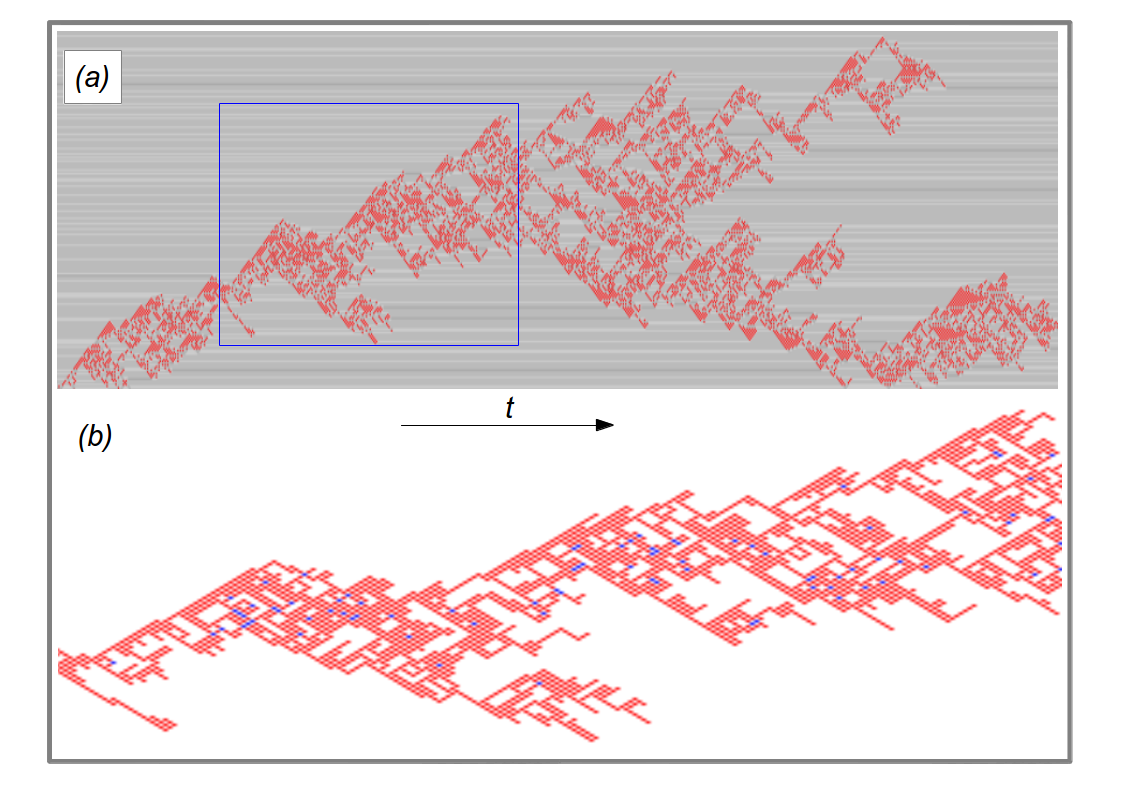}
 \par\end{centering}
 \caption{\label{ava} (Color online)  A  typical avalanche  of  boundary driven  Oslo model (time  flows  towards  right). 
 (a) The stress  profile: $z_i\le2$  are   marked  as gray (different shades) and sites with $z_i>2$ are marked 
 as red.  (b) A  portion of the avalanche,  marked as a  rectangle in (a), is zoomed to  view     single (red) and multiple (blue) topplings. }
\end{figure}

In the original version, the model is driven by adding grains of rice at the left boundary.
In terms of stresses this means that the system is driven by increasing $z_1$ by one unit.
If this leads to an instability, the entire avalanche of topplings is done before $z_1$ is 
increased again.   A typical avalanche in the  boundary driven case, starting from a single seed is shown in Fig. \ref{ava}.

We say that the pile is  bulk-driven, when we choose a random site $i\in ]1,L[$ and increase its 
stress by 1 unit. Notice that this would be a somewhat unusual drive in a real rice pile:
It would correspond to adding 1 rice grain at sites $1,2, \ldots i$ each.  We expect that avalanche size distributions
for bulk-driving will be different from those for the boundary-driven case, but some critical exponents like $D$ 
and $z$ (defined below) would be the same. 

In addition, we expect that finite size corrections will be very different. For boundary 
driving, the only large length scale is the distance $L$ to the far boundary. For bulk
driving,   another length scale comes into play: the (random)  distance of point of addition from the boundary. This is,  of course, averaged over, but typically gives rise to  much larger 
finite size corrections.

\subsection{Fixed energy version}

Finally, we shall also consider the  FES version with periodic boundary conditions. In that case no stress
can get lost. If we drive the system by adding stress, we sooner or later must reach
the critical point where avalanches never stop. On the one hand, this is the cleanest case
because finite size corrections are minimal. On the other hand, as pointed out in the 
introduction, simulations at the critical point are not trivial in this version.

In the subcritical case simulations are rather straightforward: starting with any 
initial configuration with $\langle z\rangle = Z/L < z_c$, we follow the avalanche 
(if at least one site is unstable) until it dies. After that, all sites are stable.
If $\langle z\rangle$ is sufficiently close to $z_c$, there will be some sites with 
$z_i=2$. We now trigger a new avalanche by declaring one (or several) of these sites
as unstable (if no site with $z_i=2$ exists, we increase $Z$, until we are close enough 
to the critical point). 

Notice that declaring a stable site with $z_i=2$ does not alter the NCS, hence we 
do not expect to encounter the problems mentioned in \cite{footnote3}. 

Simulations are equally straightforward in the supercritical case, where the above
procedure soon leads to an infinite avalanche. As in the BTW case \cite{Fey1}, an 
avalanche will not stop after each site has toppled once, and this will happen in 
general after $\ll O(L)$ time steps. 

On the other hand, following avalanches on large lattices until they die is not a viable 
option at the critical point, because avalanches  may not  die even after very many time
steps. In that case we have (at least) three options how to 
proceed:

(a) We  could use finite lattices and perform a finite size scaling (FSS) analysis 
\cite{Christ:2004}. This gives reasonable results, although it requires more numerical 
effort and the extrapolation $L\to\infty$ 
is associated with the usual uncertainties of any extrapolation.

(b) We  could  introduce a small amount of dissipation (i.e., with some very small probability 
$\epsilon$, Eq. (\ref{toppl}) is modified such that one of the neighbors has its stress 
not increased), and extrapolate to $\epsilon \to 0$. This was the strategy used in 
\cite{Mohanty:2002,Mohanty:2007}. While this should give cleanest results, it has the 
drawback that it requires more 
simulations and also involves an extrapolation. We did not try it in the present work.

(c) We  could  simply cut the evolution at some large time $T_{\rm max}$. This seems to be 
the strategy chosen in most previous simulations (e.g. in the BTW simulations of 
\cite{Vesp:2000}). As we shall see, results can be extremely misleading, unless 
this is done carefully.

\subsection{Initial conditions}

We know from previous simulations and from test runs that $z_c \approx 1.7326$. 
We now pick a rational number $n/m$ slightly smaller than $z_c$, e.g. $n/m = 45/26=
1.7307\ldots$. A sequence $w_{m,n}$ of $m$ digits $z_i \in \{1,2\}$ is then 
constructed such that $\sum_i z_i = n$ and that $w_{m,n}$ is as uniform as 
possible. For $(n,m) = (45,26)$ such a sequence is  
$w_{26,45} = (12^212^312^312^312^212^312^3)$ or any of its cyclic permutations. The 
initial configuration is then simply a repetition of $L/m$ such words, provided $L$ 
is a multiple of $m$.  In practice, we used   rational  approximants  closer to $z_c$, such as 
149/86 or 473/273.

\subsection{Transients}

\begin{figure}
 \begin{centering}
 \includegraphics[scale=0.3]{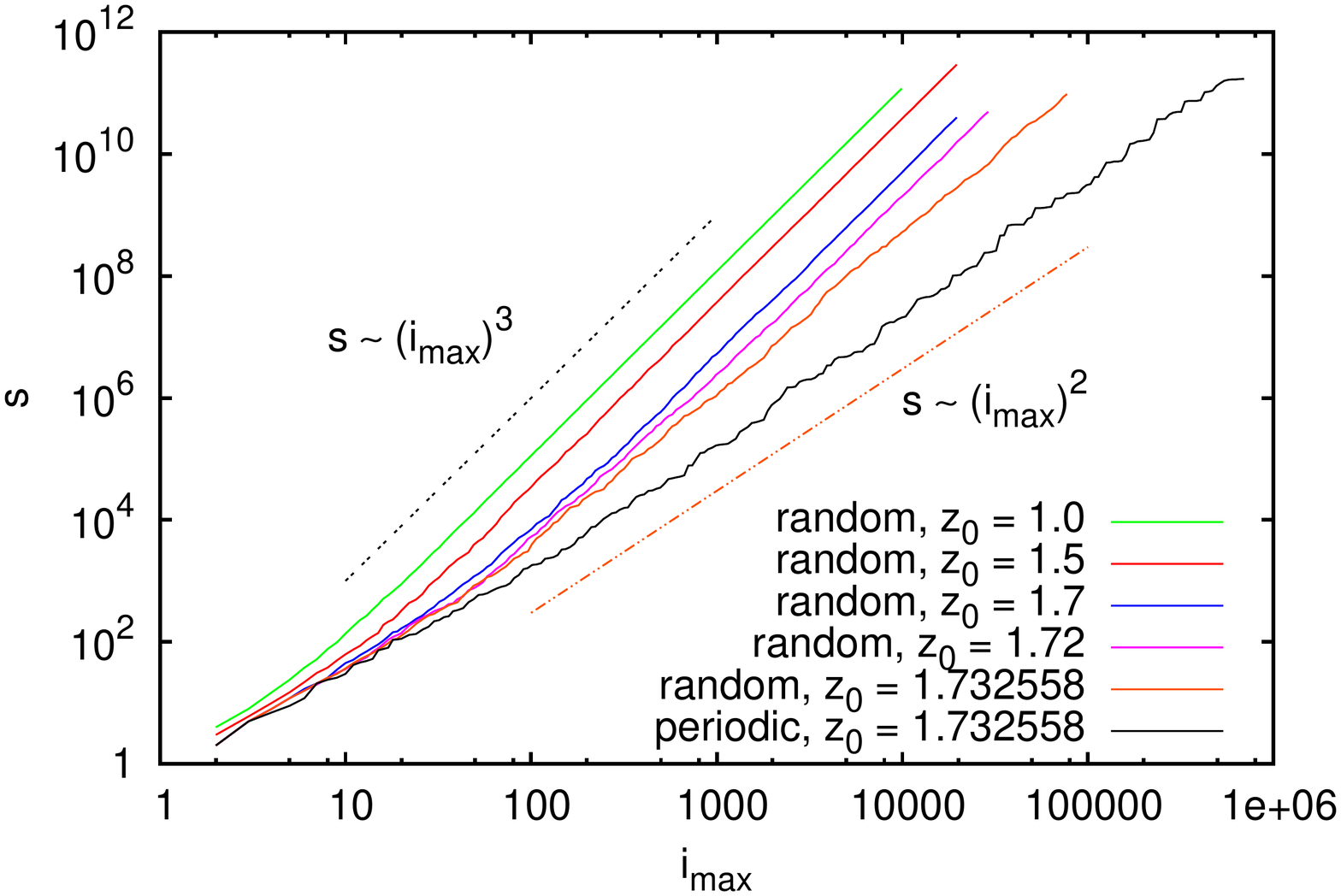} 
 \par\end{centering}
 \caption{\label{trans_time} (Color online) Number of topplings on a semi-infinite lattice driven 
   at its left end, until a site at distance $i_{\rm max}$ from this end is 
   first toppled. Each curve is based on $\approx 1000$ runs.}
 \end{figure}

First we discuss transients in the boundary-driven case. 
To see most clearly the transients, we used very large lattices 
($\geq 10^7$ sites) driven at the left boundary. We call the ``active region"
at time $t$ the part $[1,i_{\rm max}(t)]$, where $i_{\rm max}(t)$ is the 
rightmost point that had toppled at some time $t' \leq t$. We monitor
the evolution while $i_{\rm max}(t) < L$, i.e., while the active region
still spreads. In Fig.~\ref{trans_time} we show the  total number of topplings, starting from the initial time $t=0$ till the time 
the disturbances from the boundary first reach the site $i= i_{\rm max},$
for different
initial configurations. Notice that this gives a lower estimate for the 
transient CPU time, because even if the active region covers the entire 
lattice, it is still not clear whether it has the correct NCS correlations.
The top five curves are for random 1 / 2 sequences. If $z_0 = \langle z_{i,0}\rangle 
\ll z_c$, clearly $s\sim i_{\rm max}^3$. As  $z_0$ 
comes closer to $z_c \approx 1.73260$, this increase is slower, but it is still
much faster than the increase  $s\sim i_{\rm max}^2$ observed for periodic 
initial configurations. 

\begin{figure}
 \begin{centering}
 \includegraphics[scale=0.3]{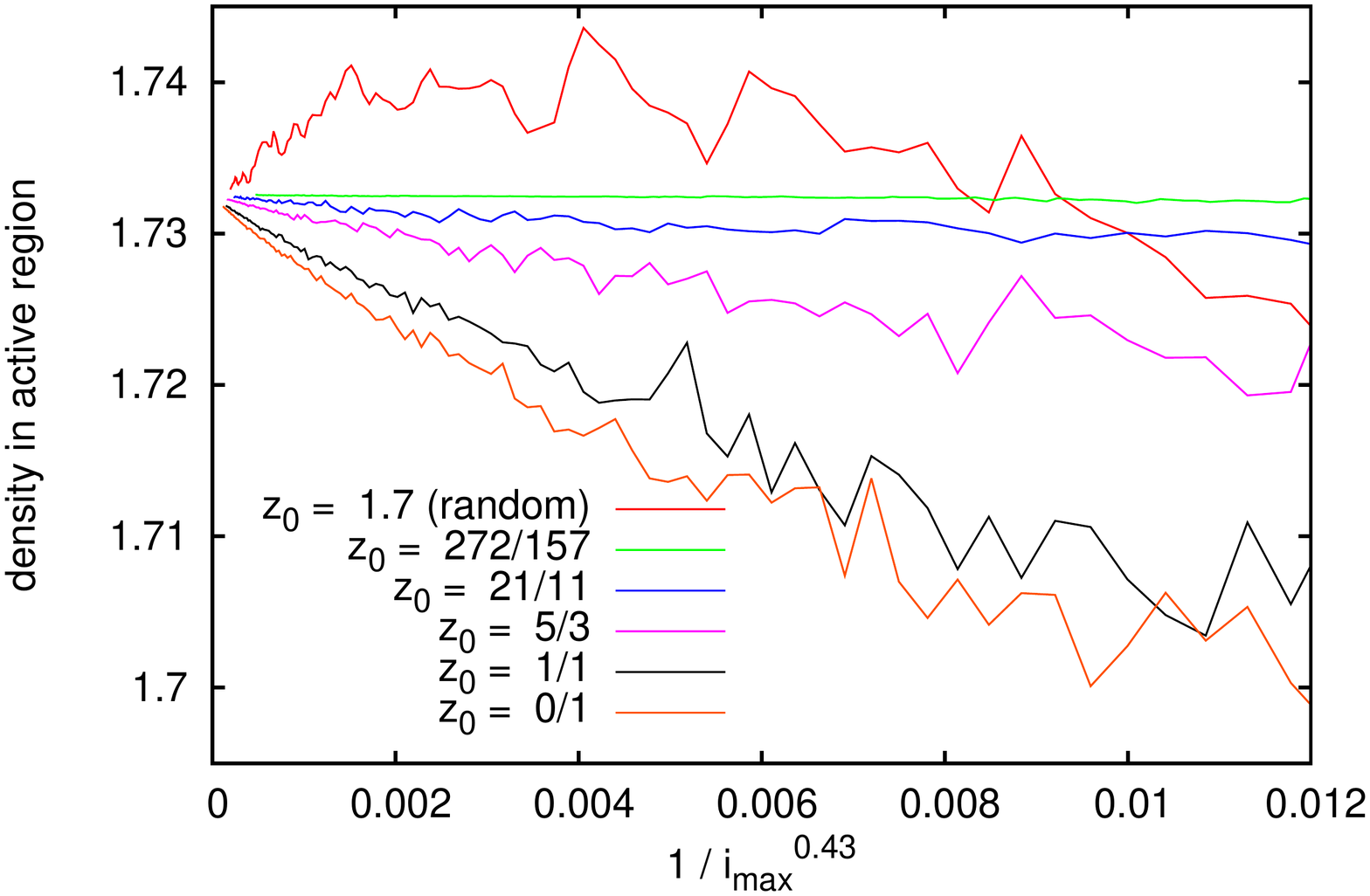} 
 \par\end{centering}
 \caption{\label{trans_fluct} (Color online) Average density in the active region, 
   plotted against an inverse power of the size $i_{\rm max}$ of the active region, 
   for different initial configurations. Each curve is based on a single run. 
   For all curves the stationary density is reached as 
   $i_{\rm max}\to\infty$, but the speed of convergence and the fluctuations depend
   strongly on the choice of initial state. Periodic initial states with density
   close to the stationary one are optimal. Notice that what looks like a horizontal
   straight line is indeed the data for $z_0 = 272/157 = 1.732484$.}
 \end{figure}

While Fig.~\ref{trans_time} suggests that periodic initial configurations lead to
much shorter transients, it could still be that the configurations at the time
when $i_{\rm max}$ is reached have much larger fluctuations and densities far from
the asymptotic one. That this is not 
true, and that periodic initial configurations lead both to  much smaller  
fluctuations and to correct densities is seem from Fig.~\ref{trans_fluct}.
There we plot the average density $\langle z\rangle_{\rm active}$ in the active
region, obtained in one single run, against an inverse power of $i_{\rm max}$. The 
lowest five curves in this 
plot correspond all to periodic initial configurations, with increasing values of 
 $z_0$. They show that both deviations from the asymptotic density and fluctuations
become smaller as the initial density approaches the stationary one. On the other 
hand, starting with a random configuration leads to huge fluctuations, even
if its density is close to the stationary one.

\section{Determination of critical exponents from numerical simulations}

In subsections A to C we shall mostly discuss simulations with open boundaries,
which are driven by adding stress at the left boundary (with the exception of Figs. 2 
to 6, which are indeed identical for open systems driven in the bulk). The fixed-energy
version is discussed in subsection D, while properties of avalanches in bulk-driven
open systems are treated in subsection E.

\subsection{The stationary state and hyperuniformity}

We will now discuss the various observables  measured in our simulations, and their analysis in terms of  the finite-size scaling theory.   
The critical exponent $\nu$ is defined in terms of the dependence of the correlation length $\xi$ 
on the distance from the critical point $\epsilon = \rho -\rho^*$, where $\rho^* $ is the critical density, by the relation
\be 
\xi \sim \epsilon^{-\nu}.
\ee

According to the finite-size scaling theory (FSS), a system with finite size $L$ at criticality will behave 
like an infinite system with finite correlation length $L$. Thus, we expect 
\be 
\langle z \rangle_L - z_c \sim L^{-1/\nu}
\ee

\begin{figure}
 \begin{centering}
 \includegraphics[scale=0.3]{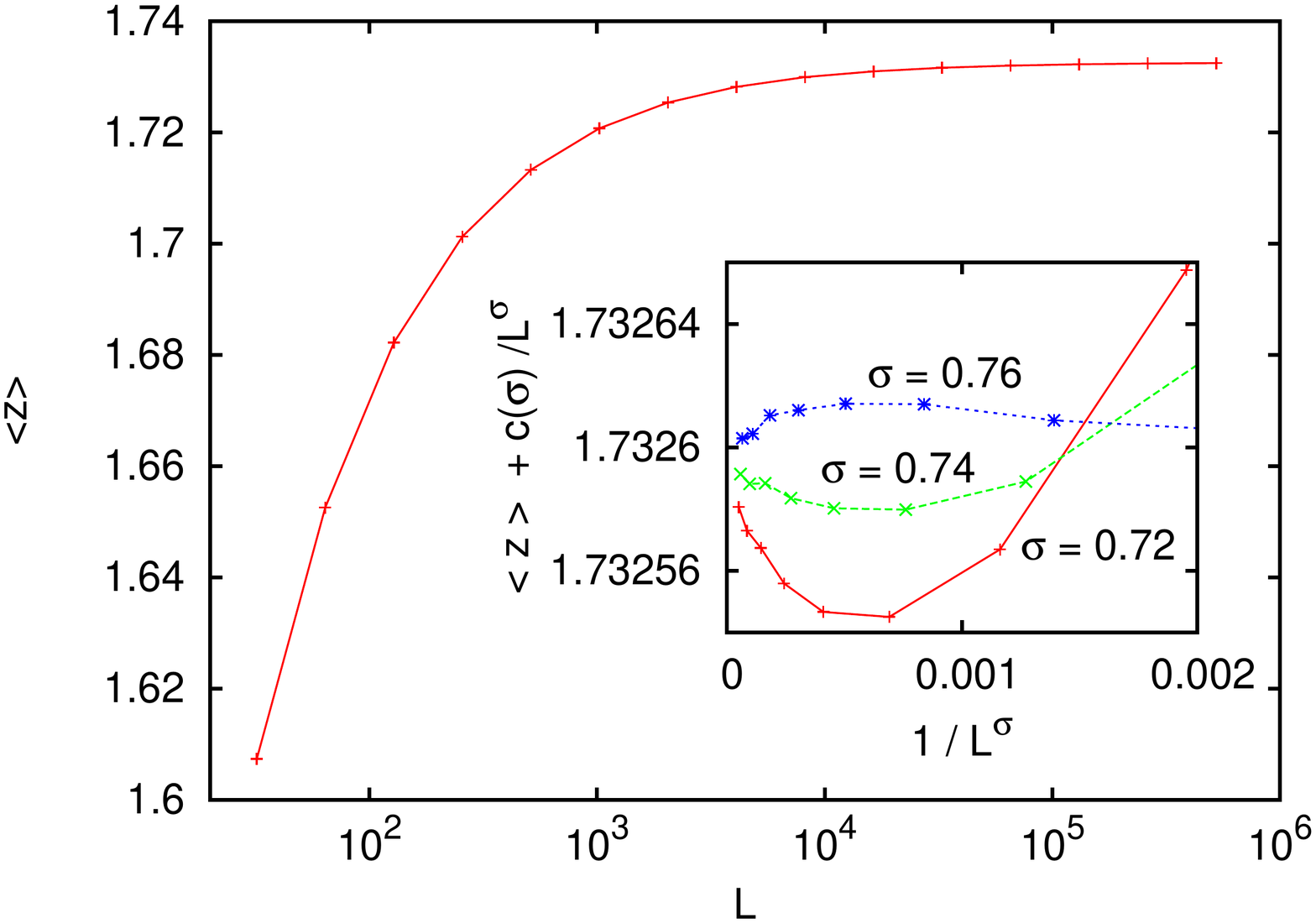}
 \vglue -5mm
 \par\end{centering}
 \caption{\label{dens} Average stress $\langle z\rangle$  of driven systems with open
   boundaries as a function of $L$. Main: log-linear plot of the raw data. Inset: The same 
   data plotted such that one can determine more precisely the finite size corrections.
   More precisely, on the $x$-axis is plotted $L^{-\sigma}$ with $\sigma$ either 0.76, 0.74, 
   or 0.72. If $\sigma$ were equal to the exponent of the leading finite size correction, 
   then the plot of $\langle z\rangle$ versus $L^{-\sigma}$ would be asymptotically a straight
   line. Since finite size corrections are very large, data would be indistinguishable from 
   straight lines in such a plot for a wide range of $\sigma$. Therefore we add to 
   $\langle z\rangle$ a term linear in $L^{-\sigma}$, such that the curves become roughly
   flat near the origin. The curve which is most straight at the origin gives then the true
   correction to scaling exponent.}
 \end{figure}

The dependence of $\langle z \rangle_L$ on $L$ provides a straightforward direct determination of $\nu$. 
In Fig.~\ref{dens} we show how the average stress $\langle z\rangle = L^{-1} \sum_{i=1}^L z_i$
depends on $L$. In the main plot we show the raw data, and in the inset a plot
which suggests the precise finite size corrections. Indeed, for reasons that will become
clear in a moment, we used in Fig.~\ref{dens} not only data obtained on boundary driven
systems, but we averaged also over systems driven in the bulk, both at random sites and also
just at the center site. The result of Fig.~\ref{dens} can be summarized as
\be
   \langle z\rangle = z_c + c / L^\sigma    \label{sig}
\ee
with 
\be
    z_c = 1.732594(4)\;, \quad \sigma =  1/\nu = \;\; 0.74(1), \label{z_c}
\ee
while the precise value of $c$ depends strongly on $\sigma$. The best previous estimates
of $z_c$ were $1.7326(3)$ for boundary driven and $1.734(2)$ for bulk
driven open systems, and $1.73260(2)$ for the FES version \cite{Christ:2004}. We are not 
aware of a previous estimate of $\sigma$.

Another way to determine $\nu$ is to look at the effects of the boundary on the  density profile.  
Let $\rho(i)$ be the mean density of particles at site $i$ in the steady state of the driven sandpile.  
From the abelian property one obtains the following  result : {\it $\rho(i)$ is independent of the way the pile is driven.} 

{\cly Proof:} Let $a_i$ be the operator corresponding to adding a particle at $i$ and letting a subsequent 
avalanche evolve until a stable state is reached again. Let $|\Phi\rangle$ be the statistically stationary 
(macro-)state obtained by driving at site $i$ with probability $p_i$. It satisfies $|\Phi\rangle = W|\Phi\rangle$
with $W = \sum_{i} p_i a_i$ . Since all $a_i$ commute due to the 
abelian property, they can be  diagonalized  simultaneously, and $|\Phi\rangle$ is an eigenvector of each $a_i$ with 
eigenvalue $1$, and  if the Markov process can reach all recurrent states, it is the only eigenvector with this property, and so  independent of the distribution $\{p_i\}$.  

\begin{figure}
 \begin{centering}
 \includegraphics[scale=0.3]{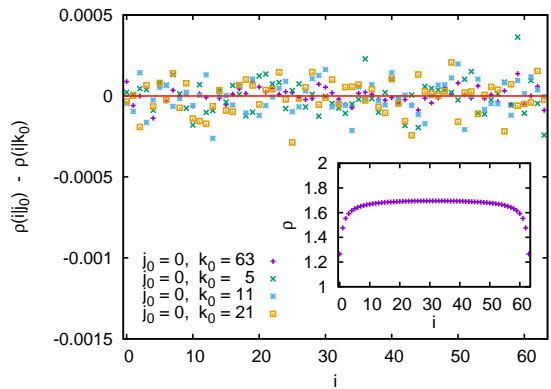}
 \vglue -5mm
 \par\end{centering}
 \caption{\label{dens_i} (Color online) The inset shows the stationary profile $\rho(i)$ for $L=64$. To
   verify that it is indeed left-right symmetric, we show in the main plot differences
   between values of $\rho(i)$ measured at the same site $i$, but for runs where the system
   was driven at different values $j_0$ resp. $k_0$. The estimated statistical error for 
   these differences was between $\pm 0.0001$ and $\pm 0.0002$.}
 \end{figure}

We have checked this directly in simulations. In Fig.~\ref{dens_i}, we plot  $\rho(i|j_0)$ the average stress at 
$i$ when the system is driven at site $j_0$. In the main plot of Fig.~\ref{dens_i} we show differences
between averages measured at the same $i$ but for different $j_0$.  These data are for a 
very small system, but the same was found also for larger $L$. This is  true also for random bulk driving, as we 
indeed verified numerically. 

\begin{figure}
 \begin{centering}
 \includegraphics[scale=0.3]{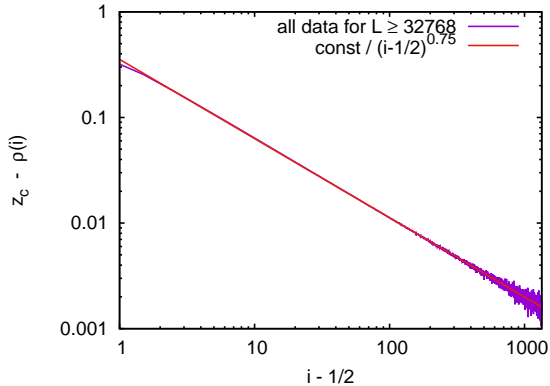}
 \vglue -5mm
 \par\end{centering}
 \caption{\label{profile} (Color online) Log-log plot of stress depletion near the end points versus
   distance from the end (more precisely from the point $x=1/2$, half way between the 
   site $i=0$ where $\rho(i)$ is zero, and the site $i=1$ where it first
   becomes positive). Lattices
   ($L\geq 2^{15}$) are sufficiently large so that finite size corrections should be 
   negligible.}
 \end{figure}

From scaling theory, we expect $\rho_i$ to differ from $z_c$ as  
\be 
\rho(i) - z_c \sim i^{-1/\nu}. 
\ee 
We show the variation of $\rho_i$ with $i$ in Fig.~\ref{profile}. 

\begin{figure}
 \begin{centering}
 \includegraphics[scale=0.3]{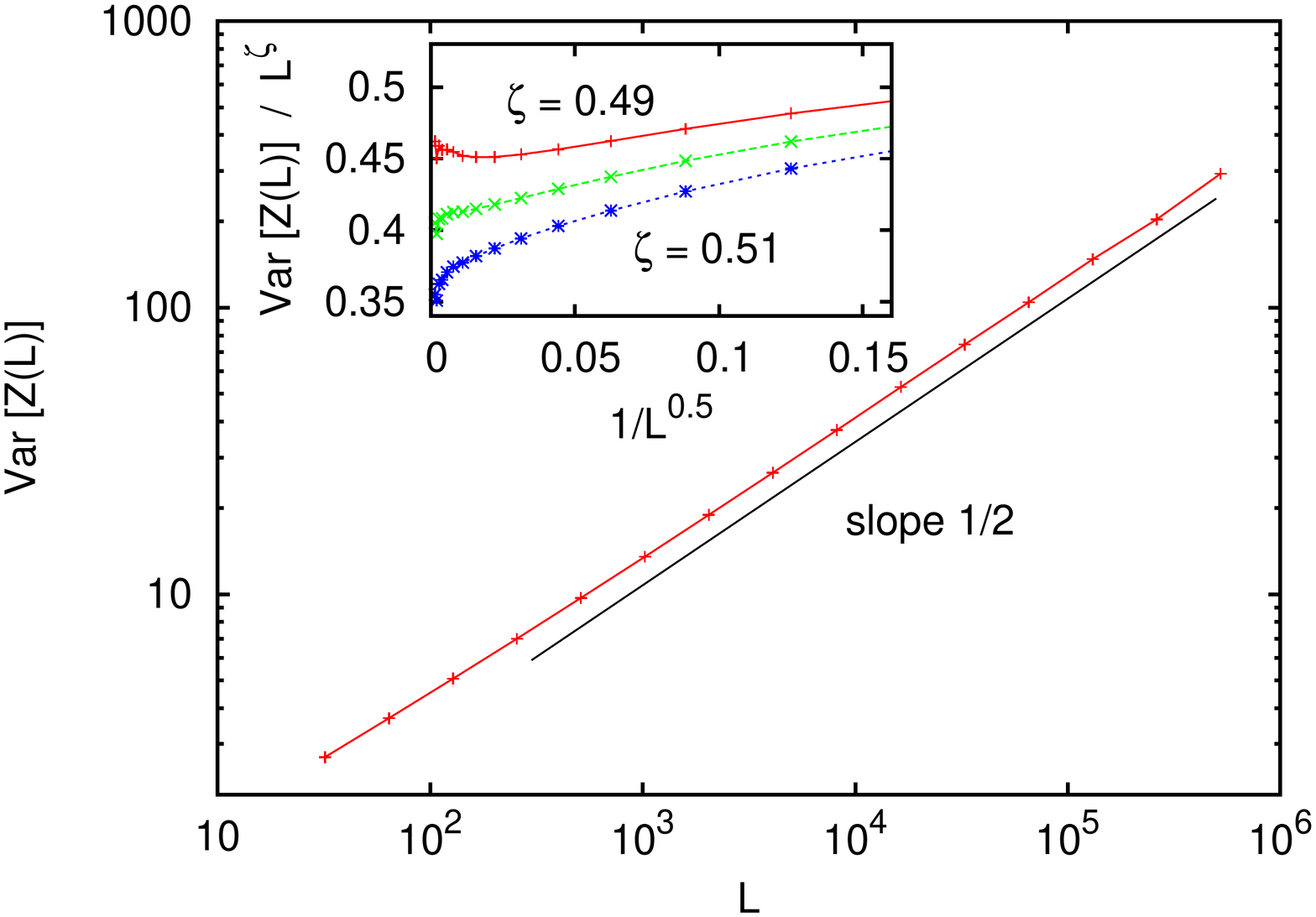}
 \vglue -5mm
 \par\end{centering}
 \caption{\label{lambda} (Color online) The main plot shows a log-log plot of ${\rm Var}[Z]$ versus $L$, 
    together with a straight line that suggests ${\rm Var}[Z]\sim L^{1/2}$ with visible
    corrections. The inset suggests that these corrections decrease roughly as $L^{-1/2}$.}
 \end{figure}

 The stress density, averaged over a finite block of size $k$, will in the critical  state show fluctuations 
of order $k^{-1/\nu}$, hence the total stress $Z$ in this block
will fluctuate by a typical amount $k \times k^{-1/\nu}$.  This gives 
\be
{\rm Var}[Z] \sim L^{\zeta} = L^{2(1-1/\nu)}
\ee
giving the  hyperuniformity exponent $\zeta = 2(1 -1/\nu)$.  The variation of ${\rm Var}[Z]$ with $k$ is 
plotted in Fig.~\ref{lambda}. We see that the value of $\zeta =1/2$ fits the data very well. Based on the results 
presented, we conjecture that $\nu$ is exactly equal to the simple fraction $4/3$.  

\subsection{Avalanche size distributions}

For the distribution of avalanches, our  clearest data come from the boundary-driven case where one adds stress at the left boundary, and lets 
it dissipate through the right one. 

Let us first discuss the avalanche size distribution $P(s,L)$, where $s$ is the 
number of topplings in an avalanche. We expect the scaling law 
\be
   P(s,L) = s^{-\tau} f(s/\varphi(L)) \times [1+\frac{a}{L^x} + \ldots],  \label{P_s}
\ee
with 
\be
   \varphi(L) = L^D \times [1+\frac{b}{L^\omega} + \ldots],       \label{phi_L}
\ee
where the last factors in both expressions correspond to finite size corrections.

 \begin{figure}
 \begin{centering}
 \includegraphics[scale=0.3]{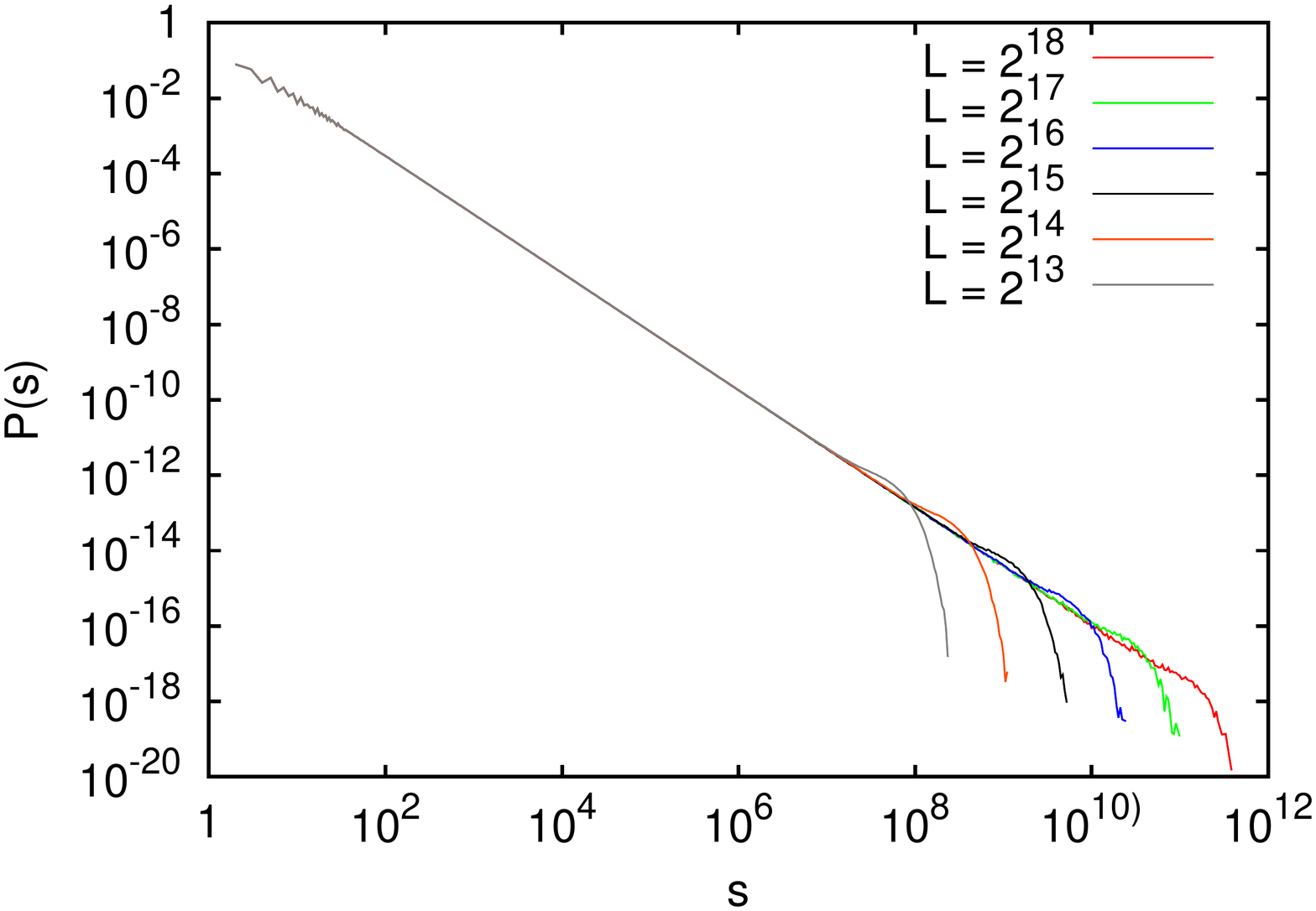} 
 \vglue -5mm
 \includegraphics[scale=0.3]{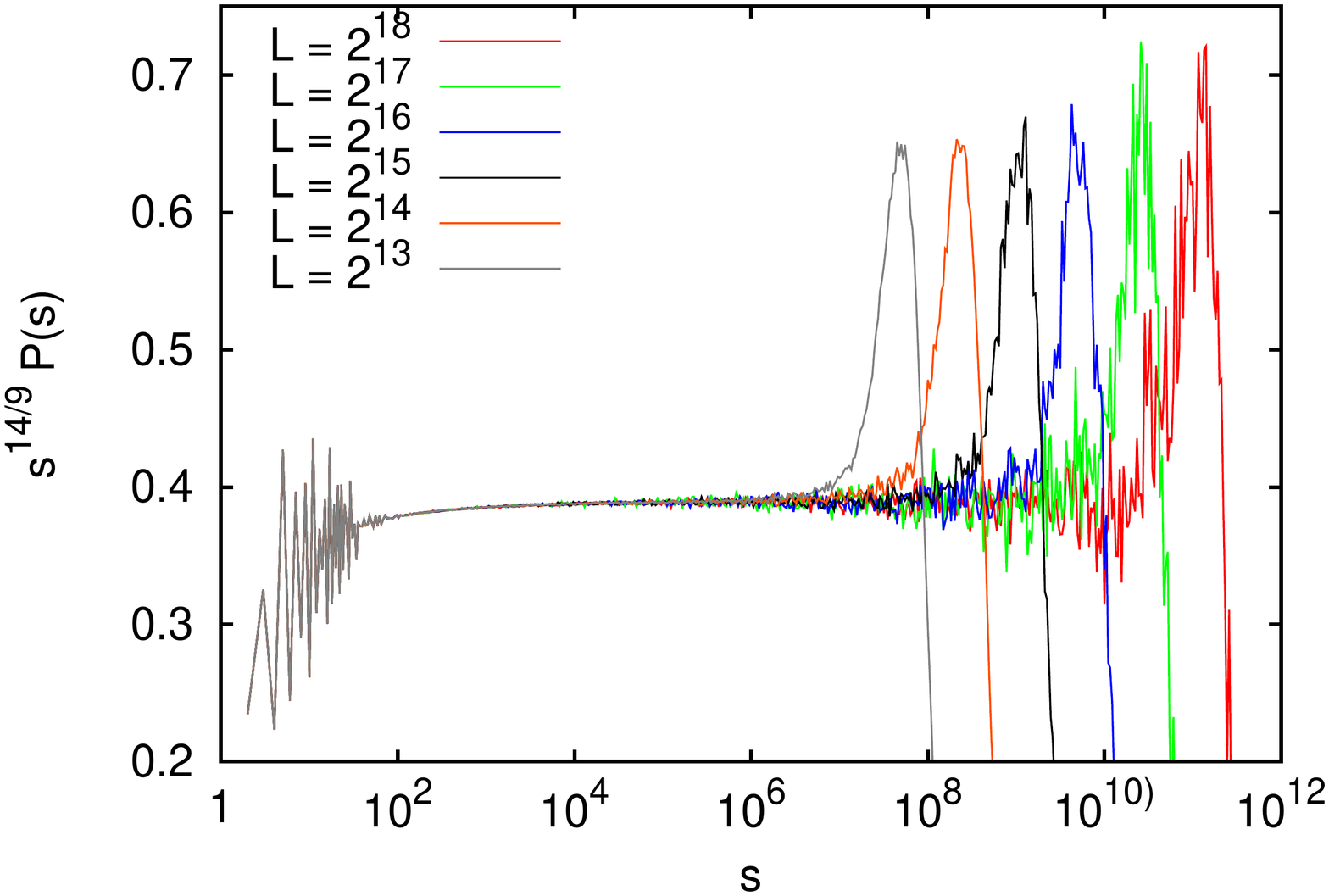} 
 \vglue -5mm
 \includegraphics[scale=0.3]{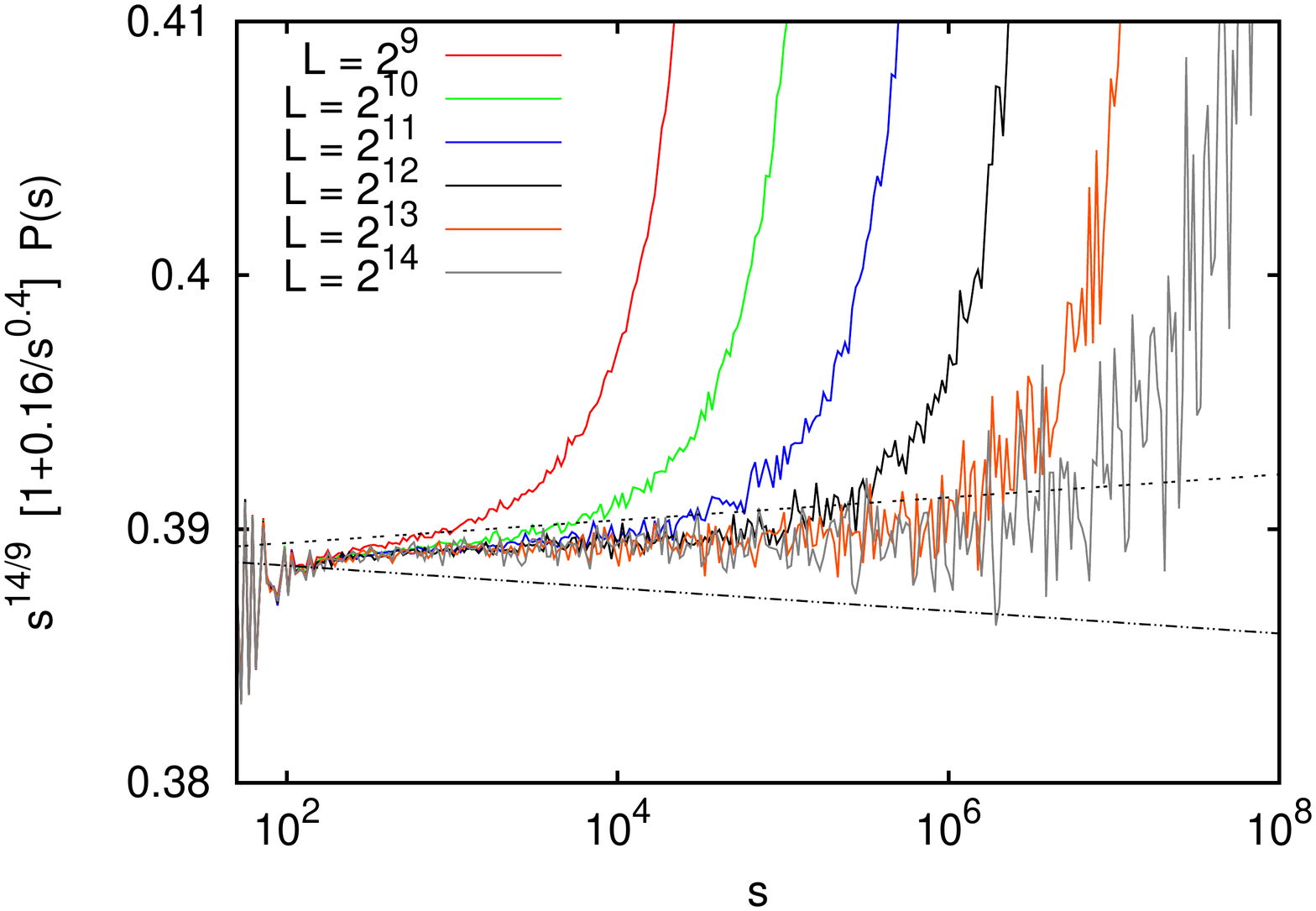} 
 \vglue -5mm
 \par\end{centering}
 \caption{\label{bdry-tau} (Color online) Avalanche size distributions for the open systems
   driven at the left boundary. Panel (a) shows the raw data (for the largest 
   lattices sizes only, to avoid overcrowding). Panel (b) shows the same data 
   multiplied by $s^{14/9}$. If this exponent is equal to $\tau$, the central 
   parts of the curves should be flat. In view of the substantial corrections
   to this, we plotted in panel (c) the data (for smaller $L$, since they have 
   less statistical fluctuations) multiplied by a further factor $(1+a/s^x)$,
   with $x = 0.4$. The two straight lines in panel (c) indicate the error 
   margins $\pm 0.0005$ of $\tau$. The numbers of avalanches used for these 
   figures range between $>2\times 10^{10}$ for $L\leq 4096$, $1.7\times 10^9$ 
   for $L=131072$, and $3.5\times 10^9$ for $L=262144$. 
   The fluctuations seen for $s<100$ are not statistical, but are 
    systematic, and are related to the  structure of the state-space of 
   recurrent stable configurations \cite{Chua-Christ}.}
 \end{figure}

In Fig.~\ref{bdry-tau} we show $P(s,L)$ for values of $L$ between 512 and $2^{18}$. 
The raw data shown in panel (a) just demonstrate the impressive range, but they are
not really informative. Multiplying the data with $s^\tau$ as in panel (b) shows 
already much more details. But it still does not allow to make a precise estimate
of $\tau$. For this we have to include finite size corrections, as in panel (c) (where
we mostly plotted data for the smallest values of $L$, which were not shown in panels 
(a) and (b)). The correction to scaling exponent is close to 1/2, and we will justify
the choice $0.4$ below. Our best estimate
of $\tau$ is $1.5556 \pm 0.0005$. This is a factor 4 more precise than the
best previous estimates $1.556 \pm 0.002$ \cite{Christ:96,Pruss:2003,Paczuski}.
It strongly suggests that $\tau = 14/9$ exactly, as conjectured in \cite{Christ:2004a} 
(a ten times more precise value was claimed 
in \cite{Christ:2004a}, but this was revised in a later paper by the same author \cite{Christ:2004}).

 \begin{figure}
 \begin{centering}
 \includegraphics[scale=0.3]{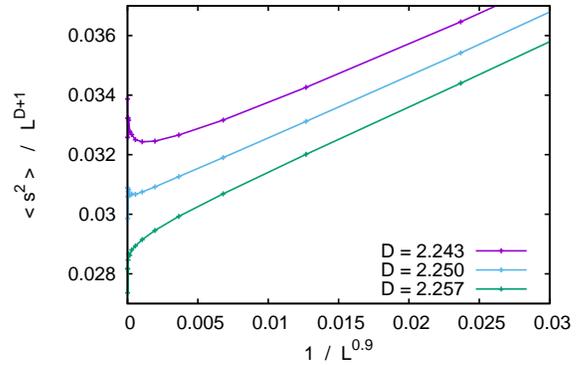} 
 \vglue -5mm
 \par\end{centering}
 \caption{\label{sk-open} (Color online) Re-scaled second moments of avalanche sizes for the 
   boundary model, plotted against $1/L^{0.9}$. The moments are divided by powers of $L$,
   such that the curve would be straight if the power were equal to $D+1$ and 
   the correction to scaling exponent were $\omega=0.9$. This is indeed the case 
   for $D=9/4$. The other two curves (with exponents $9/4 \pm 0.007$) are clearly
   sub- and supercritical.}
 \end{figure}

Using Eqs.~(\ref{P_s}), (\ref{phi_L}), and the fact that $\langle s \rangle = L$ 
(which is true exactly, without any finite size corrections) gives 
\be
   D=(2-\tau)^{-1} = 9/4\pm 0.002  \quad {\rm and} \quad \omega = x D \approx 0.9.   \label{scal-s-bdry}
\ee
Superimposing the peaks in Fig.~\ref{bdry-tau}b would give a compatible but much less 
precise estimate of $D$ because of the finite size corrections in $\varphi(L)$. 
But a more precise value, with error bars similar as those that follow from the scaling 
relation, can be obtained from higher moments of $s$. From Eq.~(\ref{P_s}) we expect 
\be
   \langle s^k \rangle \sim L^{(k+1-\tau)D} = L^{D+1} \times [1 + \frac{b'}{L^\omega}].   \label{sk}
\ee
In Fig.~\ref{sk-open} we plot $L^{y}\langle s^2 \rangle$ against $1/L^{0.9}$ for three 
values of $y$. The central curve is for $y=13/4$, and it is a perfect straight line up to 
fluctuations for the two largest lattices ($L=2^{18}$ and $L=2^{19}$). On the other
hand, the two other curves are clearly sub- and supercritical. A similar result is obtained 
from the third moment (not shown). Apart from verifying the estimate of $D$, these data
suggest very strongly that indeed the correction to scaling exponent is $\omega =0.9$.

 \begin{figure}
 \begin{centering}
 \includegraphics[scale=0.3]{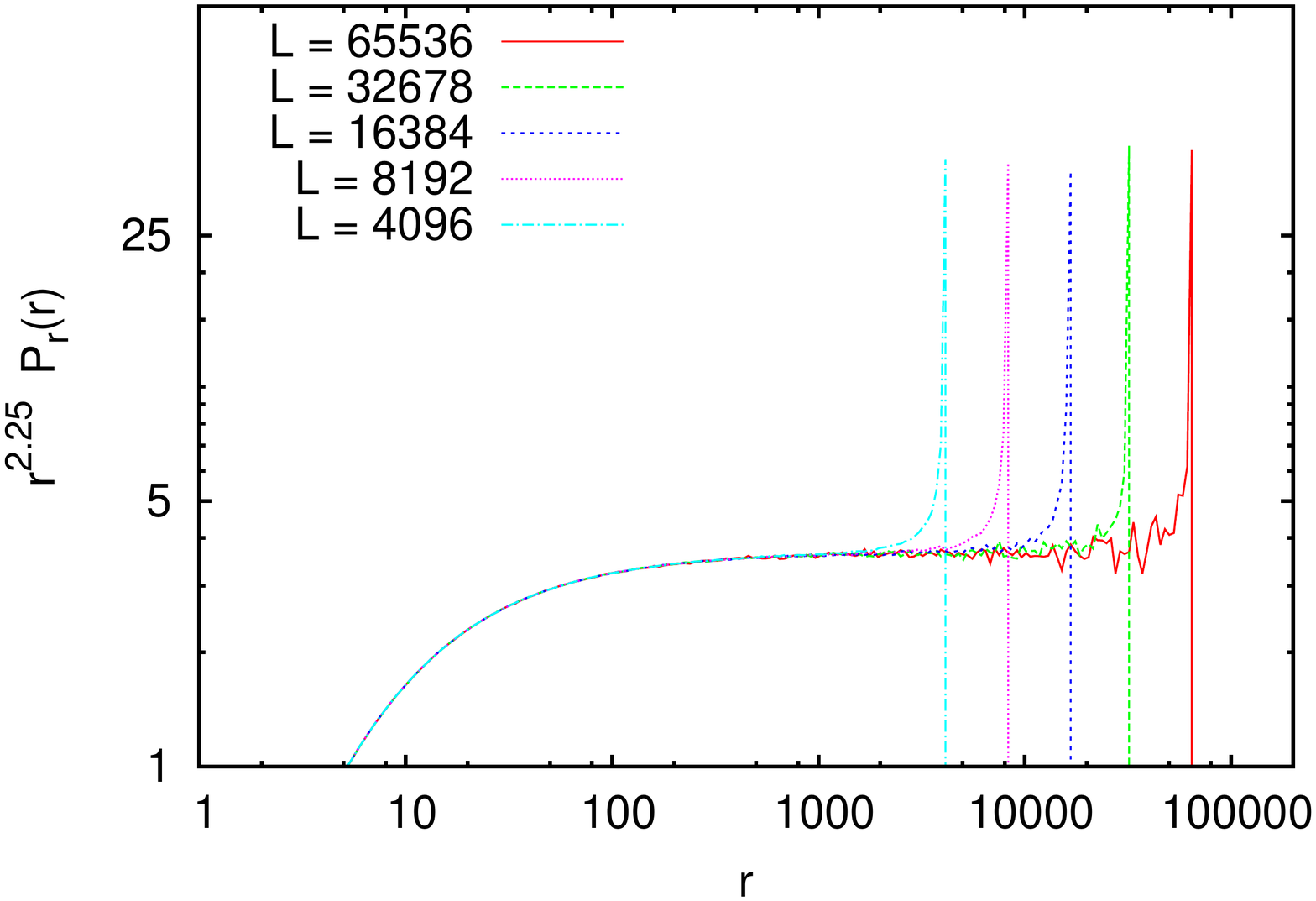} 
 \vglue -5mm
 \includegraphics[scale=0.3]{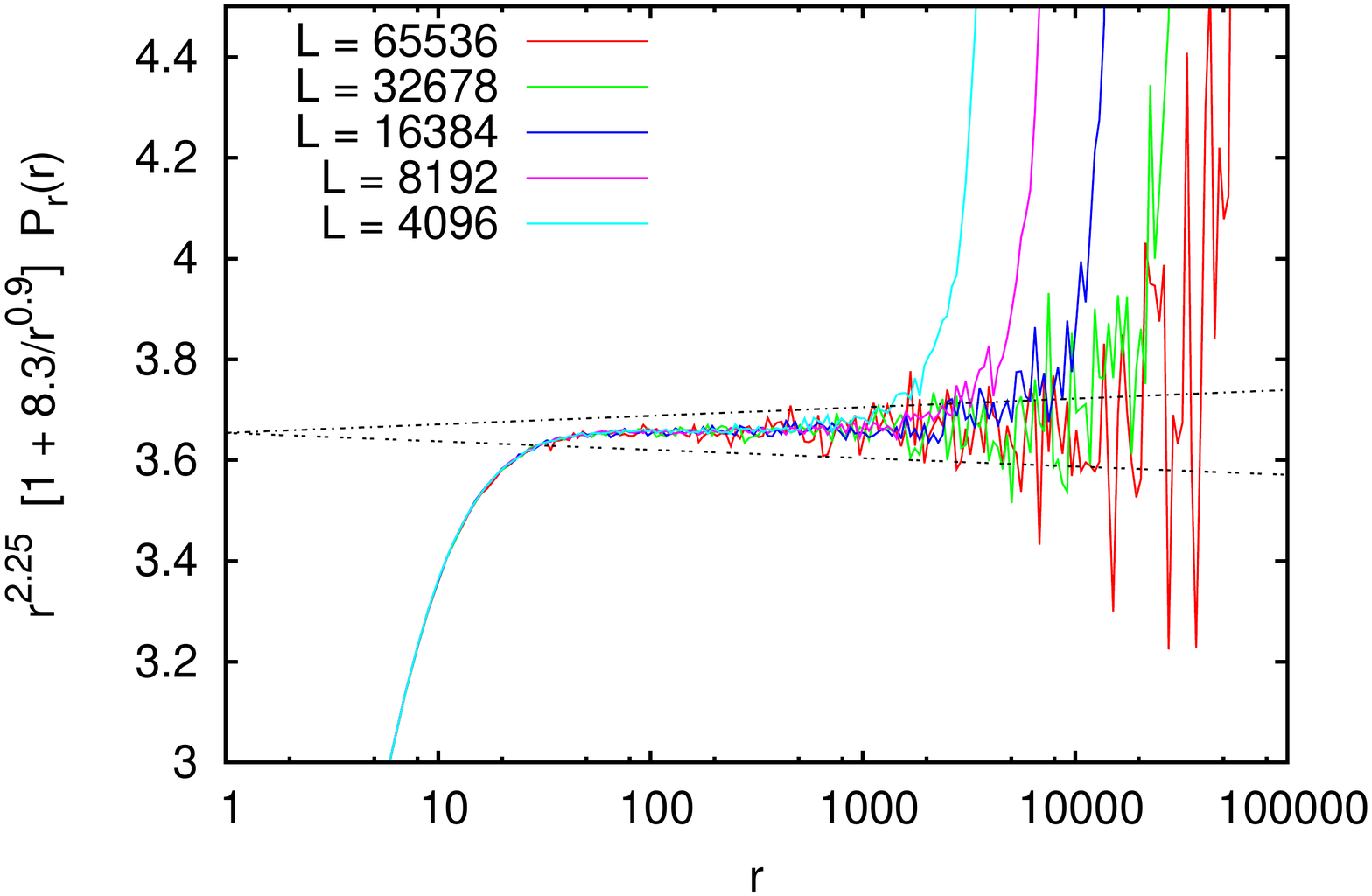} 
 \vglue -5mm
 \par\end{centering}
 \caption{\label{bdry-r} (Color online) (a) Log-log plot of $P_r(r,L)$, the spatial size distribution
   of avalanches for boundary driven systems, multiplied by $r^{9/4}$. (b) The same data, 
   but multiplied by $1+8.3/r^{0.9}$ and plotted on a log-linear plot. The straight lines
   indicate the estimated errors $\pm 0.002$}
 \end{figure}

The distribution of spatial extensions of avalanches, $P_r(r,L)$, we assume the scaling form 
\be
   P_r(r,L) \approx r^{-\tau_r} g(r/L) ,  \label{P_r}
\ee

As we have $s \sim R^D$, it is easily seen that   $\tau_r = 1+ (\tau-1) D$. For $D=9/4$, and $\tau =14/9$, we get $\tau_r = 9/4$.   

Plots analogous to Fig.~\ref{bdry-tau}b and \ref{bdry-tau}c are 
shown in Fig.~\ref{bdry-r}. This time the corrections to scaling are much bigger, but they
seem to be described again to leading order by a rational power. The value $\tau_r = 9/4$  fits our data well, with an error of $\pm 0.002$. At the same time, accepting the 
scaling $s\sim r^D$ in the scaling region, we predict the correction to scaling exponent
as $0.4 \times 9/4 = 0.9$, in perfect agreement with the data.

\subsection{Temporal evolution of avalanches}\label{open-aval-tim}

 \begin{figure}
 \begin{centering}
 \includegraphics[scale=0.29]{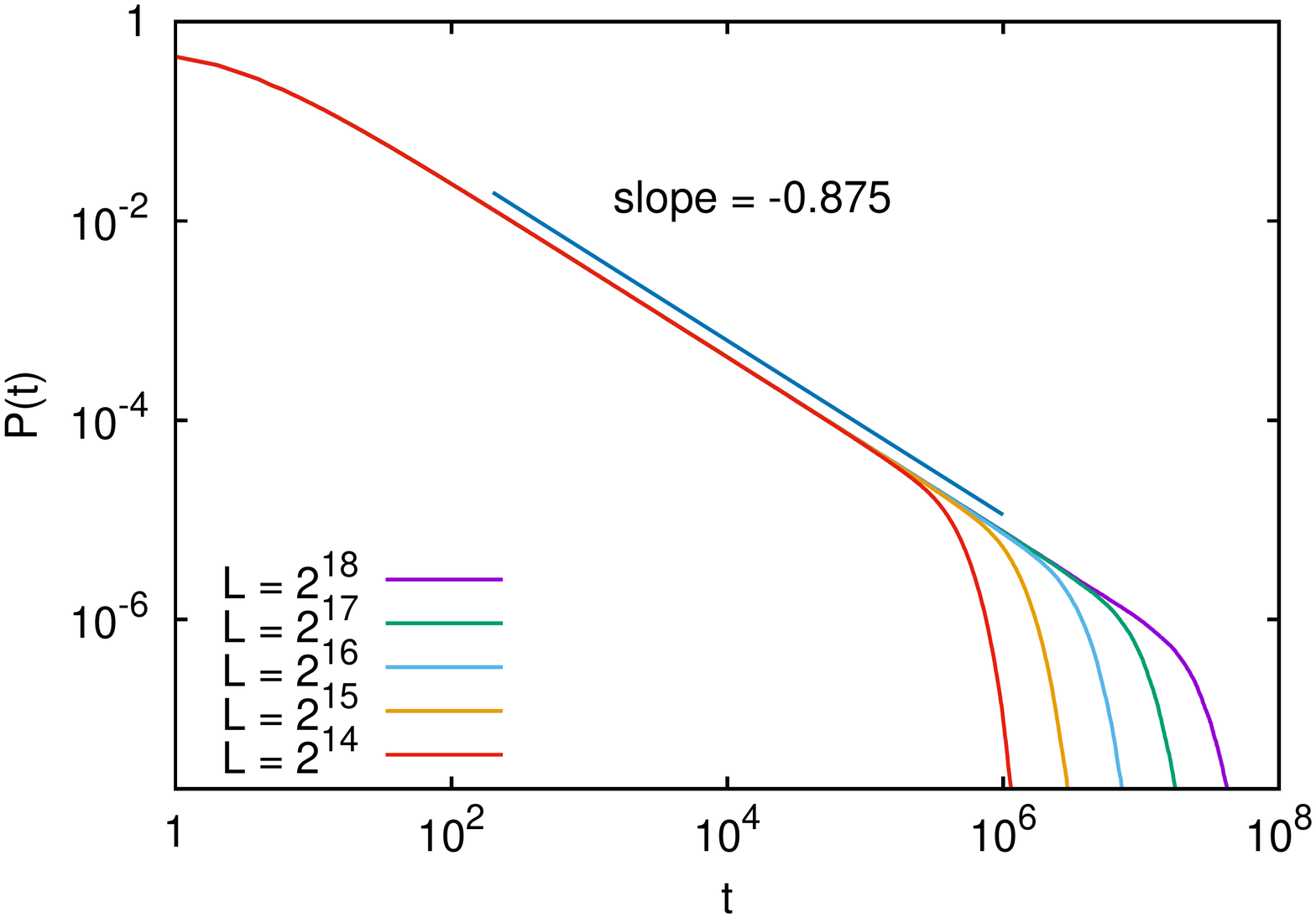}
 \vglue -6mm
 \includegraphics[scale=0.29]{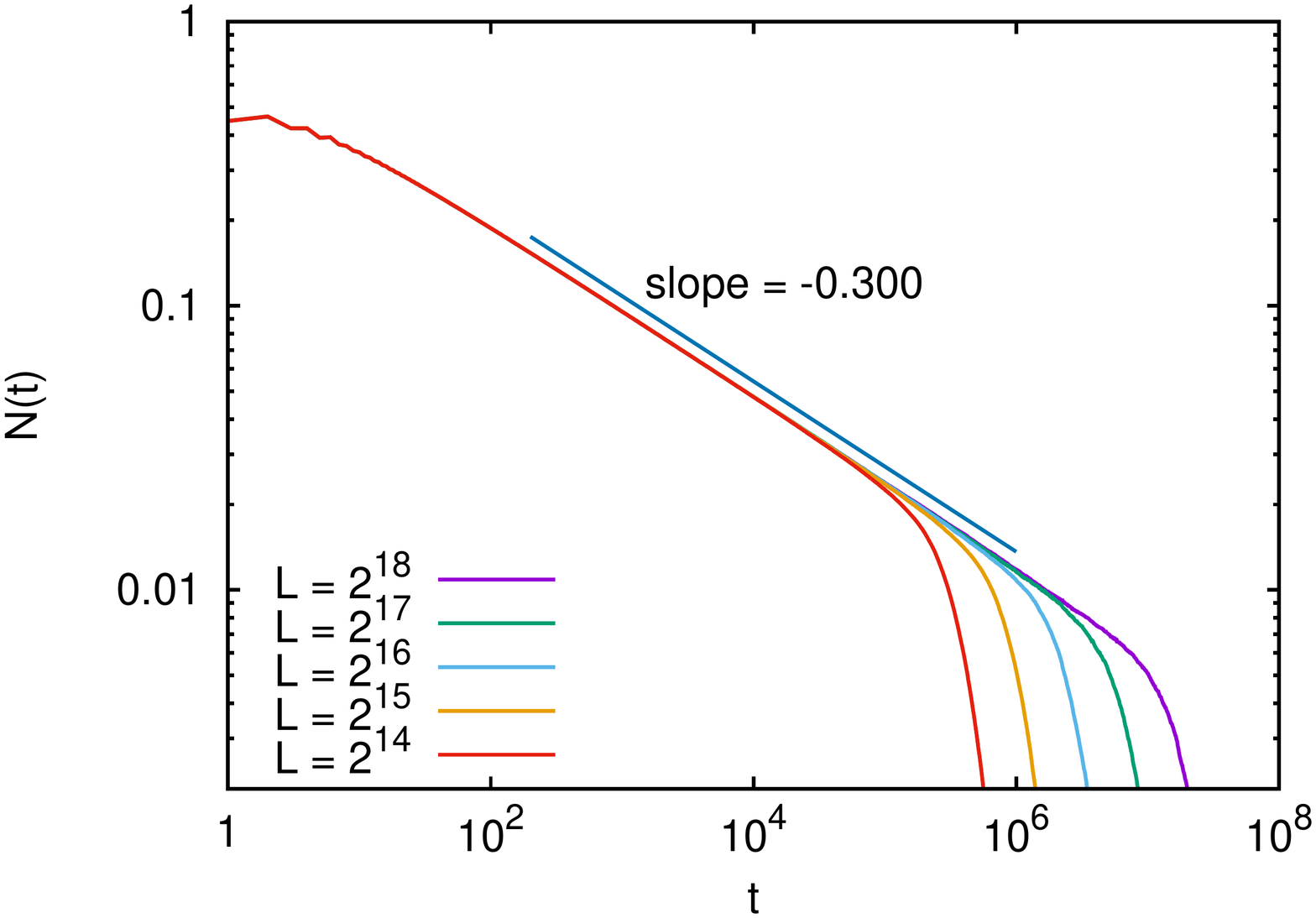}
 \vglue -6mm
 \includegraphics[scale=0.29]{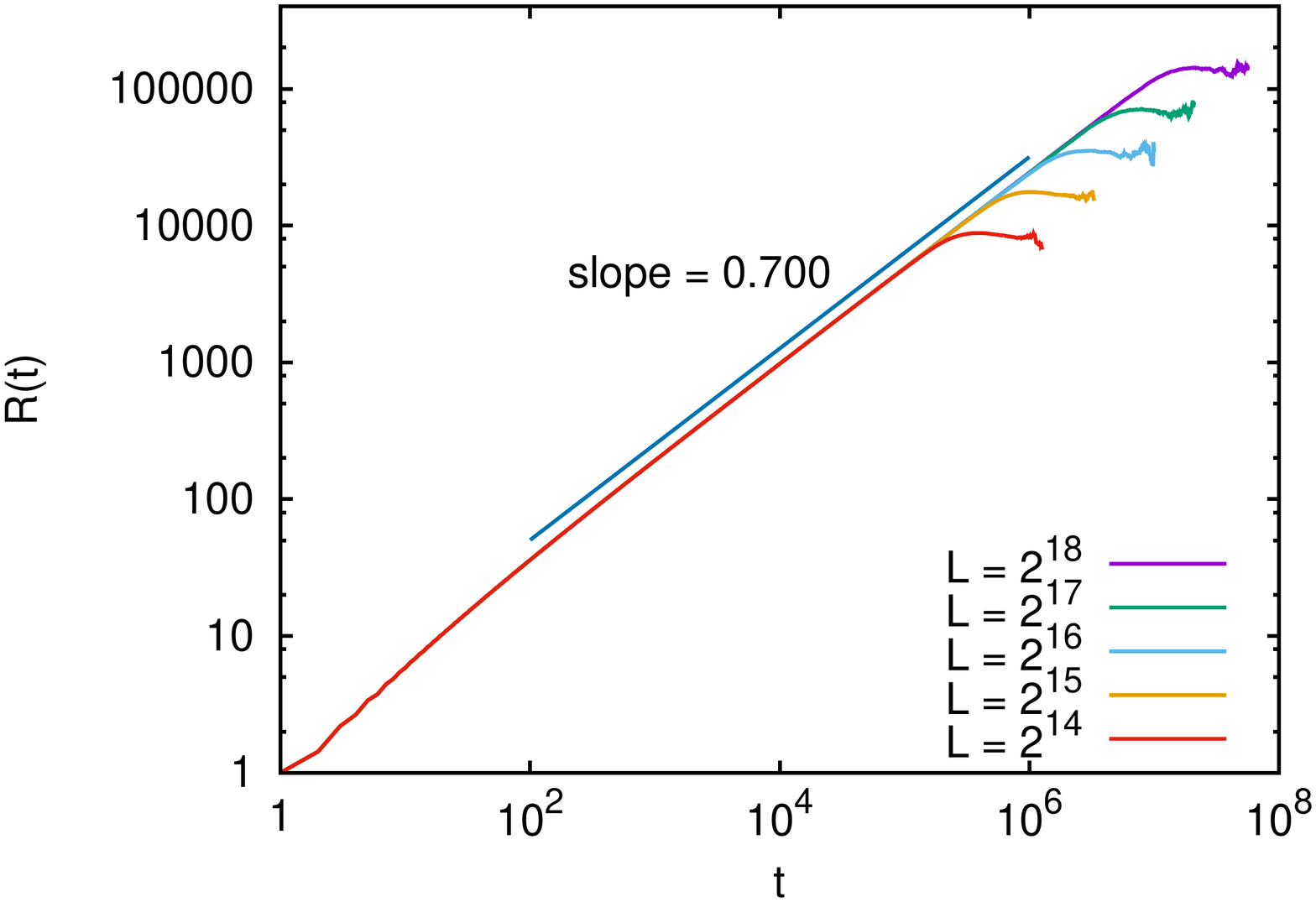}
 \vglue -6mm
 \par\end{centering}
 \caption{\label{open-PNR-t} (Color online) Log-log plot of $P_t(t,L)$ (panel a), $N(t,L)$ (panel b), and 
   $R(t,L)$ (panel c). In each plot, only data for $L\geq 16384$ are shown. The straight 
   lines indicate power laws in the central (scaling) region.}
 \end{figure}

We now discuss the time-dependent exponents of the avalanches.  We define the dynamical exponent $z$ by the relation
\be
R(t) \sim t^{1/z}
\ee
where $R(t)$ is the average distance  of topplings at time $t$ from the boundary where the avalanches were triggered. 
Other related quantities are $P_t(t)$ and $N(t)$, which are respectively the probability that the avalanche survives up to time $t$, and the average number of topplings at time $t$ in the avalanches that survive up to time $t$. We define the exponents $\eta$ and $\delta$ by the relations
\be
   P_t(t) \sim t^{-\delta} \quad {\rm and} \quad N(t) \sim t^{\eta}.
\ee

Results of these measurements are shown in 
Fig.~\ref{open-PNR-t}. They all show very clean scaling regions, with 
\be
   \delta = 7/8,  \quad \eta = -3/10,  \quad {\rm and} \quad z = 10/7.
\ee
We do not quote formal error bars, because by now we obviously conjecture that these 
rational numbers are exact, and any error estimate (which by its very nature is subjective, 
critical exponents being obtained by {\it extrapolating} data) would probably be biased 
by this conjecture. We nevertheless can say informally that plots analogous to 
Figs. \ref{bdry-tau}c and \ref{bdry-r}b suggest $\delta = 0.875(1),  \eta = -0.300(1)$,
and $1/z = 0.700(2)$.

Typical previous estimates were $\delta = 0.85(2)$, $\eta = 0.33(2)$, and $z = 1.42(2)$ 
\cite{Bona-thesis}. They were, however, made by assuming a Manna universality class and 
thus lumping together simulation results from various models. As a rule of thumb, our 
present estimates are an order of magnitude more precise than previous ones. On the 
other hand, extracting correction to scaling exponents from Fig.~\ref{open-PNR-t} was 
not very successful, because obviously more than one correction term is important in 
each case. Presumably, there are also important analytic corrections resulting from an
inherent uncertainty how to define $t$ up to an arbitrary constant of order 1.

 \begin{figure}
 \begin{centering}
 \includegraphics[scale=0.3]{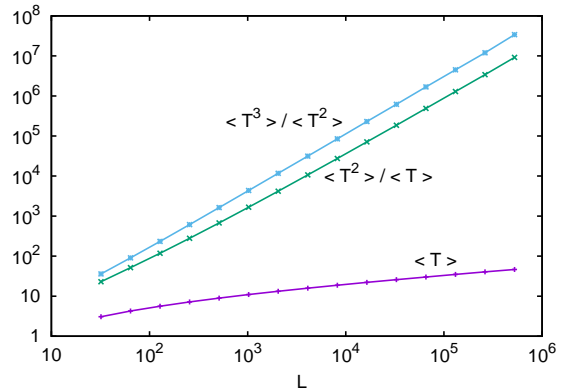}
 \vglue -5mm
 \par\end{centering}
 \caption{\label{bdry-lifetime} (Color online) Log-log plots of the average lifetime  $\langle T\rangle$
   of boundary-driven avalanches, and of moment ratios 
   $\langle T^{k+1}\rangle / \langle T^k\rangle$ for $k=1$ and $k=2$.}
 \end{figure}

Another estimate of $z$ can be obtained from the 
moments of $T$, the life time of avalanches. When defining $\langle T^k\rangle$, one 
has to specify how avalanches with different size $s$ are weighted. In Fig.~\ref{bdry-lifetime}
we show results for $\langle T\rangle, \langle T^2\rangle / \langle T\rangle$, and 
$\langle T^3\rangle / \langle T^2\rangle$, where $T = s^{-1}\sum_{{\rm topplings} j} t_j$
and $t_j$ is the time of the $j$-th toppling  in the parallel update scheme. We see no clear scaling law for 
$\langle T\rangle$, a scaling law with large finite size corrections for 
$\langle T^2\rangle / \langle T\rangle$, and finally a clean scaling with reasonably 
small finite size corrections for $\langle T^3\rangle / \langle T^2\rangle$, 
\be
   \langle T^3\rangle / \langle T^2\rangle \sim L^{1.43(1)}.
\ee
The latter is   consistent with  our conjectured exact value  $z =10/7$.   

\subsection{The Fixed-energy sandpile: closed boundaries case}

\subsubsection{Supercritical systems: The order parameter exponent}

As we said in the introduction, simulating the fixed ``energy" (i.e. stress) model is 
easiest and most straightforward away from the critical point.  In contrast, measuring 
the properties of single avalanches is non-trivial  both in the critical and in the supercritical 
phase. But estimating the 
density $\rho_{a,\infty}$ of active sites in a stationary  supercritical state, and thus the 
order parameter $\beta$ defined through
\be
   \rho_{a,\infty} \equiv \lim_{t\to\infty} \rho_a(t) \sim (\langle z\rangle - z_c)^\beta,                                  \label{beta}
\ee
is easy.  We start with a periodic configuration with the desired total stress (which implies 
also that we use for $L$ a multiple of the period). There will  be $O(L)$ sites with $z_i=2$,
half of which are declared as unstable. We then follow the evolution until stationarity of 
$\rho_a$ is reached and enough statistics is collected thereafter.

 \begin{figure}
 \begin{centering}
 \includegraphics[scale=0.30]{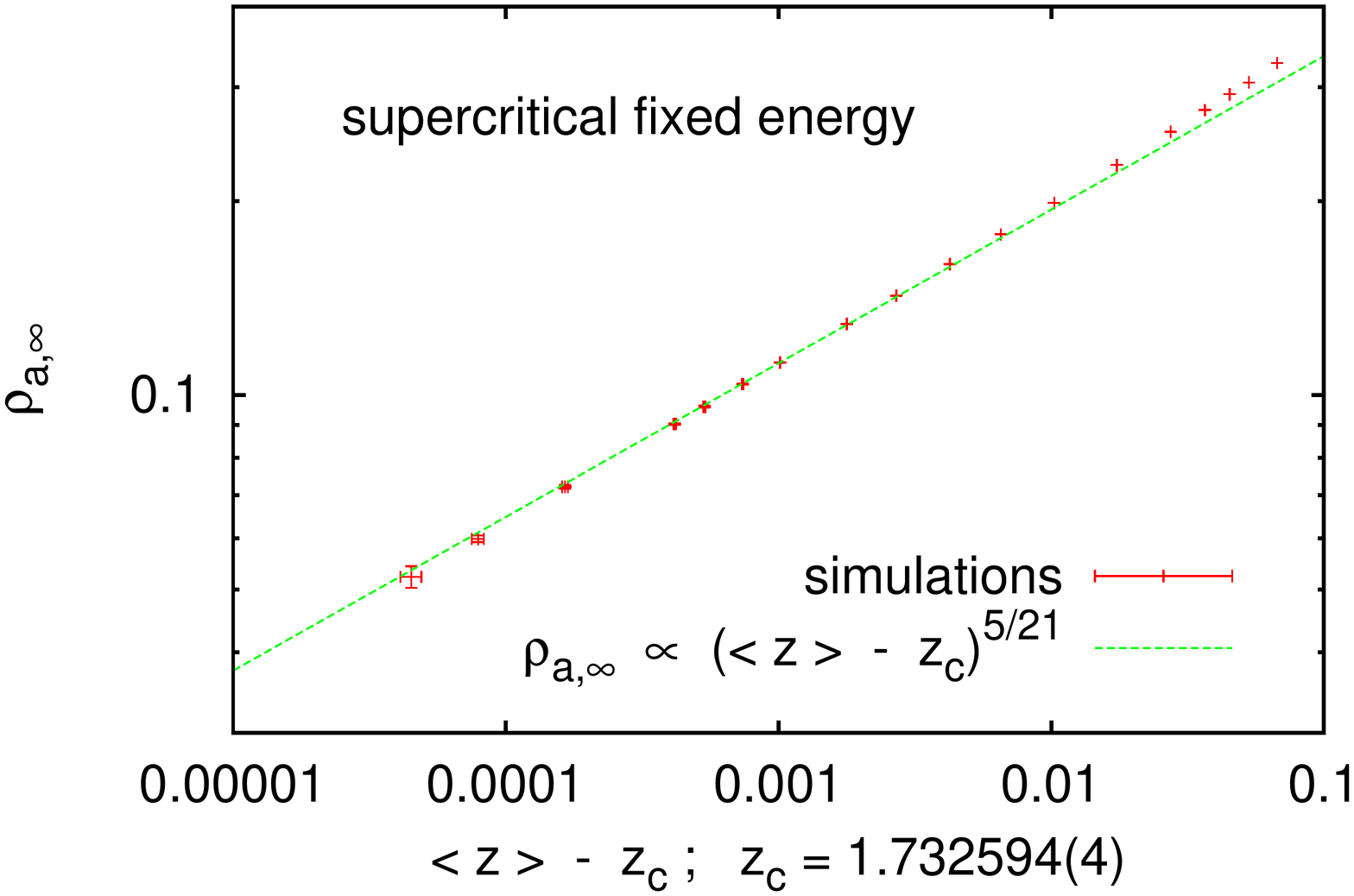}
 \vglue -5mm
 \par\end{centering}
 \caption{\label{supercrit-beta} (Color online) Log-log plot of $\rho_{a,\infty}$ against 
   $\langle z\rangle - z_c$. The straight line indicates the power law predicted by the 
   scaling theory. The best fit would indeed be obtained with an exponent $0.243(5)$. The 
   leftmost point was obtained by simulating 40 lattices with $L=10^6$ for $5\times 10^7$
   time steps.}
 \end{figure}

The approach to stationarity will be roughly exponential in the far supercritical regime,
but in the critical region it will follow a power law. In the latter region the difference
between the periodic initial state and the true NCS will become important, and we shall 
defer the discussion of this subtle case to a later subsection. Here it is sufficient to point 
out that in the worst case (i.e. closest to the critical point, where the correlation length
becomes comparable to $L$) the transient time increases as $L^z$. As we  have seen, the 
correlation length scales as $\xi \sim (\langle z\rangle - z_c)^{-\nu}$ with $\nu = 4/3$.
Thus we can use lattices of sizes up to $L\approx 10^6$, simulated over $5\times \approx 10^7$
time steps, to test Eq. (\ref{beta}) down to $\langle z\rangle - z_c \approx 0.00005$.

Fig.~\ref{supercrit-beta} shows results from such runs. Each point in this plot is obtained
from at least 40 such runs, and it was verified that the density of activity had become 
stationary. The straight line indicates the exponent 
\be
   \beta = 5/21     \label{beta_th}
\ee
that  follows from the scaling theory discussed below. The data shown in the figure would by 
themselves give a best fit $\beta = 0.243(5)$, compatible with the above. 

To obtain Eq.~(\ref{beta_th}), we notice first that FSS suggests that for finite $L$ and 
exactly at criticality $\rho_{a,L} \sim L^{-\beta/\nu}$. The number of topplings in large 
avalanches (those which dominate the higher moments of $s$) scales then as $L$ times this
density times the duration of the avalanches, 
\be
   s \sim L\times L^{-\beta/\nu} \times L^z. 
\ee
Assuming that $s\sim L^D$ with $D=9/4$ gives then
\be
    \beta = (1+z-D)\nu = 5/21.
\ee

\begin{table}
\begin{center}
\caption{Estimates of the critical exponent $\beta$ defined in Eq.~(\ref{beta}). The acronyms for 
   the various models are explained in the references.}
\label{table-beta}
\begin{tabular}{lll} \hline \hline
   0.24(3)     &  \cite{Bona-thesis}   &  overall Manna class \\ \hline
   0.42(2)     &  \cite{Dick:2001}     &  Manna    \\
   0.416(4)    &  \cite{Dickman-Tome}  &  restricted Manna   \\
   0.41(1)     &  \cite{Dick:2002}     &  restricted Manna   \\
   0.289(12)   &  \cite{Dick:2006}     &  restricted Manna   \\
   0.29(2)     &  \cite{Kockel}        &  CDP   \\
   0.28(2)     &  \cite{Ramasco}       &  CDP   \\
   0.382(19)   &  \cite{Lubeck:2004}   &  DCMM \\
   0.277(18)   &  \cite{BBBMH}         &  CCMM   \\
   0.308(2)    &  \cite{BBBMH}         &  CTTP   \\
   0.275(6)    &  \cite{BBBMH}         &  CLG     \\
   0.277(3)    &  \cite{Fiore}         &  modified CLG \\ \hline
   0.25(3)     &  \cite{Lesch}         &  qEW   \\
   0.33(2)     &  \cite{Duemmer}       &  qEW   \\
   0.250(3)    &  \cite{Kim-Choi},\cite{Song-Kim}\footnote{A more precise value was given in \cite{Kim-Choi}; the value cited here is the one given later in \cite{Song-Kim}}      &  qEW  \\ 
   0.245(6)    &  \cite{Ferrero}       &  qEW  \\ \hline
   0.396(5)    &  \cite{Lee}           &  Oslo \\  
   0.243(5)    &  present work         &  Oslo, direct fit \\
   5/21 = 0.2380... \hspace{2mm} &  present work \hspace{5mm} &  Oslo, scaling relation \\ \hline\hline
   0.2764...   & \cite{Hinrichsen}     &  DP\\ \hline\hline
\end{tabular}
\end{center}
\end{table}

There exist a large number of previous estimates of $\beta$, either for the Oslo model itself
or for other models which are supposed in the same (Manna) universality class, see 
Table \ref{table-beta}. They are all are much 
bigger, with one notable exception: $\beta = 0.24(3)$ was obtained in \cite{Bona-thesis}. 
All other estimates are supposedly more precise but outside our error bars. 
The problem in determining $\beta$ is obviously the large corrections to scaling which are
seen in Fig.~\ref{supercrit-beta}, and which require very large systems to be studied.
In Table \ref{table-beta} we quote also the value for DP. In many previous papers it was 
concluded that the Manna class has to be distinct
from DP, mainly because it has a {\it larger} value of $\beta$. We see now that the opposite
is true, and Manna $\neq$ DP because its $\beta$ is {\it smaller} than that of DP.

We include in  Table \ref{table-beta} also three estimates for the qEW model. Since the mapping
of the Oslo model onto interface pinning is such that the interface height is just the number 
of topplings, the activity density $\rho_a(t)$ is just the average speed of the interface at
time $t$. The value quoted in Table \ref{table-beta} is for the exponent (called $\theta$ in 
 \cite{Kim-Choi,Song-Kim}) that described how the speed increases in the de-pinned phase with 
the distance from the critical point, $v\sim (F-F_c)^\theta$. 

In \cite{Corral}, the relation between 
\be
   \beta = (1+z-D)/(3-D)    \label{CP}
\ee
was proposed. Although the numerical value of $\beta$ obtained thereby in \cite{Corral}
is different from ours, Eq.~(\ref{CP}) is satisfied by our exponents.  Together with
Eq.~(\ref{beta_th}) it gives 
\be
   \nu = 1/(3-D) .
\ee

 Finally, we note that for FES with deterministic toppling rules, it has been noted that the critical density at which infinite avalanches first appear, depends on the starting configuration \cite{Fey1}. For sandpiles with stochastic toppling rules, this is not a problem, and 
the SOC and FES versions of the Oslo model have the same critical  density.

\subsubsection{Subcritical single-seed avalanches}

The next easy case are isolated avalanches in the subcritical phase. We again start with
a periodic configuration (this time with $\langle z \rangle <z_c$). We declare all
sites (including those with $z_i=2$) as stable. To trigger an avalanche we simply pick
a random site among those with $z_i=2$ and declare it as unstable. This avalanche
 will be finite with probability 1 and will have also finite size, thus we can follow 
its evolution until it dies and the configuration is stable again. After that, we again
declare a random site with $z_i=2$ as unstable and repeat. 

By measuring avalanche sizes, we verified that transients are very short: Average avalanche
sizes converge within error bars to a stationary value, after each site has toppled less 
than 1000 times, even when $\langle z\rangle$ is very close to $z_c$. Thus we can again 
get good statistics for lattices with $L$ up to $10^6$. Lattices of this size are indeed 
needed in order to avoid finite size effects, if we want to measure very close to the critical 
point. 

 \begin{figure}
 \begin{centering}
 \includegraphics[scale=0.30]{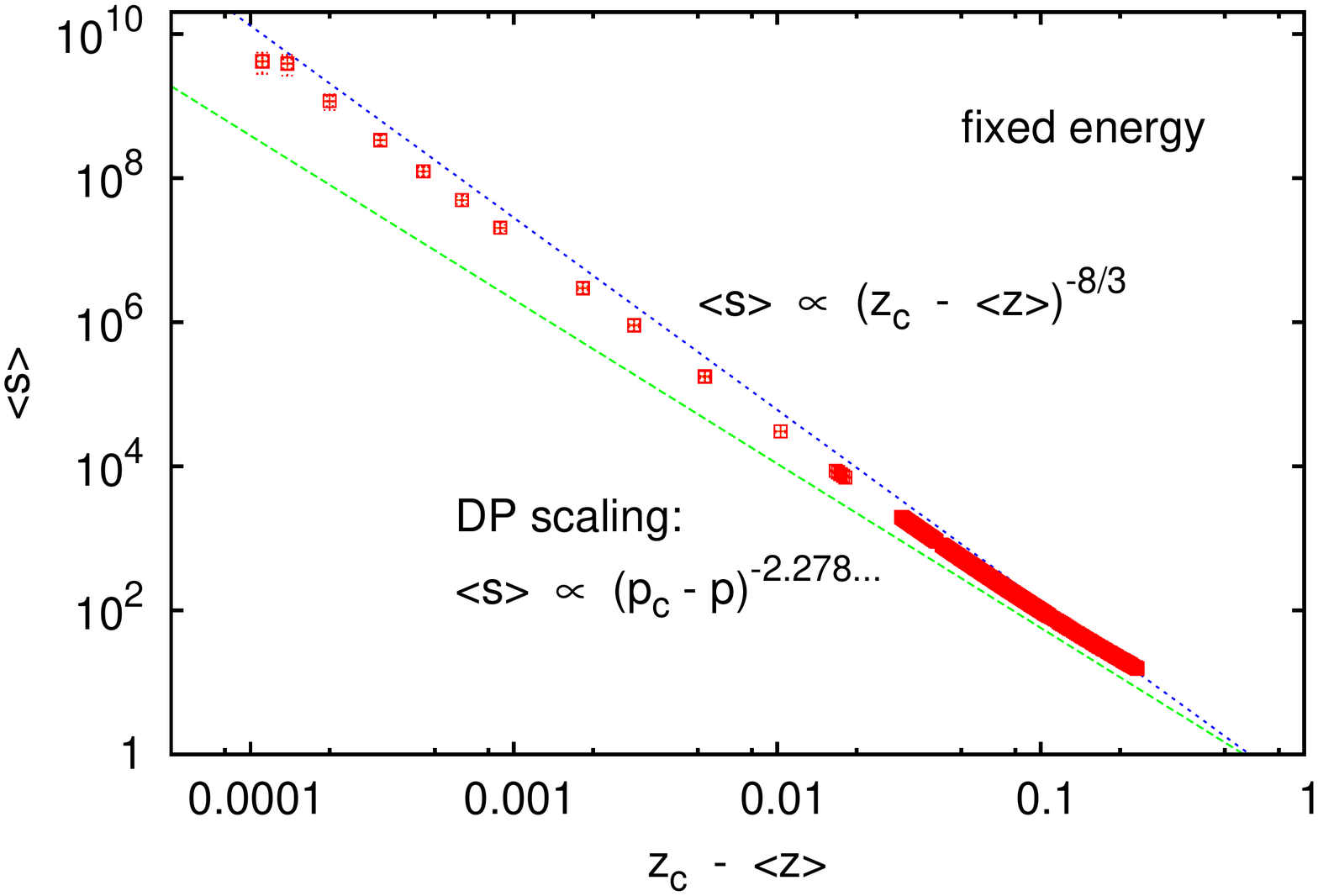}
 \vglue -5mm
 \par\end{centering}
 \caption{\label{clustersize-subcrit} (Color online) Log-log plot of $\langle s\rangle$ against
   $z_c - \langle z\rangle$ for subcritical avalanches in the FES version. 
   The upper straight line indicates the power law predicted by the
   scaling theory. The best fit would indeed be obtained with an exponent $2.68(2)$. 
   The lower straight line shows the behavior that would have been expected, 
   if the Oslo model were in the DP universality class. The leftmost point was 
   obtained by simulating 3600 avalanches on a lattice with $L=5\times 10^5$.}
 \end{figure}

Results are shown in Fig.~\ref{clustersize-subcrit}, where we plot $\langle s\rangle $ against 
the distance from the critical point. We see a clear power law in the critical region, but 
important scaling corrections when $z_c - \langle z\rangle$ becomes large. The latter could 
have suggested that the power is  the  same as  for DP, but this is actually excluded: While the DP exponent
$\gamma$, defined as
\be
   \langle s\rangle \sim (z_c - \langle z\rangle)^{-\gamma},    \label{gamma}
\ee
is 2.278 \cite{Hinrichsen}, a direct fit to our data would give $\gamma=2.68(2)$. The 
upper straight line shown in Fig.~\ref{clustersize-subcrit} represents our scaling conjecture
\be
   \gamma =  2\nu = \;\; 8/3,
\ee
which follows from $ \langle s \rangle \sim L^2$ via FSS and 
which is fully compatible with the directly measured value.

\subsubsection{Finite-size scaling: Critical avalanches on finite lattices}

Exactly at the critical point, we cannot use either of the two strategies discussed in the previous 
subsections. In this subsection we simulate single avalanches, triggered in the way described
above, on lattices of sufficiently small $L$ so that we can follow all of them until 
they die. Avalanche distributions will be discussed below, but first we shall discuss moments
of their sizes and durations.

 \begin{figure}
 \begin{centering}
 \includegraphics[scale=0.30]{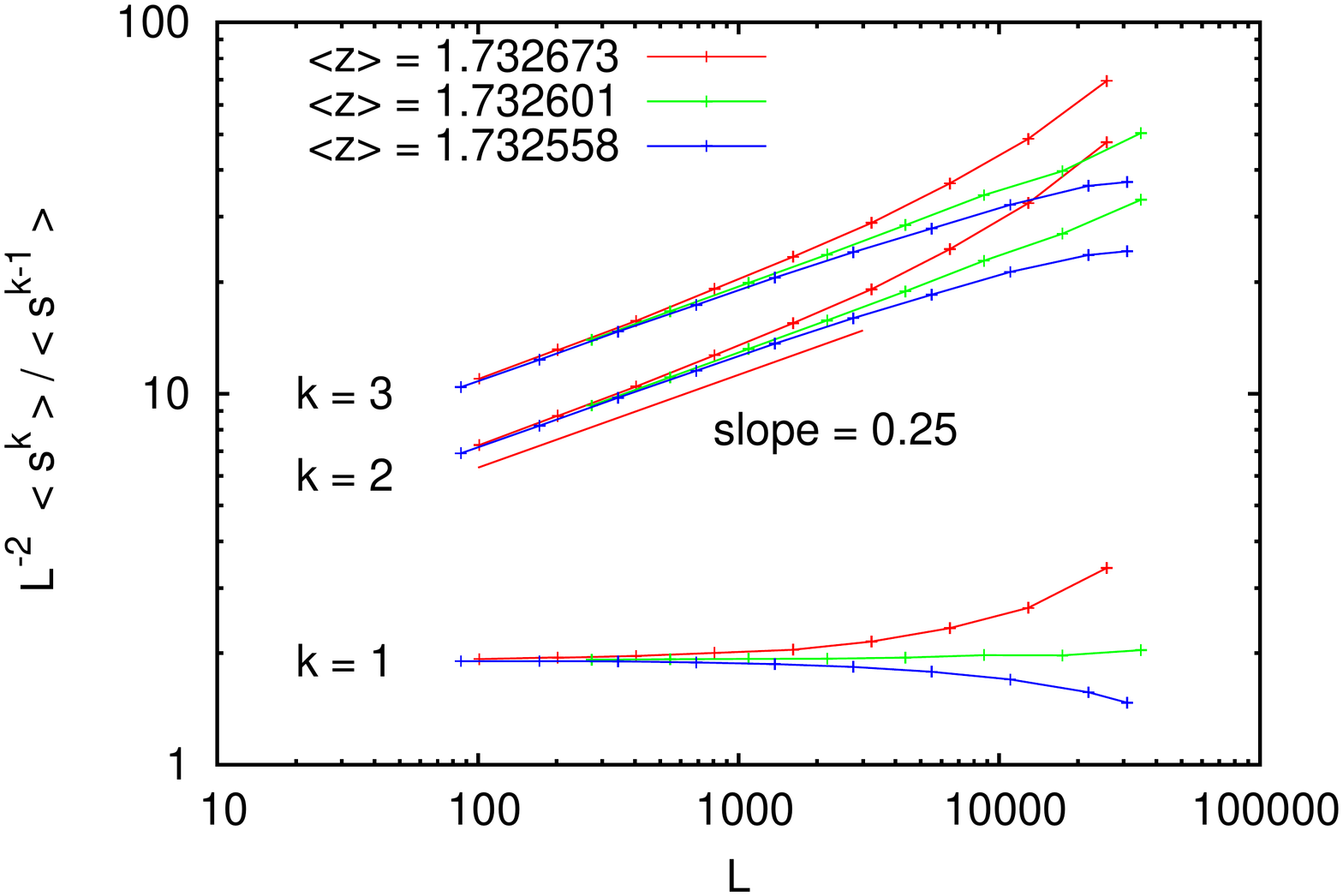}
 \vglue -5mm
 \includegraphics[scale=0.30]{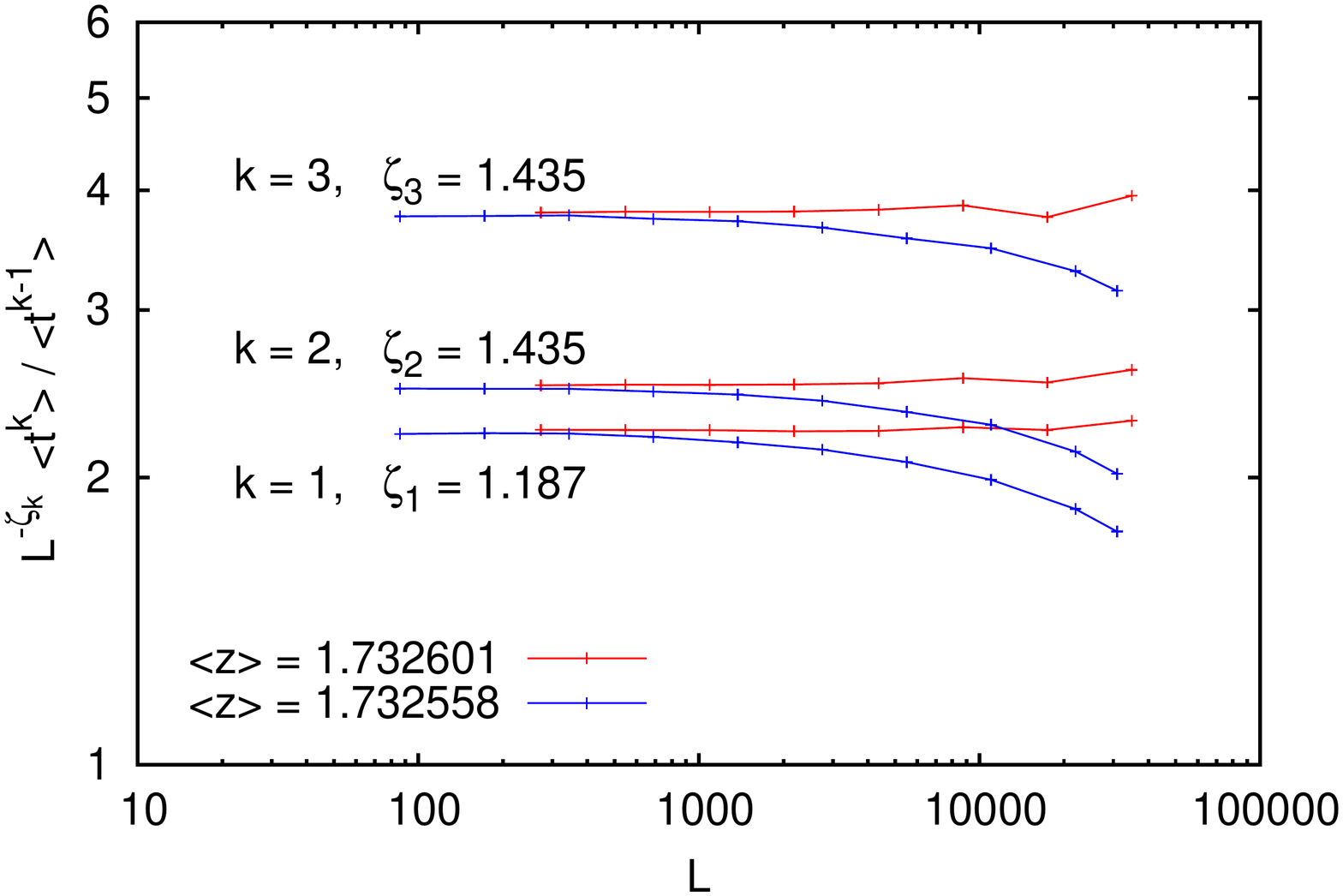}
 \vglue -5mm
 \par\end{centering}
 \caption{\label{finitesize-moments} (Color online) Panel (a): Log-log plot of $\langle s\rangle$ 
   (lowest triple of curves) and $\langle s^k\rangle/\langle s^{k-1}\rangle$ for $k=2$
   and 3 (topmost triples) against $L$. Each curve in each triple corresponds to 
   a different periodic start configuration (periods 101, 273, and 86 from top to
   bottom), and each value of $L$ is a multiple of the corresponding period. The 
   data are divided by $L^2$ which makes the central curve in the lowest triple
   horizontal and makes the central curves in the two upper triples scale as $L^{1/4}$
   (straight line). Panel (b): Analogous results size-weighted for moments of avalanche
   life times. This time only data for the two smaller values of $\langle z \rangle$
   are shown, and all data are divided by powers of $L$ which make the critical curve
   of each pair horizontal.}
 \end{figure}

Moments of the avalanche size $s$ are shown in Fig.~\ref{finitesize-moments}a, while moments of their
life times are shown in panel (b). The latter were computed as in subsection \ref{open-aval-tim}.
In panel (a) we show results for three values of $\langle z \rangle$ close to $z_c$, while results
for only two of them are shown in panel (b). 

The bottom triple of curves in panel (a) show 
$\langle s \rangle/L^2$. These values are independent of $L$ within errors for the central curve
which is essentially critical, showing that 
\be
   \langle s \rangle \sim L^2
\ee
for critical avalanches in the FES ensemble, just as it is for bulk driven avalanches on open 
lattices. This is not entirely trivial, since the argument predicting this scaling for open
systems no longer holds. The fact that we nevertheless find the same scaling in both ensembles
is a strong indication that the avalanches have the same statistical properties.

The same conclusion is reached by looking at the two upper triples of curves in panel (a) which show
the ratios $L^{-2} \langle s^k \rangle/\langle s^{k-1} \rangle$ for $k=2$ and 3. Here the 
critical curves show that all moments satisfy exactly the same critical scaling $ \langle s^k \rangle
\sim L^{2+9k/4}$ as for open systems.

\begin{figure}
 \begin{centering}
 \includegraphics[scale=0.30]{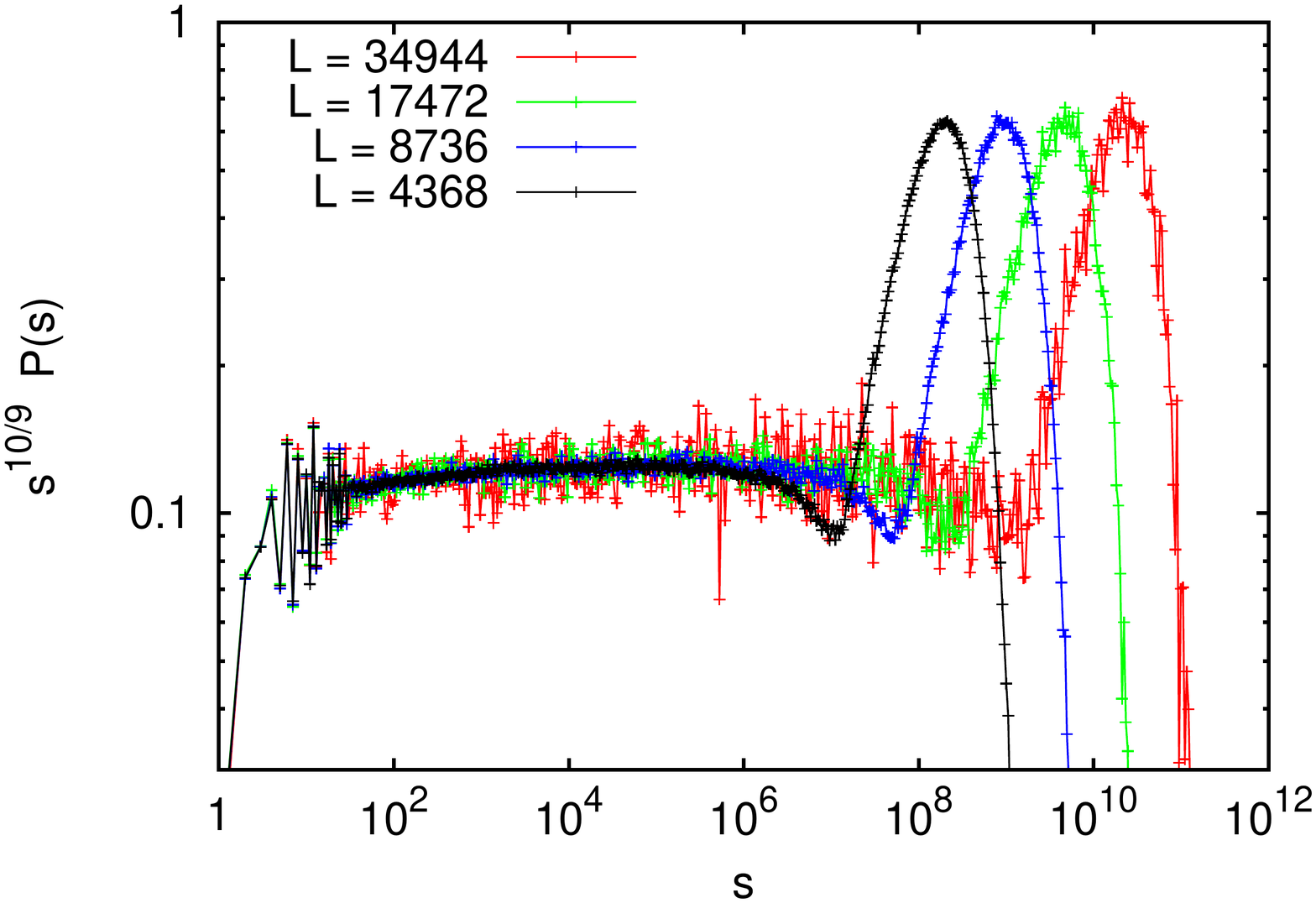}
\vglue -5mm
 \par\end{centering}
 \caption{\label{Ps-closed} (Color online) Log-log plot of the avalanche size distributions for 
   the fixed-energy version at $\langle z \rangle = 1.732601$, and for different lattice sizes
which are all multiples of 273. The actually plotted data are $s^{10/9} P(s)$.}
\end{figure}

 \begin{figure}
 \begin{centering}
 \includegraphics[scale=0.29]{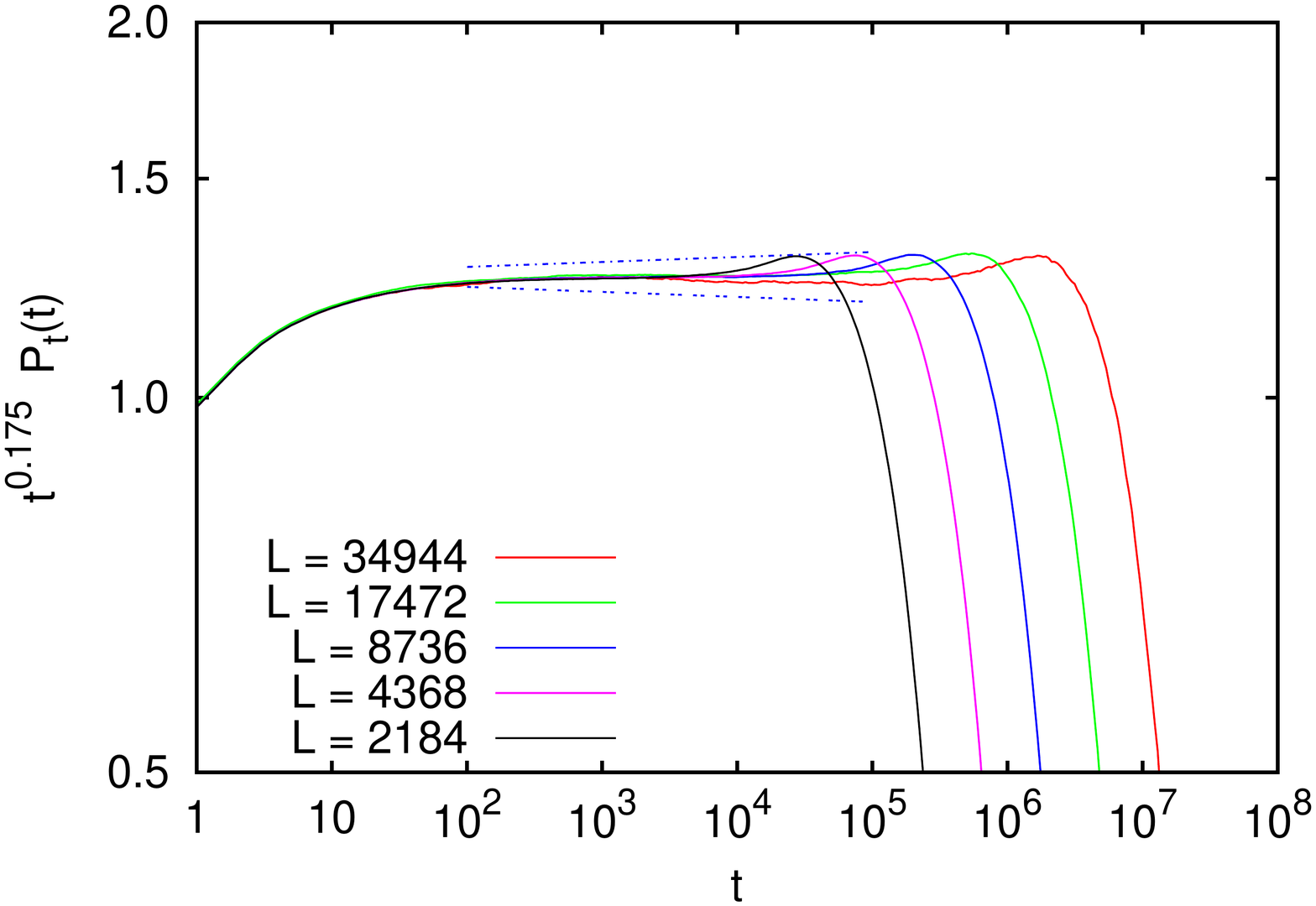}
 \vglue -6mm
 \includegraphics[scale=0.29]{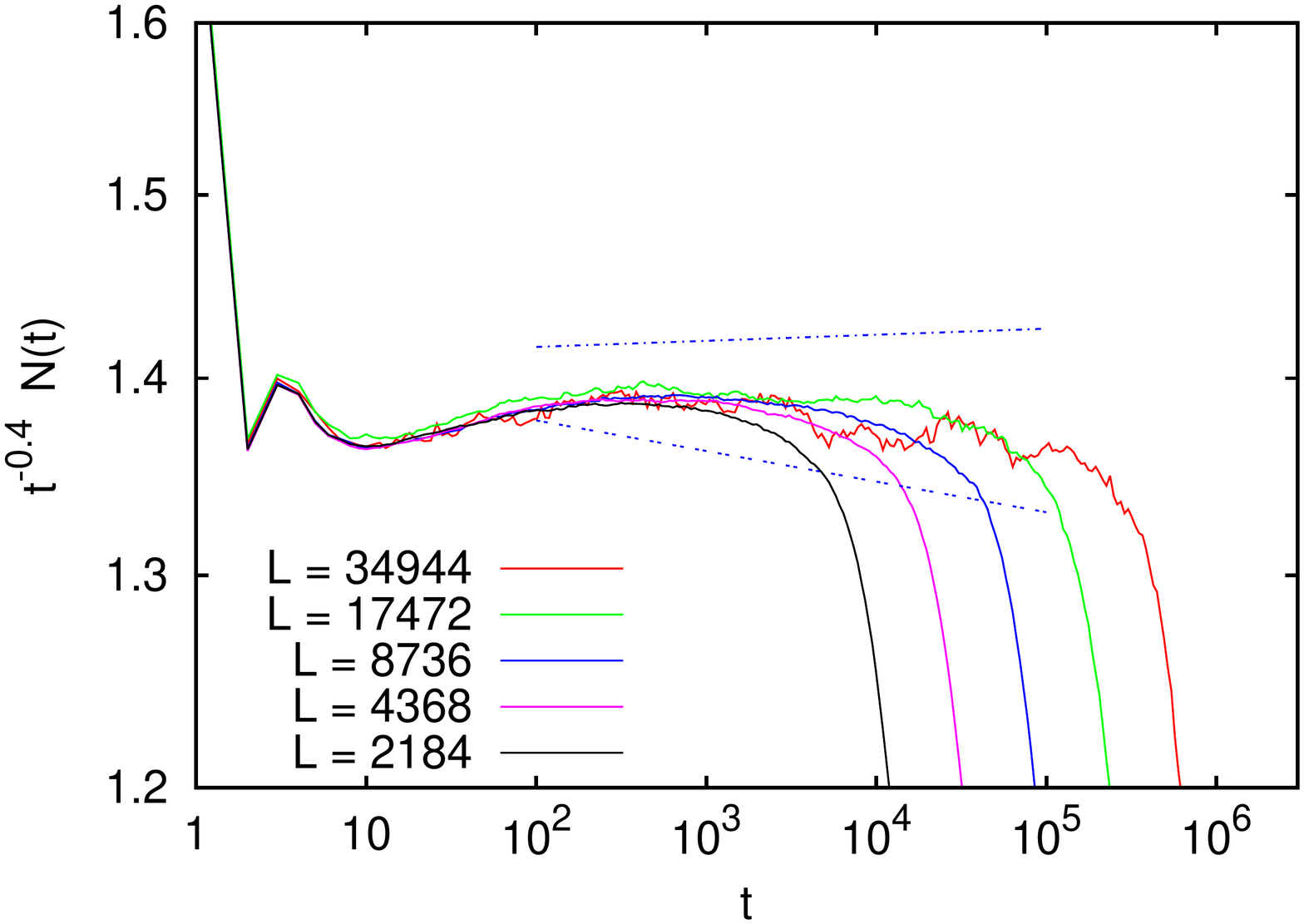}
 \vglue -6mm
 \includegraphics[scale=0.29]{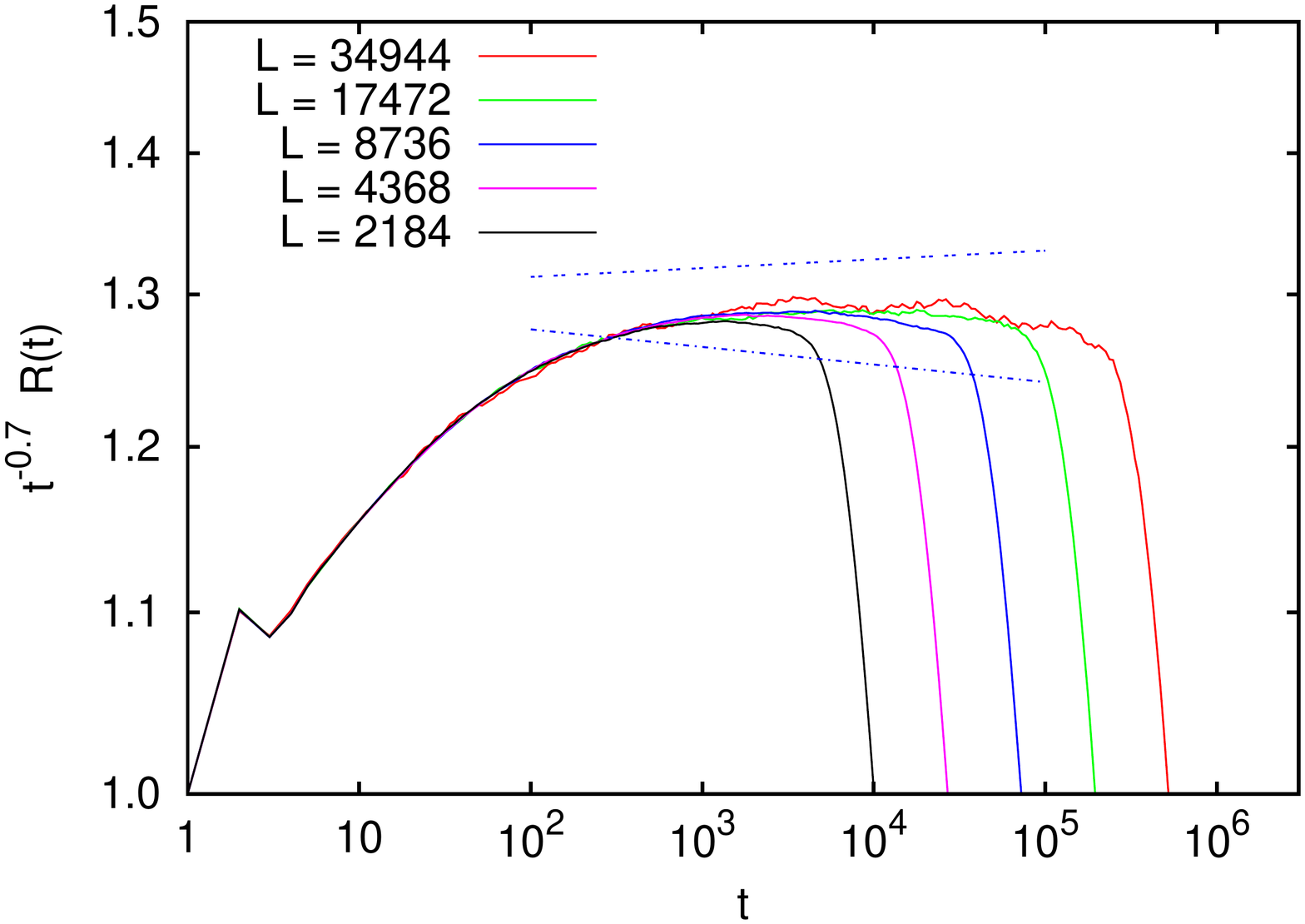}
 \vglue -6mm
 \par\end{centering}
 \caption{\label{finitew-PNR-t} (Color online) Log-log plots of $P_t(t,L)$ (panel a), $N(t,L)$ (panel b), and
   $R(t,L)$ (panel c). In each plot, the raw data were divided by the conjectured power laws, 
   i.e. the actually plotted data are $t^{3/8} P_t(t,L), t^{-2/5} N(t)$, and $t^{-7/10} R(t)$. 
   As in Fig.~\ref{Ps-closed}, $R(t)$ is the rms. distance from the first toppling. In each panel,
   the straight lines indicate the error bars mentioned in the text.}
 \end{figure}

The data for avalanche durations shown in panel (b) tell a similar story.
The two topmost pairs of curves show that $\langle T^k \rangle/\langle T^{k-1} \rangle$ with
$k\geq 2$ scale with the same power of $L$, which is within errors the same as the exponent
$z$ found also in the open case (the fitted value of $z$ now is 1.438(10), while our previous
estimate was $10/7 = 1.4286$). In agreement with the bulk-driven open case, $\langle T \rangle$
now also shows good scaling, with exponent $1.187(10)$. 

 Distributions of avalanche sizes and of the three time dependent properties $P_t(t), N(t)$, and 
$R(t)$ are shown in the  Fig. \ref{finitew-PNR-t} (we did these simulations  at $\langle z \rangle =1.732601$  before arriving at the final estimate for $z_c$ in Eq. (\ref{z_c}),
but the small  deviation from the best estimate of $z_c$  should not matter
much). For $P(s)$ we show only a plot analogous to Figs.~\ref{bdry-tau}b
and \ref{Ps-bulk}, where we divided the raw data by the supposed power law $s^{-\tau_{\rm bulk}}$, 
see Fig.~\ref{Ps-closed}. Although the scaling is not perfect, 
the improvement compared to the bulk driven case  with open boundaries shown in Fig.~\ref{Ps-bulk} is dramatic. Now 
we can argue rather convincingly that $\tau_{\rm bulk}=10/9$. The best estimate based on this plot
alone would be $\tau_{\rm bulk}=1.10(1)$, based both on the scaling region and on the heights  of 
the peaks (which should also scale as $s^{-\tau_{\rm bulk}}$).

The three panels of Fig.~\ref{finitew-PNR-t} show $t^{3/8} P_t(t), t^{-2/5} N(t)$, and $t^{7/10}R(t)$.
The actual best exponent estimates based on these plots alone would be $\delta_{\rm bulk}=0.175(3),
\eta_{\rm bulk}=0.398(3)$, and $z_{\rm bulk}=0.699(3)$, as indicated by the dashed straight lines 
in each panel.

 Within the statistical errors, the sum $\delta+\eta$ is the same as for open boundary driven 
systems,
\be
   \delta_{\rm bulk}+\eta_{\rm bulk} = \delta+\eta = 23/40.
\ee
This means that the activity $N(t)/P_t(t)$ per {\it surviving} avalanche shows the same scaling in 
both cases. It should indeed scale as the product of the activity density in the active region
(which scales as $t^{-\beta/z/\nu}$, as we shall see later) times its spatial extent (which scales
as $t^{1/z}$). Therefore
\be
   \delta+\eta = (1-\beta/\nu)/z   \label{delta-eta}
\ee

\subsubsection{Simulations involving termination of the evolution in non-stationary states}

So far we have only discussed simulations of the fixed-energy Oslo model where it was either not 
necessary to terminate the evolution because avalanches died anyhow, or where the system had 
already reached a stationary state. For simulating systems very close to the critical point,
it seems however necessary to terminate the evolution before avalanched have died or before 
stationarity is reached. As we shall see, extreme case is needed in interpreting such simulations.

 \begin{figure}
 \begin{centering}
 \includegraphics[scale=0.3]{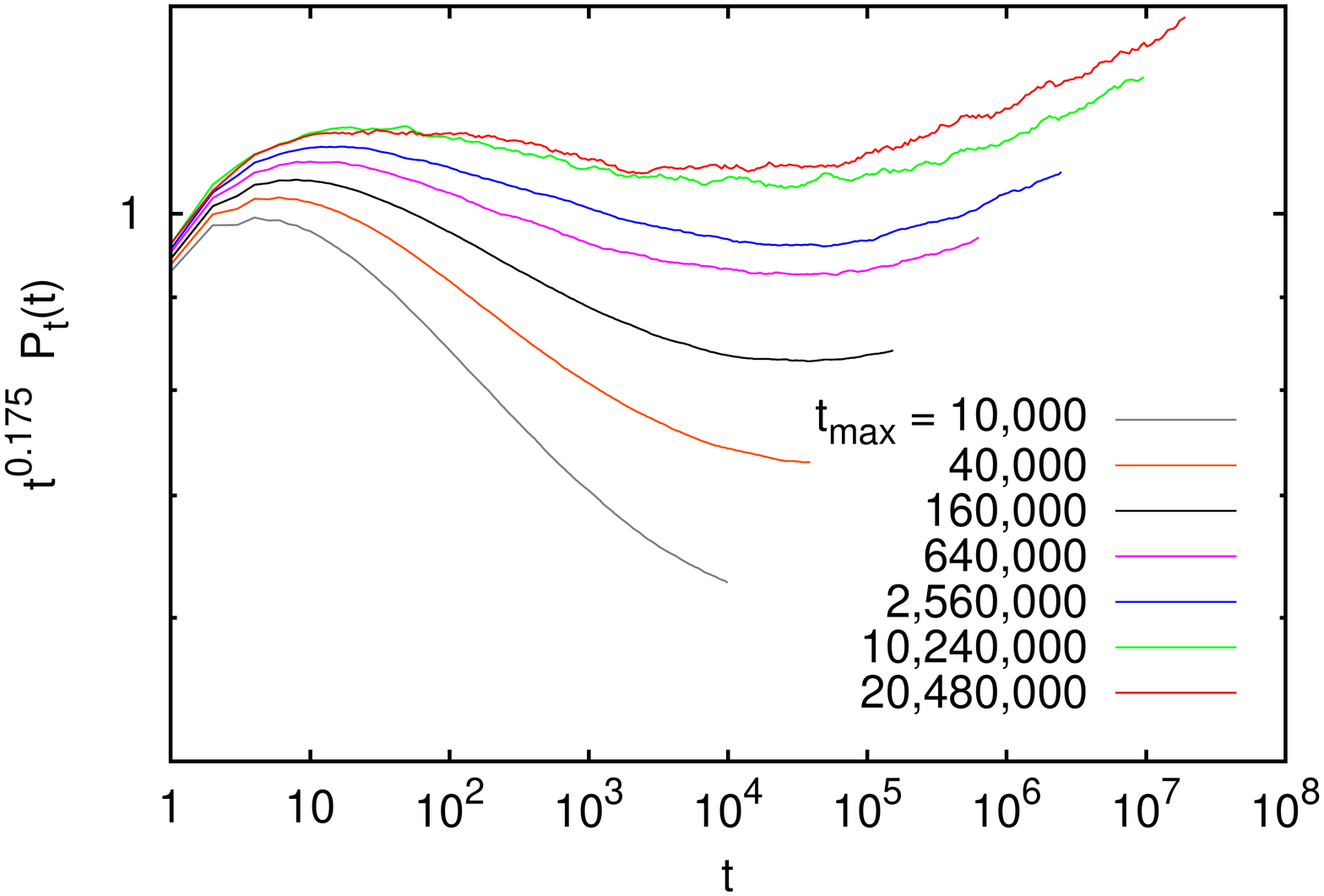}
 \vglue -5mm
 \par\end{centering}
 \caption{\label{P_t-cutoff} (Color online) Log-log plots of rescaled survival probabilities $t^{3/8} P_t(t)$,
   if the evolution of each still surviving avalanche is terminated at time $t_{\rm max}$. Here 
   the largest lattices (for the largest $t_{\rm max}$) had $L=2^{21}$. The average stress was 
   $1.732591$ (i.e., slightly subcritical), but it was verified that results were indistinguishable
   at slightly supercritical $\langle z\rangle$. From our previous simulations we would have 
   expected the curves to become horizontal for large $t$ and large $t_{\rm max}$.}
 \end{figure}

Let us first discuss simulations of single avalanches, triggered by declaring random sites with
$z_i=2$ in an otherwise stable configuration as unstable. If an avalanche survives for a time 
$> t_{\rm max}$, its evolution is cut off by declaring all sites with $z_i=2$ as stable. Since 
only very few avalanches survive until $t_{\rm max}\gg 1$, one might hope that this gives reasonable
results if $t_{\rm max}$ is sufficiently large. Indeed, this strategy is rather common in studies
of FES sandpile models. Fig.~\ref{P_t-cutoff} shows survival probabilities $P_t(t)$ on very 
large lattices and at $\langle z\rangle$ very close to criticality, for different values of the 
cutoff $t_{\rm max}$, ranging from $10^4$ to $2\times 10^7$. Since we have
multiplied the data by the factor $t^\delta$, we should have expected the curves to become horizontal
for large $t$ and large $t_{\rm max}$. They do indeed become horizontal for $1 \ll t \ll t_{\rm max}$,
but estimating $\delta$ from the data with largest $t$ would give gross inconsistencies. The same 
is true for $N(t), R(t)$, and $P(s)$. In all these cases we could get consistent results by 
first taking $t_{\rm max}\to\infty$ and only then going to large $t$, but this would not be 
very practical.

 \begin{figure}
 \begin{centering}
 \includegraphics[scale=0.3]{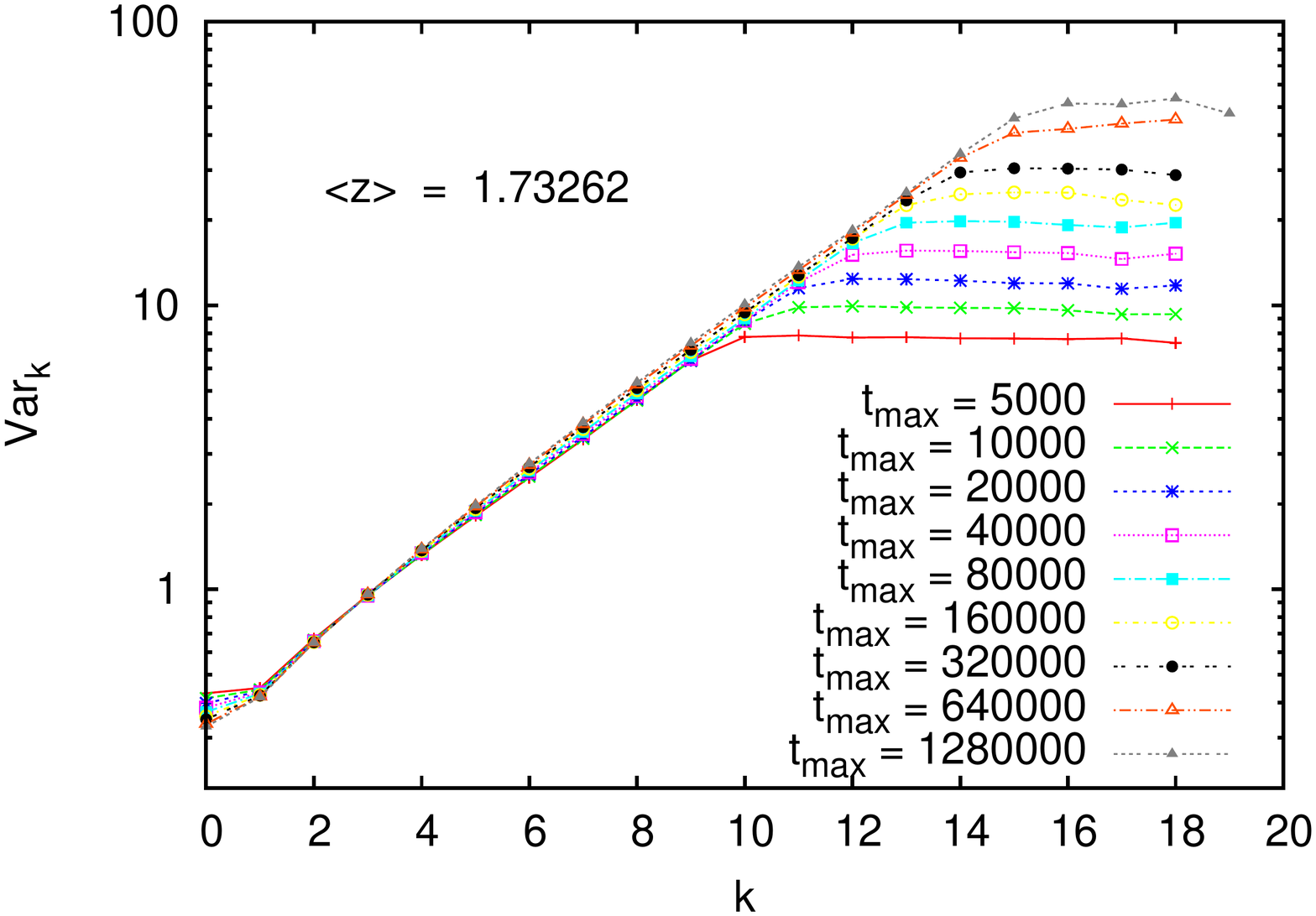}
 \vglue -5mm
 \par\end{centering}
 \caption{\label{hyper-cutoff} (Color online) Log-log plot of variances of the total stress on intervals of 
   length $2^k$, ${\rm Var}_k = {\rm Var}[Z_k]$ with $Z_k = \sum_{i=i_0}^{i_0+k-1 {\rm mod} L} z_i$.
   Each curve was obtained by terminating the evolution of avalanches at $t_{\rm max}$, and 
   the total CPU time used for each curve was roughly the same. Lattice sizes were up to $4\times
   10^6$, and $\langle z\rangle$ is very close to critical.}
 \end{figure}

The reason for this failure is that if avalanche evolution is stopped at $t_{\rm max}$, also 
the correlations in the NCS needed to make it  critical and hyperuniform cannot develop at distances 
$> t_{\rm max}^{1/z}$. Essentially,  criticality and hyperuniformity are then confined to scales $< t_{\rm max}^{1/z}$
and correlations at larger scales are those of the initial state and different from those in 
the NCS, even if the simulation is kept going
for extremely long times, see Fig.~\ref{hyper-cutoff}. Since total CPU time was roughly the 
same for each curve in Fig.~\ref{hyper-cutoff}, it seems unlikely that longer runs would 
establish critical correlations  on substantially larger scales.

Thus simulations of single avalanches where the evolution is stopped at finite times seem 
not very useful. But simulations near criticality where a finite fraction (50\%, say) of sites 
with $z_i=2$ are initially unstable are useful, and are crucial for understanding scaling. 
Let us denote by $\epsilon = \langle z\rangle-z_c$ the distance from the critical point.
Naively, one should expect that activity satisfies in this case a finite size scaling (FSS)
ansatz
\be
    \rho_a(t,L,\epsilon) \sim t^{-\theta} F(\epsilon L^{1/\nu}, \epsilon t^{1/\nu_t}) \label{FSS}
\ee
with $\nu_t = z\nu$. In order to agree with Eq.~(\ref{beta}), the scaling function $F(x,y)$ 
has to scale for $y\to 0$ and $x\to\infty$ as $y^\beta$ and furthermore $\theta = \beta/\nu_t$.
The problem is of course that we expect this to hold when the state at $t=0$ is a NCS, but 
we have no foolproof way to construct one. Even worse, a NCS would have no unstable sites. 
In studying single avalanches, it seems reasonable that declaring a single $z_i=2$ site
as unstable should be a negligible perturbation, but now we want to make a finite fraction 
unstable. It is far from obvious what effect this has, but we can turn to simulations to 
find out numerically.

 \begin{figure}
 \begin{centering}
 \includegraphics[scale=0.3]{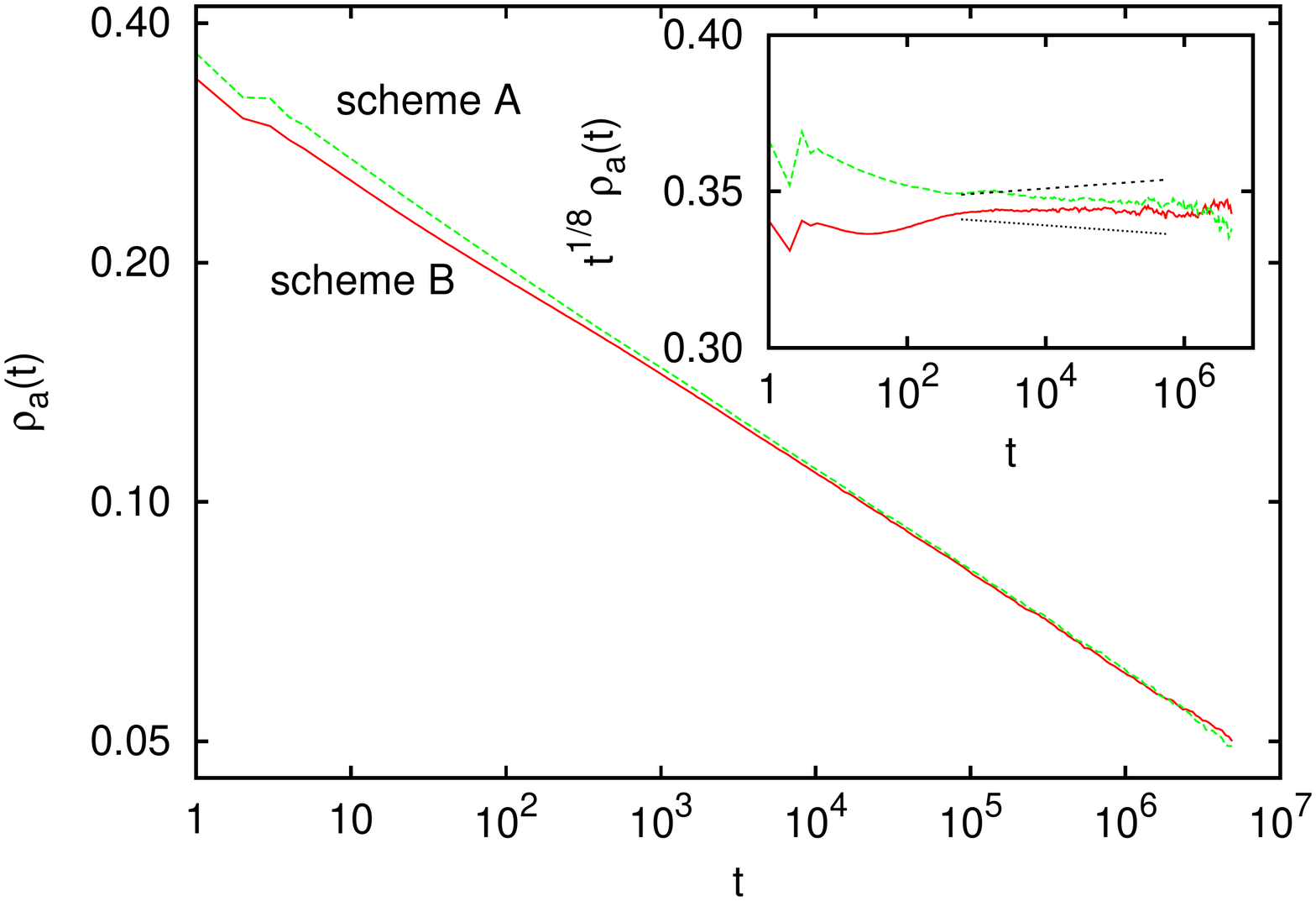}
 \vglue -5mm
 \par\end{centering}
 \caption{\label{decay-359} (Color online) Main plot: Log-log plot of $\rho_a(t)$ for 
   $\langle z\rangle = 359/208=1.732596$ and $L=524140$. For the upper curve (scheme A) 
   each run started from an periodic configuration, while the lower curve (scheme B) used 
   for each run the final state of the previous run, with half of the sites with $z_i=2$
   declared as unstable. Each curve is based on 508 runs. The inset shows the same data 
   multiplied by $t^{1/8}$. They coincide for $t>10^4$ within statistical errors. The 
   straight lines in the inset indicate the error $\pm 0.002$ of $\theta$.}
 \end{figure}

Assume we want to use Eq.~(\ref{FSS}) to estimate $\theta$ from simulations up to time 
$t_{\rm max}$, and let us assume that declaring half of the stable $z_i=2$ sites as 
unstable does not create any problem (we shall come back to this later). If we rule out
the option that we make first auxiliary runs up to $t\gg t_{\rm max}$ in order to be 
sure that we have  critical correlations  up to and beyond the needed length scale, two
options are left:
\begin{itemize} 
\item We start each run from an uncorrelated periodic configuration, hoping that correlations 
build up sufficiently rapidly so that at late times the evolution proceeds effectively in 
a NCS (scheme A);
\item We keep the configuration of the previous run and declare half of the $z_i=2$ sites
as unstable (scheme B).
\end{itemize}
If both schemes lead to the same results, it is reasonable to assume that the results are 
reliable.
   
 \begin{figure}
 \begin{centering}
 \includegraphics[scale=0.3]{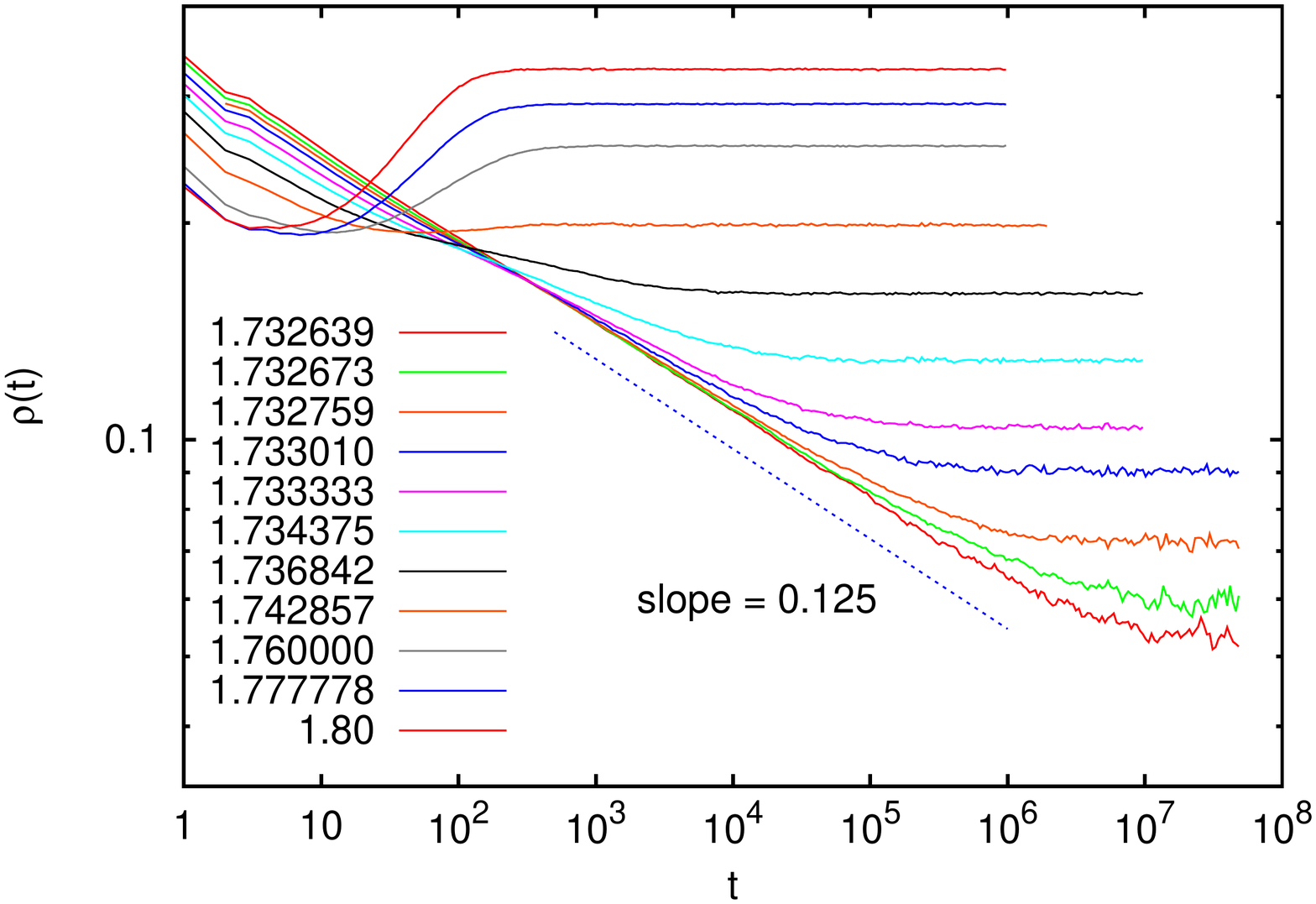}
 \vglue -5mm
 \par\end{centering}
 \caption{\label{supercrit-decay} (Color online) Activity densities for supercritical FES
   systems, large enough to have negligible finite size effects. Initial configurations were 
   chosen according to scheme A. The numbers in the legend
   indicate $\langle z\rangle$ (top curve has largest $\langle z\rangle$)}.
 \end{figure}

 \begin{figure}
 \begin{centering}
 \includegraphics[scale=0.3]{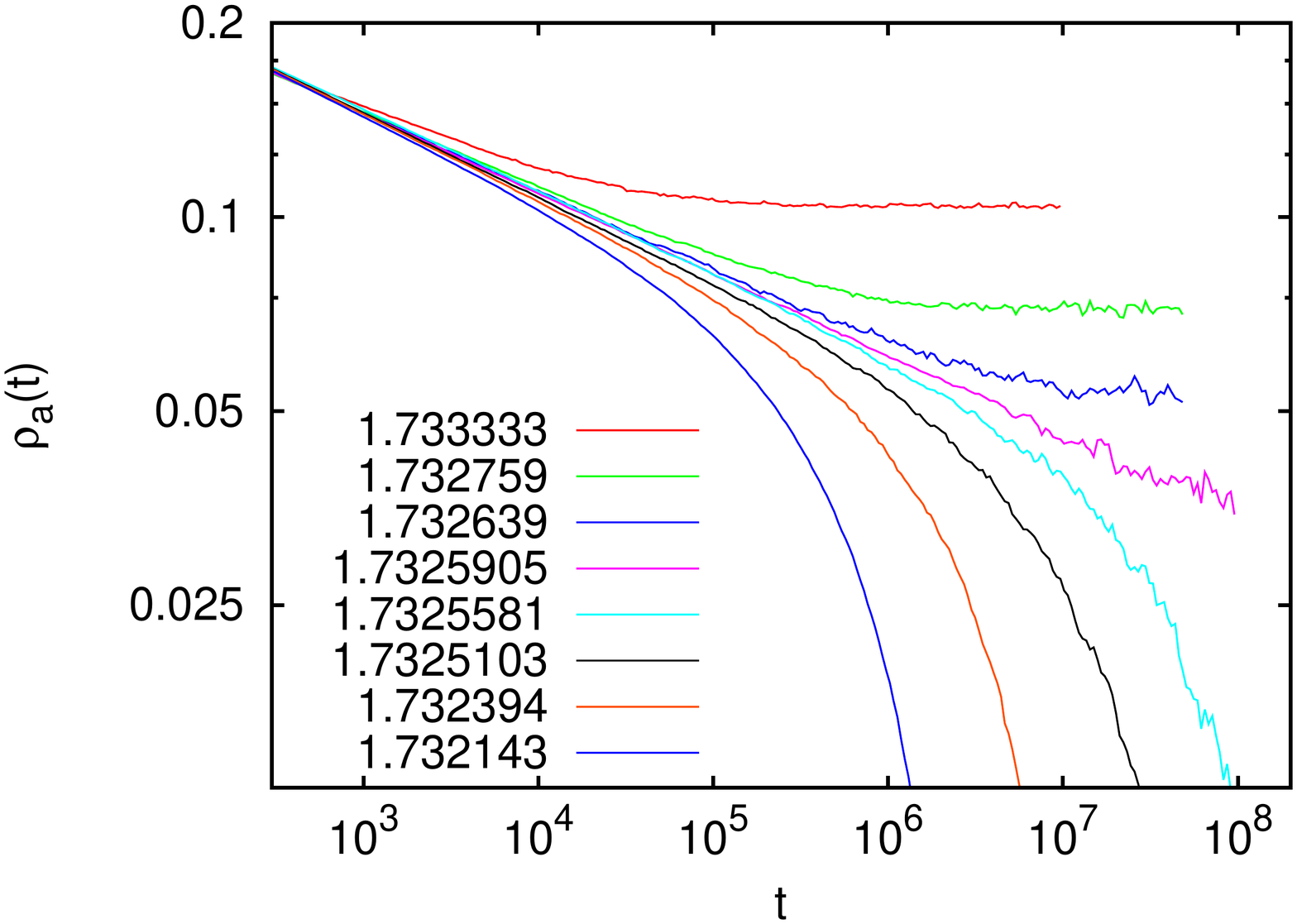}
 \vglue -5mm
 \includegraphics[scale=0.3]{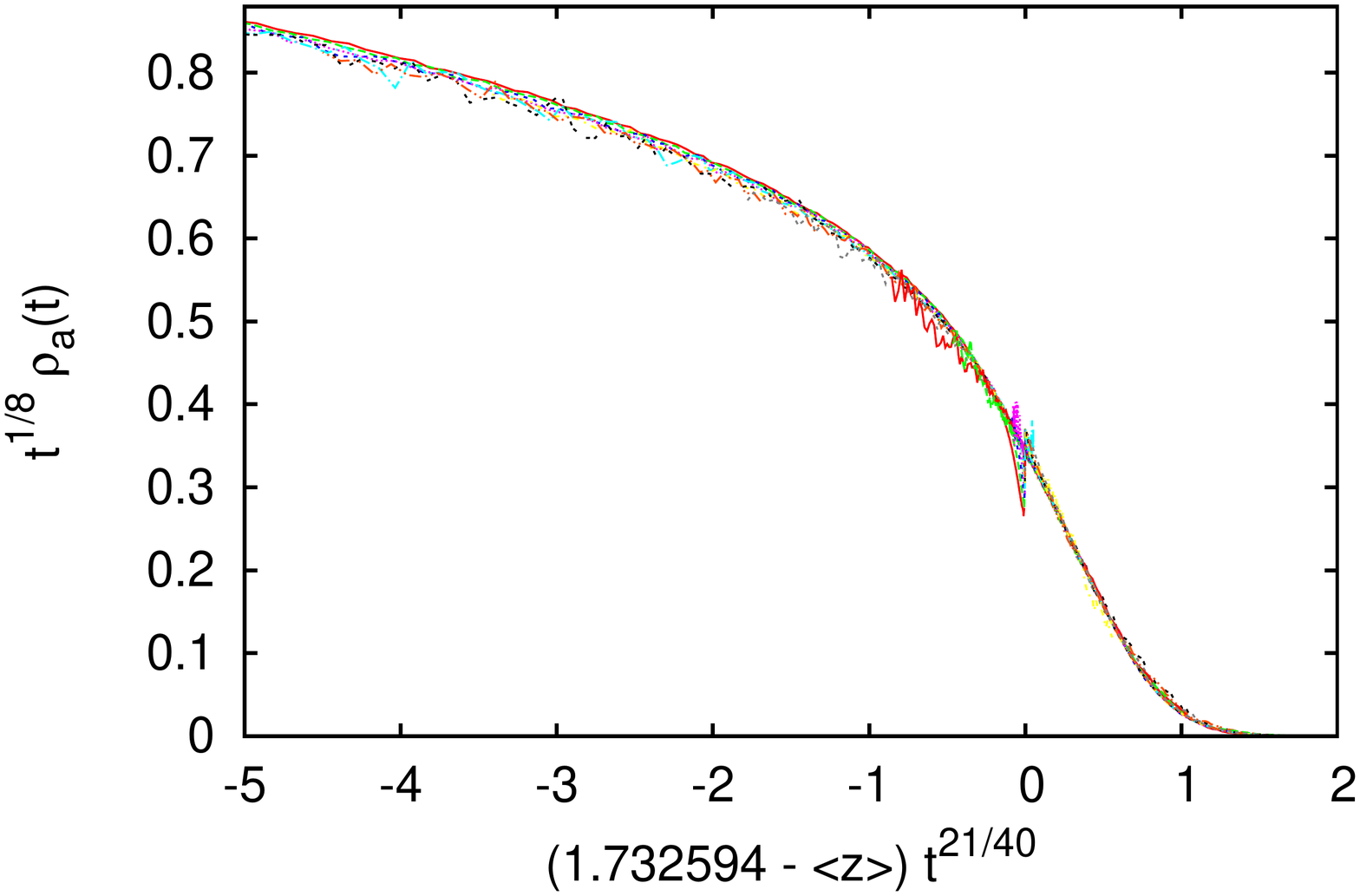}
 \vglue -5mm
 \par\end{centering}
 \caption{\label{crit-decay} (Color online) Panel a: Activity densities for near critical FES
   systems. Initial conditions used scheme A, and lattices are large enough to have negligible 
   finite size effects. The numbers in the legend
   indicate $\langle z\rangle$ (top curve has largest $\langle z\rangle$). Panel b: Data 
   collapse obtained by plotting $t^{1/8} \rho_a(t)$ against $z_c - \langle z\rangle t^{\nu_t}$
   with $\nu_t = 21/40$. There are altogether 27 curves overlaid in this plot, for 
   $\langle z\rangle$ ranging from 1.7143 to 1.7391.}
 \end{figure}

Results obtained with these two schemes are shown in Fig.~\ref{decay-359}. There we used 
a large enough lattice ($L\approx 2^{15}$) and $\langle z\rangle$ sufficiently close to 
$z_c$ that we expect a pure power law for large $t$. Both schemes
lead indeed to the same power law 
\be
   \rho_a(t) \sim t^{-\theta} \;,\quad \theta = 1/8 \pm 0.002.
\ee
On the other hand, both schemes show corrections for intermediate $t$. For scheme A they
seem to be a simple power law, but for B they are more complicated: There is a 
depletion for $10 < t < 10^3$ which indicates that declaring half of the $z_i=2$ sites
in the NCS as unstable is indeed a too violent perturbation. This is even more pronounced
for supercritical simulations, where scheme B gives very deep minima for intermediate $t$
(see Fig.~\ref{supercrit-decay}).
We thus conclude that scheme A gives, in spite of showing substantial finite-$t$ 
corrections, more reliable results that are easier to interpret. Final results are 
shown in Fig.~\ref{crit-decay}. Panel a shows $\rho_a(t)$ against $t$ for a few selected
near-critical values of $\langle z\rangle$, while panel b shows a collapse plot for 
the entire set of data in a rather wide range of $\langle z\rangle$. The structure near 
the the origin is due to the finite-time corrections mentioned above. Apart from that
we see a perfect data collapse, indicating that indeed $\theta = 1/8$ and 
\be
   \nu_t = 40/21     \label{nu_t}.
\ee
Notice that Eq.~(\ref{delta-eta}) can now be written as $\delta+\eta+\theta = 1/z$, in 
which form it is just the generalized hyperscaling relation for systems with 
multiple absorbing states proposed in \cite{Mendes}.

\subsection{Bulk driven  open systems}

 \begin{figure}
 \begin{centering}
 \includegraphics[scale=0.3]{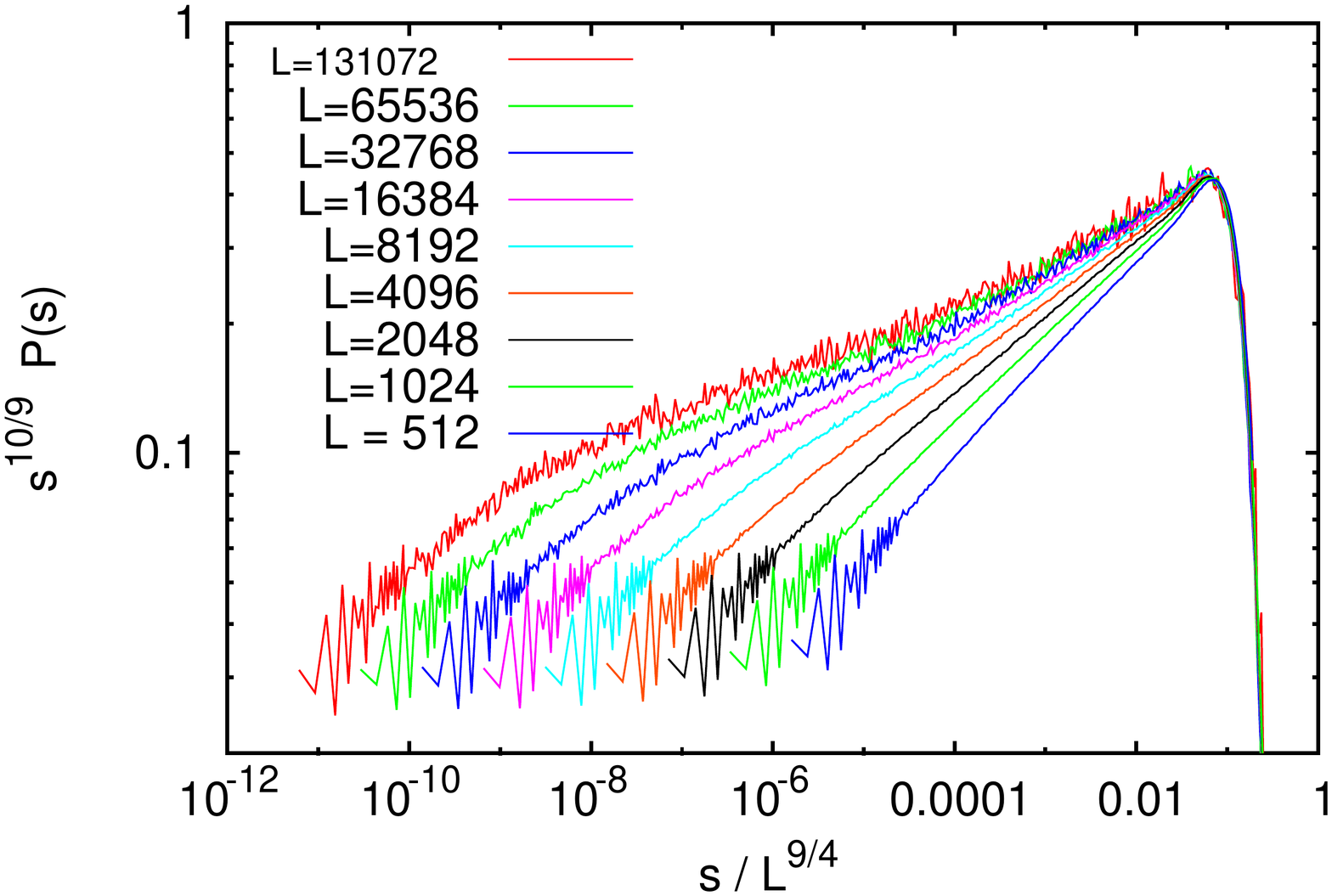}
 \vglue -5mm
 \includegraphics[scale=0.3]{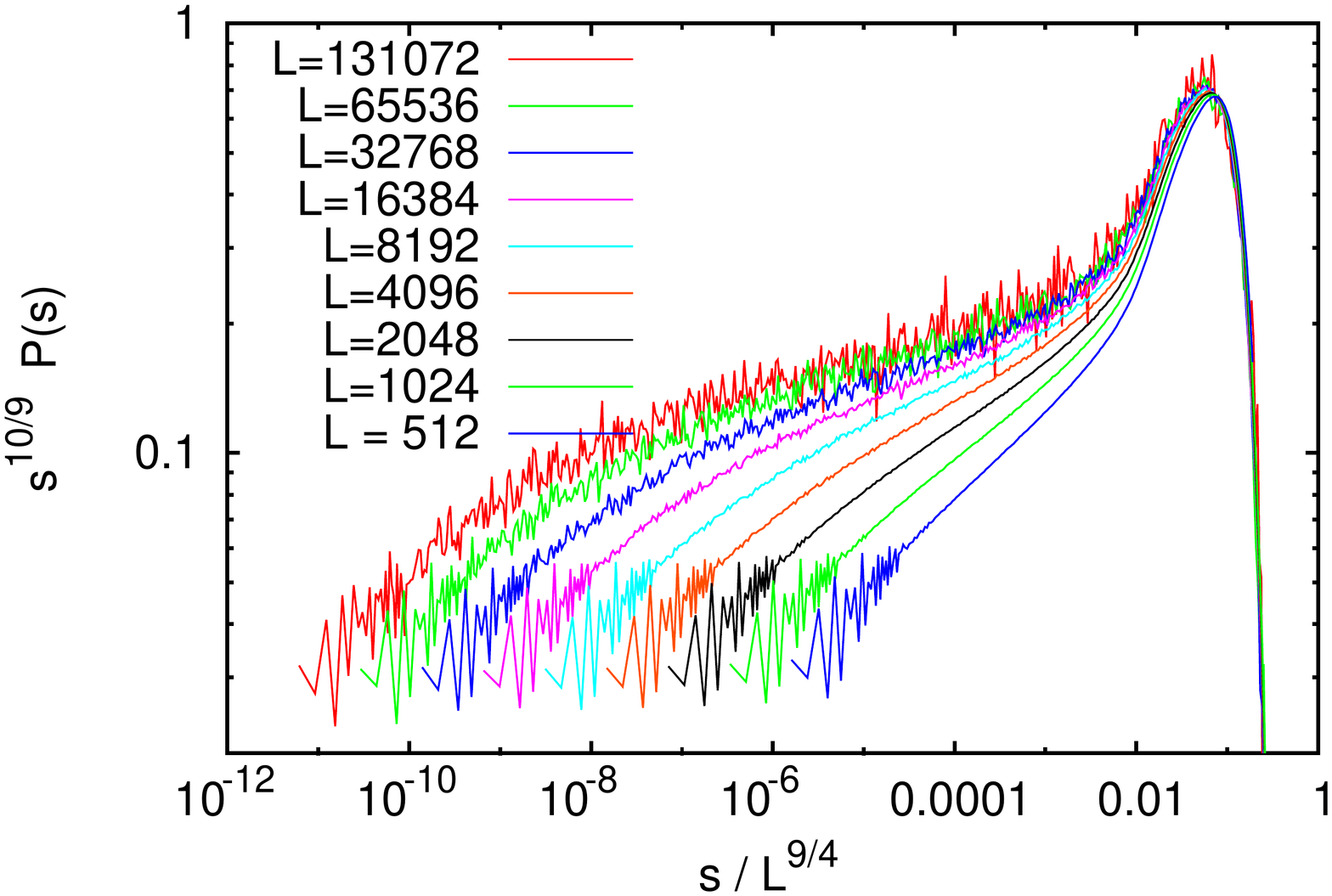}
 \vglue -5mm
 \par\end{centering}
 \caption{\label{Ps-bulk} (Color online) Log-log plot of $s^{10/9} P(s,L)$ against 
   $s/L^{9/4}$ for bulk driven open systems. In panel (a) the driving is uniformly 
   distributed over the entire region $[1,L]$. 
   while in panel (b) the center sites are driven. In both cases we see huge violations 
   of scaling, which are biggest for uniform driving.}
\end{figure}

As we just discussed, the statistical properties of the stationary state are identical to those
for boundary driving, thus we only have to discuss here the properties of avalanches.
We again expect the scaling law Eq.~(\ref{P_s}) to hold, with the same exponent $D$. But $\tau$
should now be different \cite{Christ:2004a}, because now $\langle s\rangle \sim L^2$. Assuming 
Eq.~(\ref{P_s}) we obtain 
\be
   \tau_{\rm bulk} = 2-2/D = 10/9.                          \label{tau-bulk}
\ee
This should hold for any open system with large $L$, where stress is added at sites far
from the boundaries. We shall discuss later the case where stress is added only at the center
region, but let us first discuss the case of uniform driving which was considered e.g. 
in \cite{Malthe,Christ:2004a}.

In this case,   the  stress is  not always added at sites far from the boundary, and corrections to scaling could be large, but the scaling could hold nevertheless. To test this, we plotted in 
Fig.~\ref{Ps-bulk}a $s^{10/9} P(s,L)$ against $s/L^{9/4}$. If scaling were perfect, this would
lead to a perfect data collapse. This is obviously not true (thus the fact that some avalanches
start near a boundary is not negligible), but it seems to become true in the limit $L\to\infty$.
In any case, both the position and the height of the peak scale in the right way. This means
that avalanches starting near the boundaries don't contribute to the peak, as we should 
expect.

More interesting is the case of center driving, because there we expect cleaner scaling than
in case of uniform driving -- although not as clean as for boundary driving. This is indeed
seen in Fig.~\ref{Ps-bulk}b, but the improvement over uniform driving is rather modest.

Both for center driving and for uniform driving we also measured moments of $T$. Ratios
$\langle T^k\rangle / \langle T^{k-1}\rangle$ showed now clean power laws for both $k=2$
and $k=3$, with exponent $z=1.426(5)$, consistent with  the previous estimate. But, in 
contrast to boundary driving, also $\langle T\rangle$ showed a power law with exponent
$1.19(1)$.

\begin{figure}
 \begin{centering}
 \includegraphics[scale=0.29]{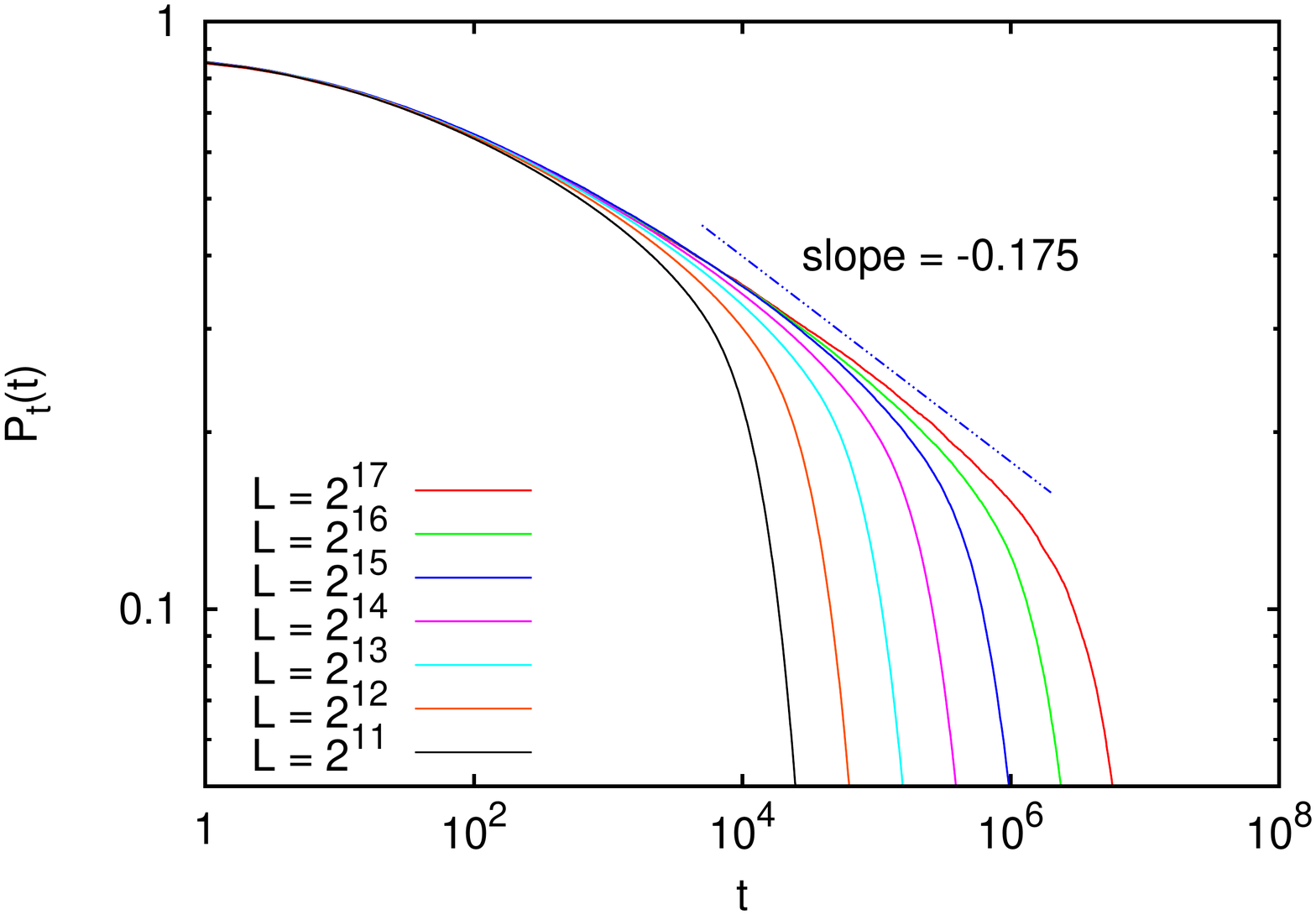}
 \vglue -6mm
 \includegraphics[scale=0.29]{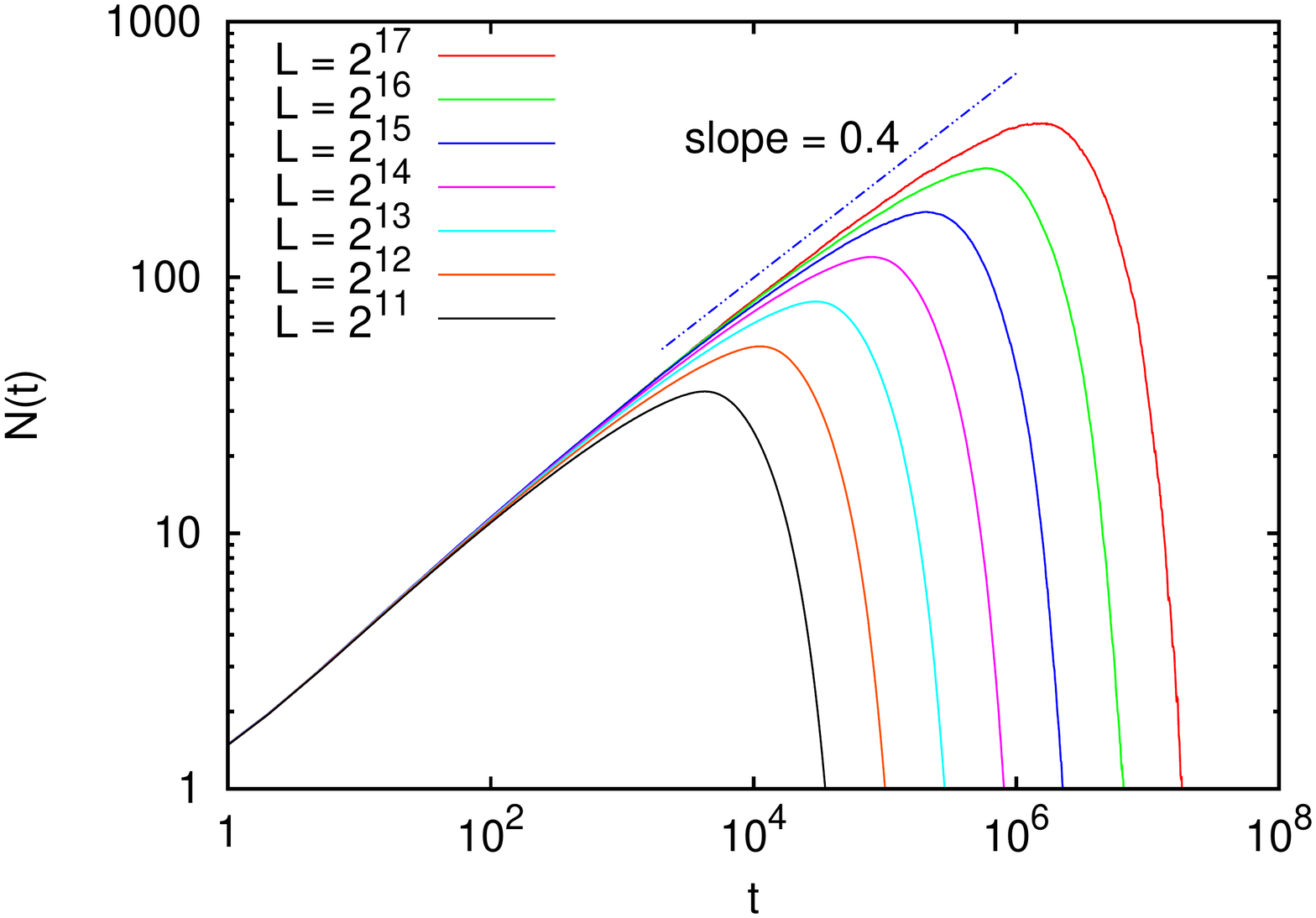}
 \vglue -6mm
 \includegraphics[scale=0.29]{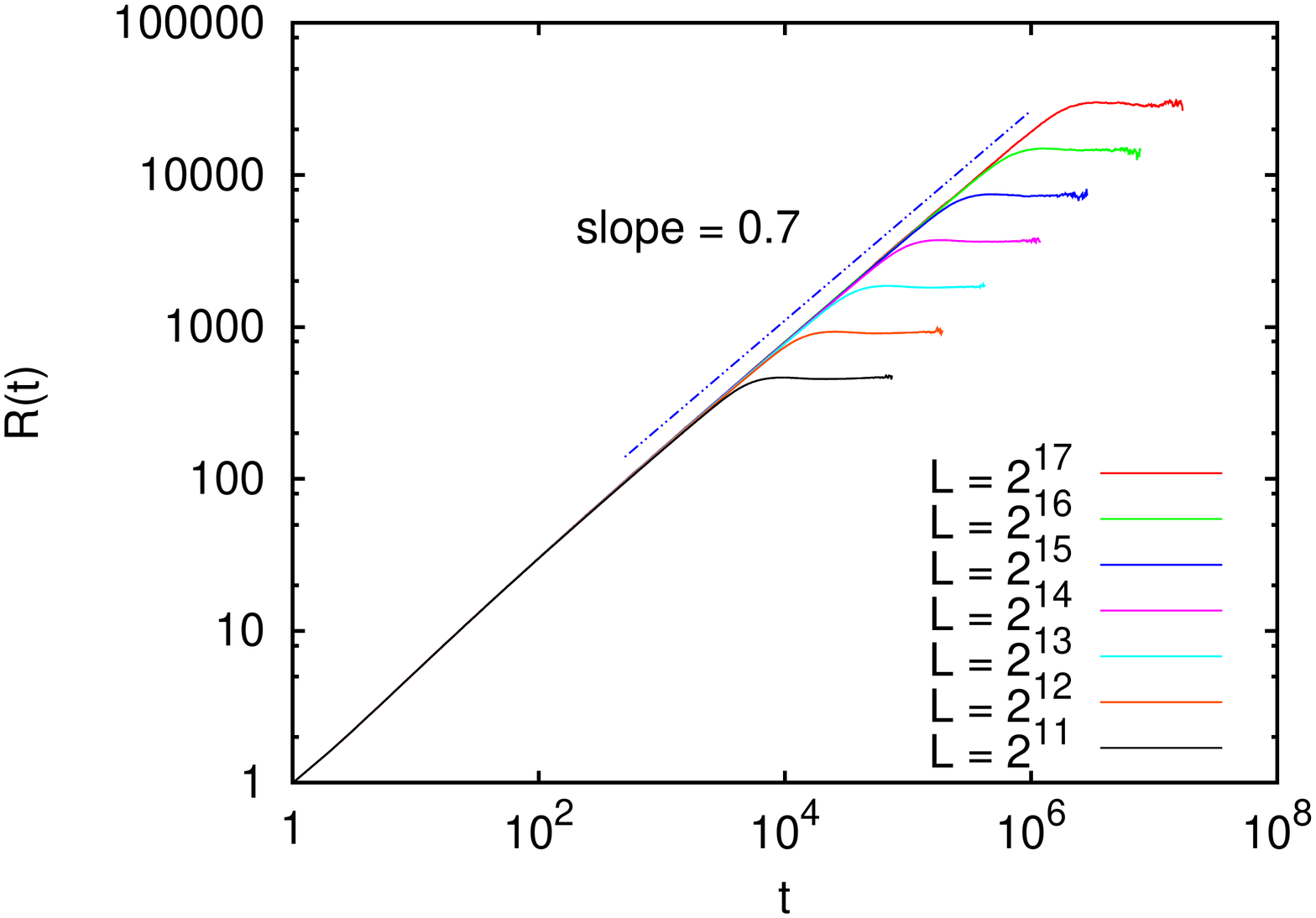}
 \vglue -6mm
 \par\end{centering}
 \caption{\label{center-PNR-t} (Color online) Log-log plot of $P_t(t,L)$ (panel a), $N(t,L)$ (panel b), and
   $R(t,L)$ (panel c) for center driven systems, similar to Fig.~\ref{open-PNR-t} for boundary
   driven avalanches. The straight lines indicate power laws in the central (scaling) region.}
 \end{figure}

Finally we show in Fig.~\ref{center-PNR-t} results for the temporal evolution of avalanches
similar to those shown in Fig.~\ref{open-PNR-t} for the boundary driven case. Again these 
data show much poorer scaling. The only curve that shows a convincing power law is that for 
$R(t)$ (which is now defined as the root mean square distance), and which clearly shows the 
same value for the dynamical exponent $z$. The exponents $\delta$ and $\eta$ are clearly
different from those for boundary driven systems. In view of the large deviations from scaling,
the estimates
\be
   \delta_{\rm bulk} = 7/40\;, \quad \eta_{\rm bulk} = 2/5
\ee
seem not very well justified by the data alone, but they  are consistent with  the fixed-energy
results presented in the last subsection.

\section {Mapping onto an interface model}     \label{surf}

Following \cite{Paczuski}, we define an interface without overhangs by 
identifying its height $H(i,t)$ with the number of topplings up to (and including)
time $t$. Alternatively \cite{Pruessner}, we could define another interface with 
height $h(i,t)$ such that $h(i,t)$ is the number of stress units {\it received} by 
topplings from its neighboring sites. The two heights are related by \cite{Pruessner}
\be
   h(i,t) = H(i-1,t)+H(i+1,t).   \label{Hh}
\ee
On the other hand, the evolution of $z_i$ can be written as
\bea
   z_{i,t} - z_{i,0} &=& h(i,t) - 2H(i,t) \nonumber \\
           &=& H(i-1,t)+H(i+1,t)- 2H(i,t) \nonumber \\
           &\equiv & \partial^2_i H(i,t),    \label{zH}
\eea
where $\partial^2_i$ is the discrete Laplacian. Finally, the number of topplings
at $(i,t)$ is just a (random) function of $z_{i,t}$,
\be
   H(i,t+1) - H(i,t) = \sigma(z_{i,t})   \label{Hsigma}
\ee
with 
\be
   \sigma(z) = 
     \begin{cases}
         0 & \quad \text{if }z \leq 1\\
         1 & \quad \text{if }z > 2\\
        0 \;\;\text{or} \;\;1 &\quad \text{with probability 1/2, if }z=2.
     \end{cases}
\ee
The last equations can be summarized as 
\be
   \partial_t H(i,t) = \partial^2_i H(i,t) +  \eta[i; H(i-1,t), H(i,t),H(i+1,t)]\label{H}
\ee
with $\eta(x,y) = \sigma(x)+x-y$. This looks formally like the qEW equation
\cite{Natter}
\be
   \partial_t H(i,t) = \partial^2_i H(i,t) +   \eta[i;H(i,t)]\label{qEW}
\ee
except that in the latter the noise $\eta$ depends only on $i$
and $H(i,t)$. In Eq. (\ref{H}), in contrast, there is explicit dependence on 
the stresses $z_{i,t}$. Note that in  Eq. (\ref{H}), unlike  the qEW
equation,  the  noise depends not only on a quenched variable at the site in question, but also on the curvature of the interface  .

Thus, the Oslo model is not exactly equivalent to the standard qEW model based on the interface $H$, 
where the noise correlations are
\be
\langle \eta(t,H)   \eta(t',H') \rangle=\delta_{t,t'}  \delta_{H,H'}.
\ee

Because $h$ is an explicit function of $H$ due 
to Eq. (\ref{Hh}), this is also true if $H$ is exchanged by $h$, in contrast
to what is claimed in \cite{Pruessner}. It is also not true that $H$ and $h$
have different scaling properties, as was claimed in \cite{Pruessner,Bona:2007}.
As should be clear from Eq. (\ref{Hh}), and as will be verified numerically, 
they show exactly the same scaling.

Let us finally discuss the interface interpretation of the original Oslo
model which is driven from its left border. As shown in \cite{Paczuski},
this corresponds to an interface that is prevented from being pinned by
pulling slowly up the leftmost point.  Consider the 
case where the interface is at its left end pulled by an amount $H(1,t) \gg L$, 
after which the pulling stops. In this case the left hand side of 
Eq.~(\ref{qEW}) vanishes, and it can be written as
\be
   H(i+1,t) = 2H(i,t) + \eta(i,H(i,t)) - H(i-1,t),
\ee
showing that  the evolution in the variable $i$, considered as a 'time'-evolution, is Markovian. Moreover, since the 
noise is assumed to be zero in average, the averaged heights satisfy 
simply $\partial^2_i \langle H(i,t)\rangle = 0$, showing that the height
profile is just linear.

\begin{figure}
\begin{centering}
\includegraphics[scale=0.3]{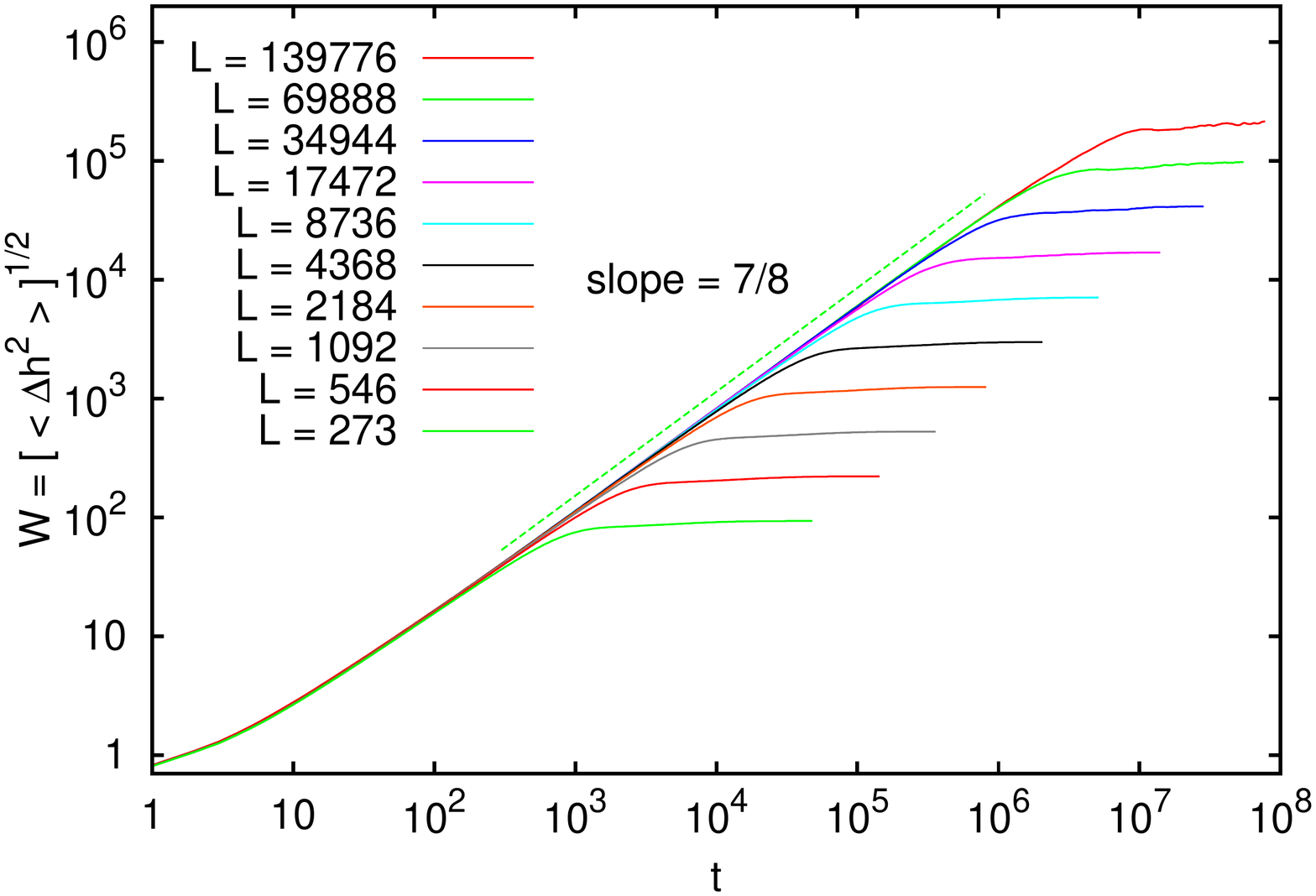}
\vglue -5mm
\includegraphics[scale=0.3]{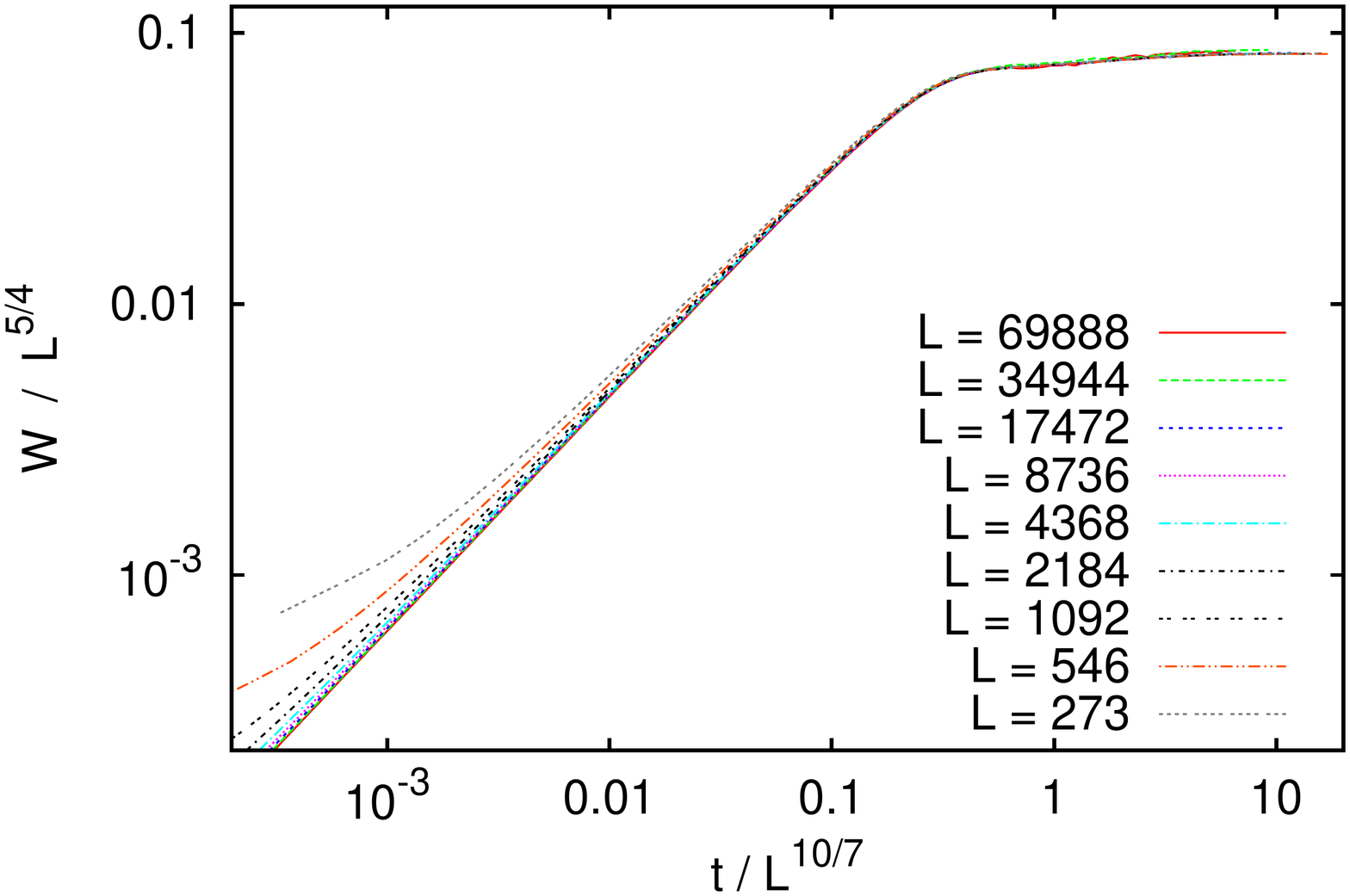}
\vglue -5mm
\par\end{centering}
\caption{\label{rough} (Color online) Roughness $W$ of interfaces constructed from 
the FES Oslo model, on lattices with density $\langle z\rangle = 473/273 = 1.732601$ and 
lattice sizes which are multiples of 273. Panel (a) shows the raw data, while panel (b)
shows a data collapse obtained by plotting $L^{-\alpha_{\rm qEW}}W$ against $t/L^z$. In this and 
the following two figures, averages are taken over all runs, those which are still active
at time $t$ and those which had already died. If we had excluded the latter from the 
averages, interfaces would be slightly less rough (roughness increases sharply immediately
before interfaces get pinned), but this would not affect their scaling (while it would 
affect the scaling of {\it average} height).}
\end{figure}

We have already discussed the main idea of the mapping on the qEW model, and we have already 
given the numerical value of the exponent, called usually $\theta$ in the qEW model, that
describes how the average interface velocity increases with the distance from the critical 
point. Here we shall discuss more relations of this type. An annoying problem in doing so
is the fact that equivalent exponents are given different names in the Oslo and qEW 
interpretations. We shall deal with it by adding a subscript ``qEW" to all qEW exponents,
e.g. $v \sim (F-F_c)^{\theta_{\rm qEW}}$ (notice that $z$ and $\nu$ are defined in the same
way in the Oslo and qEW models). In the following we shall discuss only the behavior 
exactly at the critical point, which we approximate to sufficient precision as 
$\langle z\rangle = 473/273 = 1.732601$. As we said in Sec.~2, there are two slightly 
different mappings from the Oslo model onto the interface model. In the first, $h(i,t)$ is 
the number of stress units received at site $i$ up to (and including) time $t$, while in 
the other $H(i,t)$ is the number of topplings.

Let us first discuss how the global roughness of an interface $h$ of base $L$ increases at 
the critical force with time. The roughness is defined as the square root of 
$W^2 = \langle (\Delta h(t))^2\rangle$, where brackets stand for an ensemble average and 
\be
   (\Delta h(t))^2 = \frac{1}{L}\sum_i h(i,t)^2 - [\frac{1}{L}\sum_i h(i,t)]^2
\ee

Results are shown in Fig.~\ref{rough}. They demonstrate the well known behavior 
\be
   W(t,L) = t^{\beta_{\rm qEW}} f(t/L^z) = L^{\alpha_{\rm qEW}} \tilde{f}(t/L^z)
\ee
with $\beta_{\rm qEW} = 1-\theta = 7/8$ and $\alpha_{\rm qEW}=z\beta_{\rm qEW} = 5/4$. The 
most precise previous estimates of the exponents were in \cite{Song-Kim} and \cite{Ferrero}. 
They are, respectively, $\beta_{\rm qEW} = 0.871(3)\; [0.872(3)],\; 
\alpha_{\rm qEW} = 1.250(3)\; [1.250(3)]$ and $z = 1.440(15)\; [1.443(7)]$, 
and are somewhat less precise than our estimates.

\begin{figure}
\begin{centering}
\includegraphics[scale=0.3]{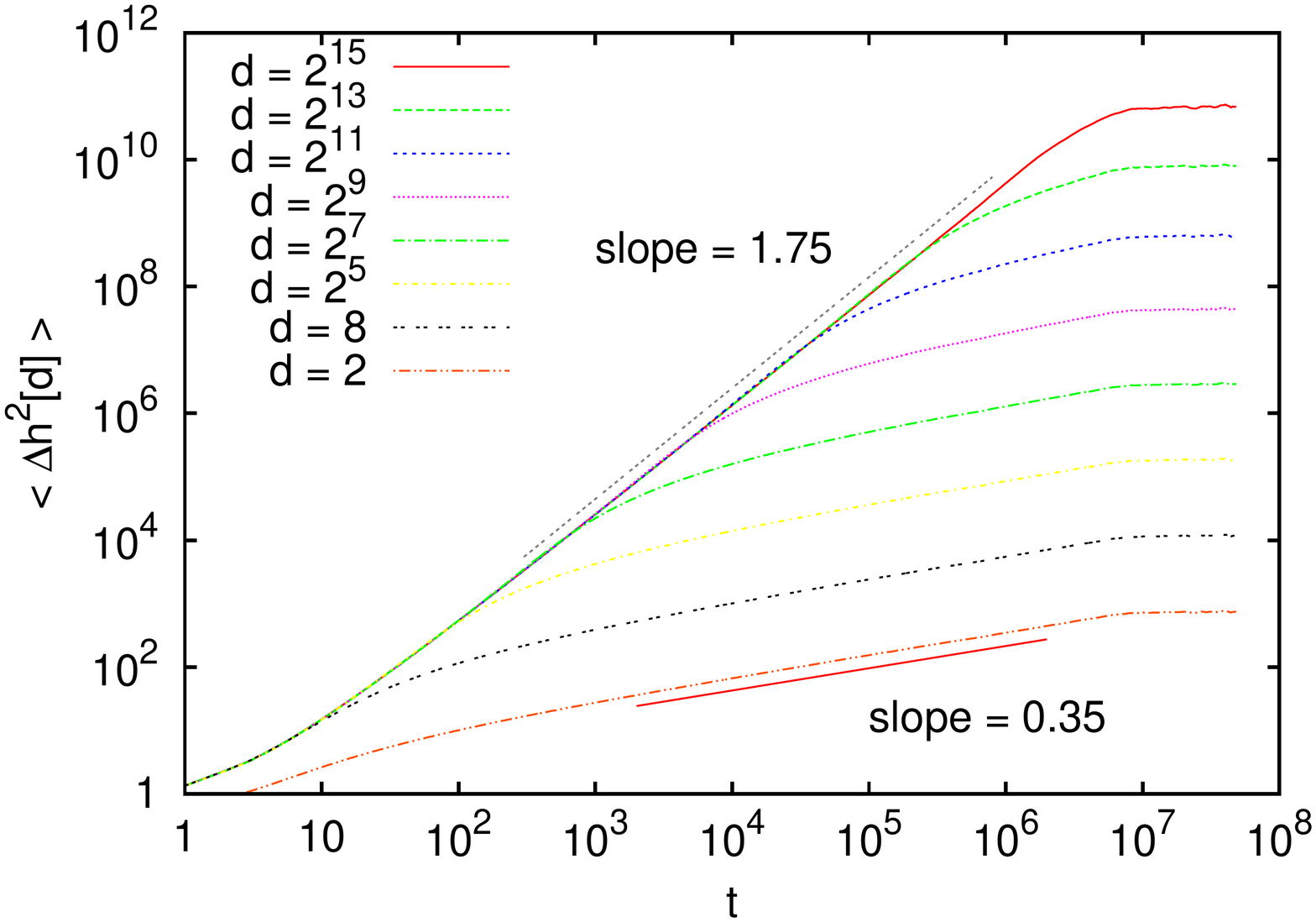}
\vglue -5mm
\includegraphics[scale=0.3]{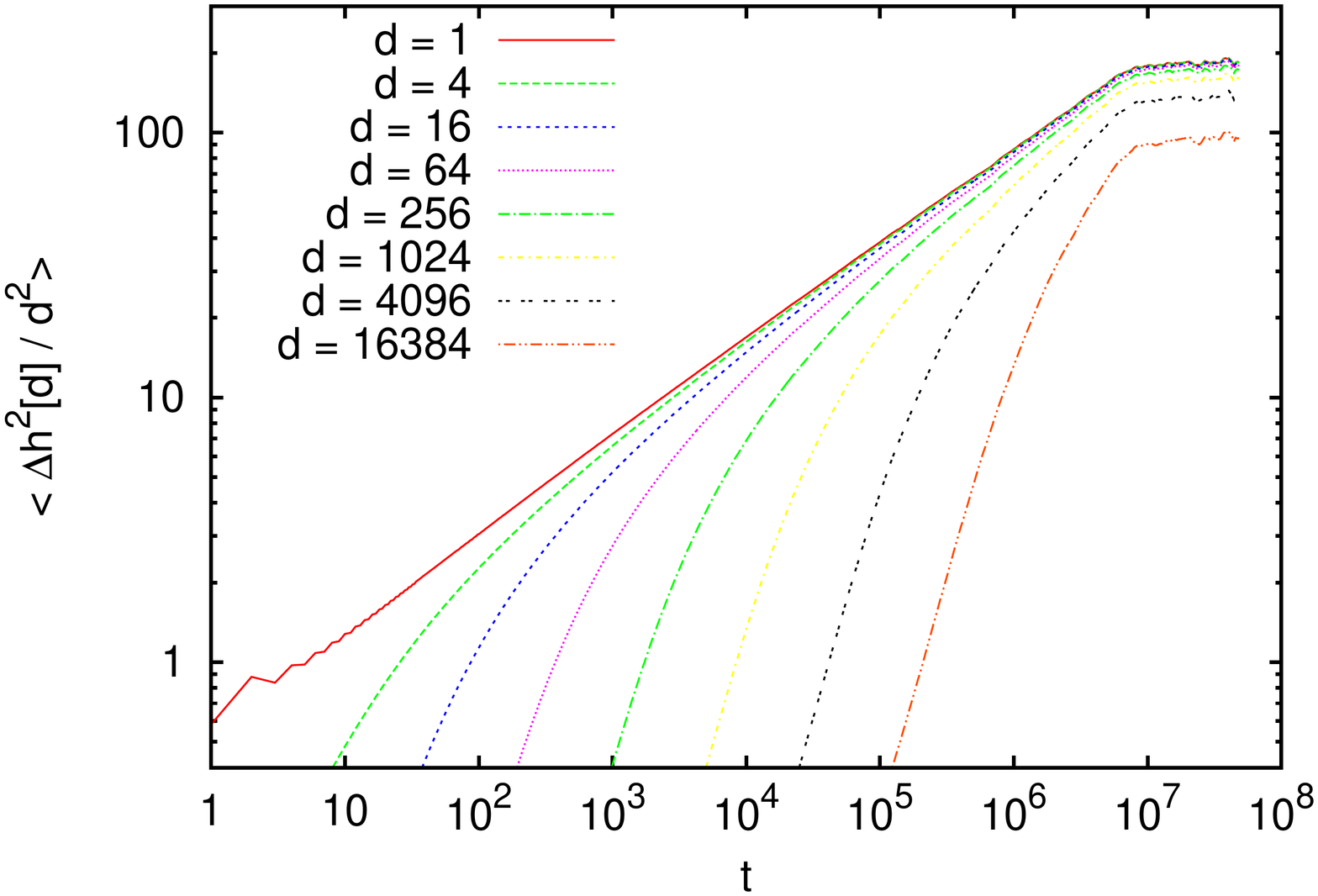}
\vglue -5mm
\includegraphics[scale=0.3]{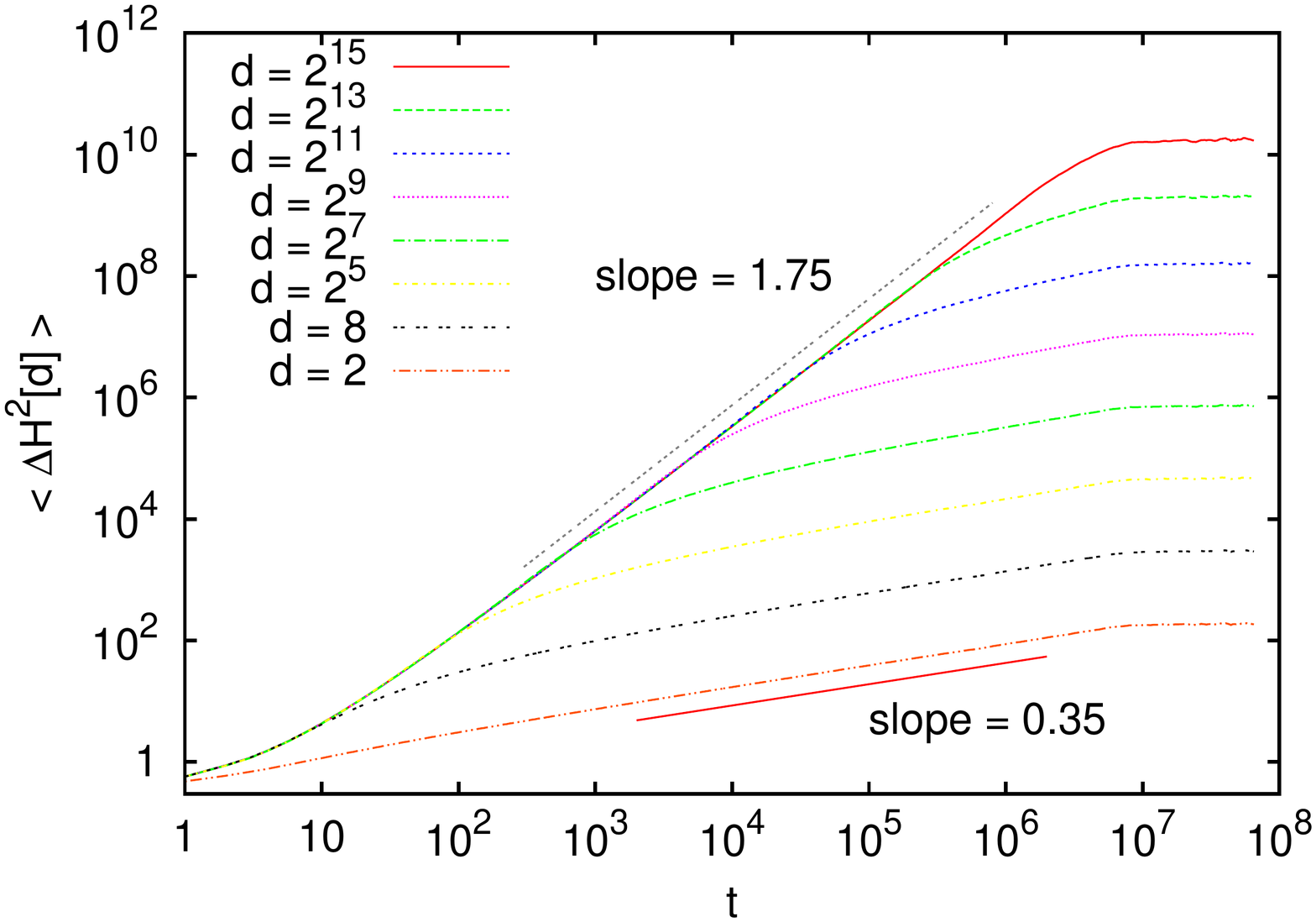}
\vglue -5mm
\par\end{centering}
\caption{\label{local-rough} (Color online) (a) Log-log plot of local squared roughnesses of interfaces 
constructed from the FES Oslo model, on lattices with density 
$\langle z\rangle = 473/273 = 1.732601$ and size $L=131040$. The straight lines indicate the 
anomalous roughening power laws. Panel (b) shows the same data, but divided by $d^2$. According
Eq.~(\ref{localrough}) one should see a data collapse for $d \ll t^{1/z} \ll L$. There is 
indeed a perfect collapse for $t\approx 10^6$, if the largest distances $d$ are discarded.
Panel (c), finally, is analogous to panel (a) but shows data for the alternative interface
definition that is based on number of {\it topplings} instead of {\it received} units of 
stress.}
\end{figure}

Since $\alpha_{\rm qEW}>1$ (critical qEW interfaces are ``superrough"), the local slope of a 
critical interface cannot be bounded \cite{Lesch-Tang}, and local roughnesses (i.e. 
roughnesses measured on a length scale $d \ll L$ must still increase for times at which 
an interface of total length $d$ would already be pinned. Thus the
roughness of a part of length $d$ of an interface of base $L\gg d$ satisfies ``anomalous
scaling" \cite{Sarma}
\be
   W(d;t,L) \sim
     \begin{cases}
         t^{\beta_{\rm qEW}} & \quad \text{for }\;  t \ll d^z\\
         d^{\alpha_{\rm loc}} L^{\alpha_{\rm qEW}-\alpha_{\rm loc}}  & \quad \text{for }\;  d \ll L \ll t^{1/z} \\
         d^{\alpha_{\rm loc}} t^{(\alpha_{\rm qEW}-\alpha_{\rm loc})/z} & \quad \text{for }\; d \ll t^{1/z} \ll L,
     \end{cases}
\label{localrough}
\ee
where $\alpha_{\rm loc}$ is a new exponent which for consistency must be $\leq 1$. In 
\cite{Sarma}, $\alpha_{\rm qEW}-\alpha_{\rm loc}$ was called $\kappa/2$, but we shall follow \cite{Lopez,Bona:2007}
and define 
\be
   \kappa = \frac{\alpha_{\rm qEW}-\alpha_{\rm loc}}{z},
\ee
so that the scaling for intermediate times reads now 
\be
    W  \sim d^{\alpha_{\rm loc}} t^\kappa \quad \text{for }\; d \ll t^{1/z} \ll L.  \label{intermed}
\ee

If the Oslo model is in the qEW universality class, the interfaces it is mapped onto must also
satisfy these scaling relations. In order to test this, we show local roughnesses of $h(i,t)$ in 
panels (a) and (b) of Fig.~\ref{local-rough}, while local roughnesses of $H(i,t)$ are shown in 
panel (c). All three panels in Fig.~\ref{local-rough} show data for $L=131040$ and $\langle z
\rangle = 473/273$. Panel (a) shows a log-log plot of the square of the roughness against $t$.
Each curve corresponds to a distance $d = 2^k,\; k=1,3,5,\ldots,15$ over which the roughness is 
computed. It is defined as 
$W^2(d;t,L) = \langle [h(i,t)-h(i+d,t)]^2\rangle$. We see clearly the two scaling laws 
$W(d;t,L) \sim t^{\beta_{\rm qEW}}$ for $t < d^z$ with $\beta_{\rm qEW}=0.875$, and Eq. (\ref{intermed}) with 
$\kappa = 7/40 \pm 0.01$. Notice that the latter holds for all $d$, even for $d=1$ (not shown).
Panel (b) shows the same data, but in order to see more clearly the value of $\alpha_{\rm loc}$ 
we divided the roughness by $d$. We see a perfect data collapse for $t \gg 1$ and $d \ll t^{1/z}$,
which implies that 
\be
   \alpha_{\rm loc} = 1,
\ee
in agreement with our estimate $\kappa = 7/40$.
Finally, panel (c) of Fig.~\ref{local-rough} is completely analogous to panel (a), except that
it shows roughnesses of the interface $H(i,t)$ which is obtained from the number of topplings
instead of the number of units received by a site. It clearly demonstrates that both interface 
definitions lead to the same scaling.

Our result $\alpha_{\rm loc} = 1$ is obtained in many different 1-d interface models
\cite{Sarma,Lopez}, but to our knowledge it cannot be derived analytically for the qEW model.
Previous numerical estimates \cite{Lopez-Rod} agree with it. On the other hand, our finding that 
the interfaces $H(i,t)$ and $h(i,t)$ satisfy the same scaling disagrees with claims made in 
\cite{Pruessner,Bona:2007}. In particular, we find for both the same exponent $\kappa$. This
value agrees with what is called ``A scaling" in \cite{Bona:2007}. In contrast to claims 
made in \cite{Bona:2007}, the Oslo model seems incompatible with what is called there ``B scaling".

\section{Conclusions}

Part of the motivation for the present work was the observation that natural critical state in
 the Oslo model is models are hyperuniform. On the one hand this suggests that transients in 
 simulations  could  be cut short by starting from very uniform initial configurations. 
This was indeed found, and it allowed the simulation of much bigger systems than previously possible.

On the other hand it suggested that -- if the hyperuniformity is strong enough -- the conserved 
field in sandpile models can be considered as rigid and non-fluctuating, in which case these
models would be in the same universality class as directed percolation.  We find that this is not so (see also \cite{Dick:2015}). Instead,  we find compelling evidence  that the 
1-dimensional Oslo model is in the same universality class as the qEW (or linear interface) 
model. This had been a long-standing conjecture, but it had been repeatedly doubted due to 
contradictory numerical results. One main reason for these numerical problems was precisely
that hyperuniformity had not been taken into account.  In a forthcoming paper \cite{forthcoming}, 
we will  discuss
some other Oslo  type models with directed particle transfer rules  on two-dimensional lattices, which  turn out to  correspond to an Edwards-Wilkinson interface model {\it with annealed noise}.

An unexpected outcome of this work is that the vastly improved simulations (made possible in
part by judicious choices of initial conditions) suggest that the critical exponents of the 1-d Oslo model (and, more 
importantly, also the 1-d qEW model) are simple rational numbers. For some exponents this had
already been conjectured before, but not (to our knowledge) for the dynamical and  
hyperuniformity exponents, and for the exponent $\sigma$ (see Eq. (\ref{sig}) describing
the stress profile in the case of open boundaries.  Also, these exponents fall outside the infinite series of discrete rational exponents recently found for 1-d stochastic models \cite{schutz}.  Of course, in the same study, well-behaved models where the dynamical exponent  is the golden mean, i.e. an irrational value  have also been discussed. So, while the critical exponents do not {\em have to be } rational, we note that most soluble models so far have found rational critical exponents.
Showing that these conjectured values are actually correct  remains a challenge.

Our finding that the 1-d Oslo model is in the qEW universality class suggests of 
course that the same could be true for other  stochastic sandpile models, and for SOC models
with conserved fields in higher dimensions.  This does not invalidate our earlier argument about instability of the Manna model fixed point under suitable perturbation to DP.  All this says is that adding this kind of perturbation (say  adding a small  probability of $z_c$ being set to $3$, $z_c =2$ is the deterministic sandpile) does not constitute such a  relevant perturbation. One possibility is that  if  in the toppling process, we add randomness also in where the transferred particles may go ( as in the original Manna model), then the critical behavior of the model may change. This can occur, because with such randomness present, there is a much larger variation of the number of topplings at different sites in an avalanche.
In fact,  other stochastic 1-d sandpile models (like the Maslov-Zhang model \cite{Maslov-Zhang} or
the continuous Manna model \cite{BBBMH}) appear to have critical exponents different from  to the Oslo model studied here. 
Further studies are needed to clarify this point.

\begin{acknowledgments}
P.G. thanks the Leverhulme trust for financial support, and the University of Aberdeen --
where part of this work was done -- for a most pleasant stay.   DD’s research is supported partially 
by the Indian Department of Science and Technology via grant DST-SR/S2/JCB-24/2005.

\end{acknowledgments}


\begin{thebibliography}{99}
\bibitem{footnote3} In the `random subtraction' version of the fixed-energy BTW model in \cite{Chessa}, e.g., 
   simulations were done at slightly subcritical densities (so that avalanches died), and new avalanches were 
   triggered by moving one of the sand grains to an random arbitrarily far away site.   It is not clear that the steady state produced this way is the same  as what would be produced if the grain was moved only to nearest neighbor site.
\bibitem{footnote4} The latter are the most ``natural" b.c.'s if we concentrate on the $z_i$. 
   Although a different b.c. at the right boundary was used in \cite{Oslo}, we call Eq.~(\ref{toppl})
   `the original Oslo model', since the difference is irrelevant for scaling.
\end{thebibliography}
\end{document}